\documentclass[aps,twocolumn,superscriptaddress,showpacs]{revtex4-1} 
\usepackage{verbatim}
\usepackage{multirow}
\usepackage{graphicx}

\begin{document}
\bibliographystyle{apsrev}
\title{ Partial-wave analysis of $\overline K N$ scattering reactions}
\author{H. Zhang}
\author{J. Tulpan}
\author{M. Shrestha}
\author{D.~M.~Manley}
\affiliation{Department of Physics, Kent State University, Kent, OH 44242-0001}

\begin{abstract}
We investigate the two-body reactions $\overline K N\rightarrow \overline K N$, $\overline K N\rightarrow \pi \Lambda$, and $\overline K N\rightarrow \pi \Sigma$ via single-energy partial-wave analyses in the center-of-mass (c.m.) energy range 1480 to 2100 MeV. The partial-wave amplitudes for these reactions thus extracted were constrained by a multichannel energy-dependent model satisfying unitarity of the partial-wave $S$-matrix. We obtain excellent predictions of differential cross sections, polarizations, polarized cross sections, and cross sections for these reactions from a global energy-dependent solution.

\end{abstract}
\maketitle
\section{Introduction and Motivation} 
 Information on hyperon resonances is generally not as extensive as for nucleon resonances. The study of $\overline K N\rightarrow \overline K N$, $\overline K N\rightarrow \pi \Lambda$, and $\overline K N\rightarrow \pi \Sigma$ could lead to the better understanding of $\Lambda^*$s and $\Sigma^*$s.

     Most previous partial-wave analyses (PWAs) of $\overline K N\rightarrow \overline K N$, $\overline K N\rightarrow \pi \Lambda$, and $\overline K N\rightarrow \pi \Sigma$ {\cite{Armenteros1969, Conforto1971, Horn1975_1, Hemingway1975, Baillon1975,  Gopal1977}}, were based on the assumption that  partial-wave amplitudes could be represented by a simple sum of resonant and  background terms. Such an assumption violates  unitarity of the partial-wave $S$-matrix. In this work, we report on our investigation of the reactions $K^-p\rightarrow K^-p$ and $K^-p\rightarrow \overline K^0n$, $K^-p\rightarrow \pi^0\Lambda$, and $K^-p\rightarrow\pi^+\Sigma^-$, $K^-p\rightarrow\pi^0\Sigma^0$, and $K^-p\rightarrow\pi^-\Sigma^+$ via single-energy analyses and a subsequent energy-dependent analysis.  All available differential cross section, polarization, polarized cross section, and cross-section data up to a maximum c.m.\ energy of about 2.1 GeV were fitted. In order to ensure that our amplitudes had a relatively smooth variation with energy, we introduced several constraints that will be described in detail below. The determination of resonance parameters from our subsequent energy-dependent analysis is discussed in Ref. \cite{manoj2013}.

   \section{Formalism and Fitting Procedures}
        Here, we summarize the formalism for the single-energy partial-wave analyses. 
        The differential cross section $\rm d\sigma /\rm d\Omega$ and polarization $P$ for unpolarized scattering of spin-0 mesons off spin-$\frac12$ nucleons are given by \cite{bransden73}
               \begin{equation}
               \frac{{\rm d}\sigma}{{\rm d}\Omega} = {\lambdabar}^2(|f|^2+|g|^2)~,
                \end{equation}
                \begin{equation}
                P\frac{{\rm d}\sigma}{{\rm d}\Omega} =2{\lambdabar}^2\rm Im(fg^\ast)~,
                \end{equation}
        where $\lambdabar = {\hbar}/{k}$,
         with $k$ the magnitude of c.m.\ momentum of the incoming meson.
         Here, $f = f(W,\theta)$ and $g = g(W,\theta)$ are the usual spin-non-flip and spin-flip amplitudes at c.m.\ energy $W $ and meson c.m.\ scattering angle $\theta$. In terms of partial waves, $f$ and $g$ can be expanded as
       
       \begin{equation}
        f(W,\theta) = \sum_{l=0}^{\infty} [(l+1)T_{l+} + lT_{l-}]P_l(\cos\theta)~,
        \end{equation}
        \begin{equation}
        g(W,\theta) = \sum_{l=1}^{\infty} [T_{l+} - T_{l-}]P_l^1(\cos\theta)~,
        \end{equation}
        where $l$ is the initial orbital angular momentum, $P_l(\cos\theta)$ is a Legendre polynomial and $P_l^1(\cos\theta) = \sin\theta \cdot {\rm d} P_l(\cos\theta)/{\rm d}(\cos\theta$). The total angular momentum for the amplitude $T_{l+}$ is $J=l+\frac12$, while that for the amplitude $T_{l-}$ is $J=l-\frac12$.
              For the initial $\overline K N$ system, we have $I = 0$ or $I =1$ so that the amplitudes $T_{l\pm}$ can be expanded in terms of  isospin amplitudes as 
         \begin{equation}
               T_{l\pm} = C_{0}T^{0}_{l\pm} + C_{1}T^{1}_{l\pm}~,
       \end{equation}
       \newline
       where $T^I_{l\pm}$ are partial-wave amplitudes 
        with isospin $I$  and total angular momentum $J = l\pm\frac12$ with $C_I$ the appropriate isospin Clebsch-Gordon coefficients for a given reaction.
       For $K^- p \rightarrow K^- p$, for example, we have $C_{0} = {\frac12}$ and $C_{1}={\frac12}$.
\newline

        The total $K^- p$ cross section is given by $\sigma_{\rm total} = 4\pi {\lambdabar}^2 {\rm Im} f(W, 0)$, or
\begin{equation}\label{eq:Sigma_total}
\sigma_{\rm total} = 4\pi \lambdabar^2 \sum_{l=0}^{\infty}[(l+1) {\rm Im}T_{l^+} + l{ \rm Im}T_{l^-}],
\end{equation}
where here the $T_{l\pm}$ are partial-wave amplitudes for elastic $\overline KN$ scattering.

  The integrated cross section  for a particular two-body reaction is
\begin{equation}
\sigma=4\pi {\lambdabar}^2 \sum_{l=0}^{\infty} [(l+1) |T_{l^+}|^2+l |T_{l^-}|^2] .
\end{equation}

  Tables I, II, and III summarize  the available quantity and types of data in each energy bin for the three reactions $\overline K N\rightarrow \overline K N$, $\overline K N\rightarrow \pi \Lambda$, and $\overline K N\rightarrow \pi \Sigma$, respectively.   
   Single-energy fits were performed separately for (i)  $K^-p\rightarrow K^-p$ and $K^-p\rightarrow \overline K^0n$, for (ii) $K^-p\rightarrow \pi^0\Lambda$, and for (iii) $K^-p\rightarrow\pi^+\Sigma^-$, $K^-p\rightarrow\pi^0\Sigma^0$, and $K^-p\rightarrow\pi^-\Sigma^+$. In each case the available data were analyzed in  c.m.\ energy bins of width 20 MeV. This choice of bin width was appropriate because the data for smaller widths had unacceptably low  statistics and for larger  widths, some amplitudes varied too much over the energy spread of the bin.

\begin{table*}[htbp]
\caption{Summary of database for $\overline{K} N \rightarrow \overline{K} N$. Column 1 lists the central energy $W_0$ of each energy bin, columns 2 and 3 list the number of differential cross-section data points in each bin for $K^-p\rightarrow K^-p$ and $K^-p\rightarrow \overline K^0n$, respectively, column 4 lists the number of polarization data points for $K^-p\rightarrow K^-p$ in each bin, column 5 lists the number of polarized cross-section data points for $K^-p\rightarrow K^-p$ in each bin, column 6 lists the number of integrated cross-section data points for $K^-p\rightarrow K^-p$ and $K^-p\rightarrow \overline K^0n$ in each bin, column 7 lists the number of $K^-p$ total cross-section data points in each bin, column 8 lists the total number of data points for all kinds of data in each bin, and column 9 lists the references for the measurements referred to in columns 2-5.}
\begin{center}
\begin{ruledtabular}
\begin{tabular}{ccccccccc}
\multicolumn{1}{c}{$W_{0}$} & \multicolumn{1}{c}{{${d\sigma}/{d\Omega}$ } } & \multicolumn{1}{c}{${d\sigma}/{d\Omega}$ { } }& \multicolumn{1}{c}{{\mbox{$P$}}}& \multicolumn{1}{c}{{\mbox{$P$}${d\sigma}/{d\Omega}$ }} & \multicolumn{1}{c}{\multirow{2}{*}{$\sigma$}}& \multicolumn{1}{c}{\multirow{2}{*}{$\sigma_{\rm total}$}} & \multicolumn{1}{c}{ Total} & \multicolumn{1}{c}{\multirow{2}{*}{ References}} \\ 
\multicolumn{1}{c}{\rm (MeV)} & ($K^- p$) &($\overline{K}^0 n$) & ($K^- p$)& ($K^- p$)& & & \multicolumn{1}{c}{ No.} &  \multicolumn{1}{c}{} \\

\hline
1480	&	 57&	 24&	&		&15	&	5&   96	&	 \cite{Mast1976}\\
1500	&	114&	120&	&&		23	&	8&  257		& \cite{Mast1976}\\
1520	&	100&	100&	&&		28	&  10&  228		& \cite{Mast1976}\\
1540	&	178&	120&	&&		25	&	8&  323		& \cite{Armenteros1970, Mast1976}\\
1560&	117&	100&	&&		14	&	5&  231		& \cite{Armenteros1970, Alston1978_1, CrystalBall2005}\\
1580&	78&	112&	&	&	10	&	3&   200			& \cite{Armenteros1970, Alston1978_1, CrystalBall2005}\\
1600&	79&	116&	&	&	14	&	6&   209			& \cite{Armenteros1970, Alston1978_1, CrystalBall2005}\\
1620&	147&	120&	&&		14	&	6&  281		& \cite{Armenteros1970, Adams1975, Alston1978_1, CrystalBall2005}\\
1640	&	113&	148&	&&		14	&	6&  275		& \cite{Armenteros1970, Adams1975, Alston1978_1, CrystalBall2005}\\
1660	&	149&	132&	&&		19	&	6&  300		& \cite{Armenteros1970, Adams1975, Alston1978_1, CrystalBall2005}\\
1680	&	194&	210&	&&		30	&	9&  434		& \cite{Armenteros1968, Armenteros1970, Conforto1971, Adams1975, Alston1978_1, CrystalBall2005}\\
1700	&	150&	112&	&&		11	&	6&  283		& \cite{Armenteros1968, Armenteros1970, Conforto1971, Adams1975, Alston1978_1}\\
1720	&	150&	150&	&&		20	&	6&  320		& \cite{Armenteros1968, Conforto1971, Adams1975, Alston1978_1}\\
1740	&	216&	241&	&27&		29	&	6&    513		 & \cite{Armenteros1968, Albrow1971, Conforto1971, Jones1975, Adams1975, Alston1978_1}\\
1760&	176&	193&	&26&		25	&	4&    420		 & \cite{Armenteros1968, Albrow1971, Conforto1971, Jones1975, Adams1975, Alston1978_1}\\
1780	&	185&	196&	&27&		34	&	9&    442		 & \cite{Armenteros1968, Andersson1970, Conforto1971, Jones1975, Conforto1976, Alston1978_1}\\
1800&	132&	79&	&52&		18	&	4& 281			& \cite{Armenteros1968, Conforto1971, Jones1975, Conforto1976}\\
1820&	146&	60&	&27&		18	&	5& 251			& \cite{Armenteros1968, Albrow1971, Conforto1971, Conforto1976}\\
1840&	185&	80&	&27&		23	&	9& 315			& \cite{Armenteros1968, Andersson1970, Conforto1971, Conforto1976}\\
1860&	227&	100&	&30&		22	&	4&    379		 & \cite{Armenteros1968, Andersson1970, Conforto1971, Griselin1975, Conforto1976}\\
1880&	266&	120&	&28&		21	&	6&    435		 & \cite{Armenteros1968, Albrow1971, Conforto1971, Griselin1975, Conforto1976}\\
1900	&	175&	60&	&56&		18	&	6& 309			& \cite{Armenteros1968, Andersson1970, Albrow1971, Conforto1971, Griselin1975, Conforto1976}\\
1920	&	146&	60&	&27&		17	&	3& 250			& \cite{Andersson1970, Griselin1975, Conforto1976}\\
1940&	110&	80&	&30&		18	&	5& 238			& \cite{Albrow1971, Griselin1975, Conforto1976}\\
1960&	64&	40&	23&&		14	&	4&  141		& \cite{Daum1968, Griselin1975, Conforto1976}\\
1980&	34&	20&	&&		11	&	3&    65		& \cite{Griselin1975, Abe1975}\\
2000&	23&	20&	23&&		9	&	3&   75		& \cite{Daum1968, Abe1975}\\
2020	&	23&	&	23&&		12	&	5&    58		& \cite{Daum1968}\\
2040	&	54&	&	22&&		10	&	4&    86		& \cite{Daum1968, Abe1975}\\
2060&	23&	&	23&&		10	&	3&    56		& \cite{Daum1968}\\
2080	&	22&	&	22&&		9	&	3& 53		& \cite{Daum1968}\\
2100&	53&	&	22&&		7	&	2& 83		& \cite{Daum1968, Abe1975}\\
2120	&	46&	&	23&46&		12	&	5&    104		& \cite{Daum1968, Andersson1970}\\
2140&	23&	&	23&&		5	&	2& 51		& \cite{Daum1968}\\
2160&	32&	&	&&		14	&	5&  46		& \cite{Abe1975}\\
\end{tabular}
\label{Table:KN}
\end{ruledtabular}
\end{center}
\end{table*}

\begin{table*}[htbp]
\caption{Summary of database for $\overline{K} N \rightarrow \pi \Lambda$. Column 1 lists the central energy $W_0$ of each energy bin, column 2 lists the number of differential cross-section data points in each bin for $K^-p\rightarrow \pi^0\Lambda$, column 3 lists the number of polarization data points in each bin, column 4 lists the number of polarized cross-section data points in each bin, column 5 lists the number of integrated cross-section data points in each bin, column 6 lists the total number of data points for all kinds of data in each bin, and column 7 lists the references for the measurements referred to in columns 2-4.}
\begin{center}
\begin{ruledtabular}
\begin{tabular}{ccccccc}
\multicolumn{1}{c}{$W_{0}$} & \multicolumn{1}{c}{\multirow{2}{*}{${d\sigma}/{d\Omega}$}} & \multicolumn{1}{c}{\multirow{2}{*}{\mbox{$P$}}} & \multicolumn{1}{c}{\multirow{2}{*}{\mbox{{$P$}${d\sigma}/{d\Omega}$}}} & \multicolumn{1}{c}{\multirow{2}{*}{$\sigma$}} & \multicolumn{1}{c}{ Total}  & \multicolumn{1}{c}{\multirow{2}{*}{ References}} \\
\multicolumn{1}{c}{\rm (MeV)} & & & & & \multicolumn{1}{c}{ No.} & \multicolumn{1}{c}{} \\
\hline
1480	&		&  &		&	4	&	4			&		 \cite{Baldini1988}\\
1500&		&  &		&	8	&	8			&		 \cite{Baldini1988}\\
1520	&		&  &		&	9	&	9			&		 \cite{Baldini1988}\\
1540&	40	&  &	16	&	7	&	63			&		 \cite{Armenteros1970}\\
1560&	76	&16&	26	&	4	&	122			&	 \cite{Armenteros1970, CrystalBall2005}\\
1580&	56	&16&	19	&	2	&	93			&	 \cite{Armenteros1970, CrystalBall2005}\\
1600	&	56	&16&	20	&	5	&	97			&	 \cite{Armenteros1970, CrystalBall2005}\\
1620	&	56	&16&	20	&	4	&	96			&	 \cite{Armenteros1970, CrystalBall2005}\\
1640&	81	&32&	20	&	6	&	139			&	 \cite{Armenteros1970, Baxter1973, CrystalBall2005}\\
1660	&	74	&16&	19	&	9	&	118			&	 \cite{Armenteros1970, Baxter1973, CrystalBall2005}\\
1680	&	114	&22&	28	&	13	&	177			&	 \cite{Armenteros1968, Armenteros1970, Baxter1973, CrystalBall2005}\\
1700	&	58	&7 &		8	&	13	&	86				&\cite{Armenteros1968, Armenteros1970, Baxter1973}\\
1720	&	101	&27&		&	9	&	137			&	 \cite{Armenteros1968, Baxter1973, Jones1975}\\
1740&	185	&65&		&	12	&	262			&	 \cite{Armenteros1968, Baxter1973, Jones1975}\\
1760	&	138	&54&		&	11	&	203			&	 \cite{Armenteros1968, Baxter1973, Jones1975}\\
1780&	160	&88&		&	10	&	258			&	 \cite{Armenteros1968, Jones1975, Conforto1976}\\
1800	&	80	&46&		&	4	&	130			&	 \cite{Armenteros1968, Jones1975, Conforto1976}\\
1820	&	60	&35&		&	4	&	99			&	 \cite{Armenteros1968, Conforto1976}\\
1840&	80	&36&		&	5	&	121			&	 \cite{Armenteros1968, Conforto1976}\\
1860&	100	&26&		&	7	&	133			&	 \cite{Armenteros1968, Griselin1975, Conforto1976}\\
1880	&	120	&44&		&	7	&	171			&	 \cite{Armenteros1968, Griselin1975, Conforto1976}\\
1900	&	60	&18&		&	5	&	83			&	 \cite{Armenteros1968, Griselin1975, Conforto1976}\\
1920	&	80	&21&		&	7	&	108			&	 \cite{Berthon1970, Griselin1975, Conforto1976}\\
1940	&	100	&23&		&	8	&	131			&	 \cite{Berthon1970, Griselin1975, Conforto1976}\\
1960	&	60	&24&		&	5	&	89			&	 \cite{Berthon1970, Griselin1975, Conforto1976}\\
1980&	40	&6&		&	5	&	51	&			 \cite{Berthon1970, Griselin1975}\\
2000	&	40	&13&	&	4	&	57	&			 \cite{Berthon1970, Griselin1975}\\
2020	&	20	&9&		&	4	&	33		&	 \cite{Berthon1970}\\
2040	&	20	&8&		&	2	&	30		&	 \cite{Berthon1970}\\
2060&	30	&6&		&	3	&	39		&	 \cite{Berthon1970, London1975}\\
2080&	40	&8&		&	3	&	51		&	 \cite{Berthon1970, London1975}\\
2100&	30	&&		&	2	&	32		&	 \cite{London1975}\\
2120	&	70	&11&	&	4	&	85		&	 \cite{Berthon1970, London1975}\\
2140&		& &		&		&	0	   	 &	NA\\
2160	&	40	&9&		&	2	&	51		&	 \cite{Berthon1970}\\
\end{tabular}
\label{Table:Lambda}
\end{ruledtabular}
\end{center}
\end{table*}

\begin{table*}[htbp]
\caption{Summary of database for $\overline{K} N \rightarrow \pi \Sigma$. Column 1 lists the central energy $W_0$ of each energy bin, columns 2, 3, and 4 list the number of differential cross-section data points in each bin for $K^-p\rightarrow \pi^+\Sigma^-$, $K^-p\rightarrow \pi^0\Sigma^0$, and $K^-p\rightarrow \pi^-\Sigma^+$, respectively, column 5 and 6 list the number of polarization data points for $K^-p\rightarrow \pi^0\Sigma^0$ and $K^-p\rightarrow \pi^-\Sigma^+$, respectively, in each bin, column 7 and 8 list the number of polarized cross-section data points for $K^-p\rightarrow \pi^0\Sigma^0$ and $K^-p\rightarrow \pi^-\Sigma^+$, respectively, in each bin, column 9 lists the number of integrated cross-section data points for $K^-p\rightarrow \pi^+\Sigma^-$, $K^-p\rightarrow \pi^0\Sigma^0$, and $K^-p\rightarrow \pi^-\Sigma^+$ in each bin, column 10 lists the total number of data points for all kinds of data in each bin, and column 11 lists the references for the measurements referred to in columns 2-8.}
\begin{center}
\begin{ruledtabular}
\begin{tabular}{ccccccccccc}
\multicolumn{1}{c}{$W_{0}$} & \multicolumn{1}{c}{{${d\sigma}/{d\Omega}$}} & \multicolumn{1}{c}{{${d\sigma}/{d\Omega}$}} & \multicolumn{1}{c}{{${d\sigma}/{d\Omega}$}} & \multicolumn{1}{c}{{\mbox{$P$}}}& \multicolumn{1}{c}{{\mbox{$P$}}} & \multicolumn{1}{c}{{\mbox{{$P$}${d\sigma}/{d\Omega}$}}}& \multicolumn{1}{c}{{\mbox{{$P$}${d\sigma}/{d\Omega}$}}} & \multicolumn{1}{c}{{$\sigma$}} & \multicolumn{1}{c}{ Total}  & \multicolumn{1}{c}{{ References}} \\ 
\multicolumn{1}{c}{\rm (MeV)} &{($\pi^+ \Sigma^-$})  &{($\pi^0 \Sigma^0$}) & {($\pi^- \Sigma^+$}) & {($\pi^0\Sigma^0$})& {($\pi^- \Sigma^+$})& {($\pi^0\Sigma^0$})& {($\pi^- \Sigma^+$})& & \multicolumn{1}{c}{ No.} &  \multicolumn{1}{c}{} \\
\hline
1480&		&		&		&&&		& &	13	&	13		&	 \cite{Baldini1988}\\
1500	&		&		&		&&&		&	&21	&	21		&	 \cite{Baldini1988}\\
1520	&		&		&		&&&		&	&23	&	23		&	 \cite{Baldini1988}\\
1540	&	40	&	19	&	40	&&&	19	&18	&22	&	158		&	  \cite{Armenteros1970}\\
1560	&	60	&	39	&	60	&9&&	30	&30&	12	&	240		&	  \cite{Armenteros1970, CrystalBall2005, CrystalBall2008_1}\\
1580&	40	&	29	&	40	&9&	&20	&20	&6	&	164		&	  \cite{Armenteros1970, CrystalBall2005, CrystalBall2008_1}\\
1600&	40	&	29	&	40	&9&	&20	&20	&10	&	168		&	  \cite{Armenteros1970, CrystalBall2005, CrystalBall2008_1}\\
1620	&	40	&	29	&	40	&9&	&20	&20	&13	&	171		&	  \cite{Armenteros1970,CrystalBall2005, CrystalBall2008_1}\\
1640	&	40	&	48	&	40	&18&&	20	&20&	11	&		&	  \cite{Armenteros1970, Baxter1973, CrystalBall2005, CrystalBall2008_1}\\
1660	&	40	&	49	&	40	&9&	&20	&	19&17	&	194		&	  \cite{Armenteros1970, Baxter1973, CrystalBall2005, CrystalBall2008_1}\\
1680	&	80	&	49	&	80	&9&	&30	&29	&25	&	302		&	  \cite{Armenteros1968, Armenteros1970, Baxter1973, CrystalBall2005, CrystalBall2008_1}\\
1700	&	40	&	30	&	40	&&	&10	&9	&16	&	145		&	  \cite{Armenteros1968, Armenteros1970, Baxter1973}\\
1720&	76	&	30	&	75	&&10&		&&	15	&	206	&	  \cite{Armenteros1968, Baxter1973, Jones1975}\\
1740&	154	&	20	&	157	&&24&		&&	24	&	379		&	  \cite{Armenteros1968, Baxter1973, Jones1975}\\
1760	&	114	&	20	&	114	&&25&		&&	20	&	293		&	  \cite{Armenteros1968, Baxter1973, Jones1975}\\
1780	&	148	&		&	147	&&34&		&&	24	&	353		&	  \cite{Armenteros1968, Jones1975, Conforto1976}\\
1800	&	72	&		&	74	&&22&		&&	10	&	178		&	  \cite{Armenteros1968, Jones1975, Conforto1976}\\
1820&	60	&		&	60	&&15&		&&	11	&	146		&	  \cite{Armenteros1968, Conforto1976}\\
1840	&	80	&		&	80	&&13&		&&	14	&	187		&	  \cite{Armenteros1968, Conforto1976}\\
1860&	100	&		&	100	&&11&		&&	18	&	229		&	  \cite{Armenteros1968, Griselin1975, Conforto1976}\\
1880&	120	&		&	120	&&24&		&&	18	&	282		&	  \cite{Armenteros1968, Griselin1975, Conforto1976}\\
1900&	60	&		&	60	&&14&		&&	8	&	142		&	  \cite{Armenteros1968, Griselin1975, Conforto1976}\\
1920	&	80	&		&	79	&&15&		&&	12	&	186		&	  \cite{Berthon1970, Griselin1975, Conforto1976}\\
1940	&	99	&		&	100	&&14&		&&	17	&	230		&	  \cite{Berthon1970, Griselin1975, Conforto1976}\\
1960	&	60	&		&	59	&&17&		&&	10	&	146		&	  \cite{Berthon1970, Griselin1975, Conforto1976}\\
1980	&	40	&		&	38	&&	&		&	&7	&	85		&	  \cite{Berthon1970, Griselin1975}\\
2000	&	40	&		&	40	&&	&	&	&9	&	89	&		  \cite{Berthon1970, Griselin1975}\\
2020	&	19	&		&	19	&&	&	&	&5	&	43		&	  \cite{Berthon1970}\\
2040&	20	&		&	16	&&	&	&	&2	&	38		&	  \cite{Berthon1970}\\
2060	&	19	&	10	&	18	&&	&	&	&5	&	52		&	  \cite{Berthon1970, London1975}\\
2080	&	20	&	10	&	19	&&	&	&	&5	&	54		&	  \cite{Berthon1970, London1975}\\
2100	&		&	26	&		&&	&	&	&3	&	29		&	  \cite{London1975}\\
2120	&	39	&	8	&	37	&&	&	&	&4	&	88		&	  \cite{Berthon1970, London1975}\\
2140	&		&		&		&&	&	&		&	0       		    &	  NA\\
2160&	36	&		&	35	&&	&	&	&5	&	76		&	  \cite{Berthon1970}\\
\end{tabular}
\label{Table:Sigma}
\end{ruledtabular}
\end{center}
\end{table*}

\begin{figure}[htpb]
\scalebox{0.35}{\includegraphics{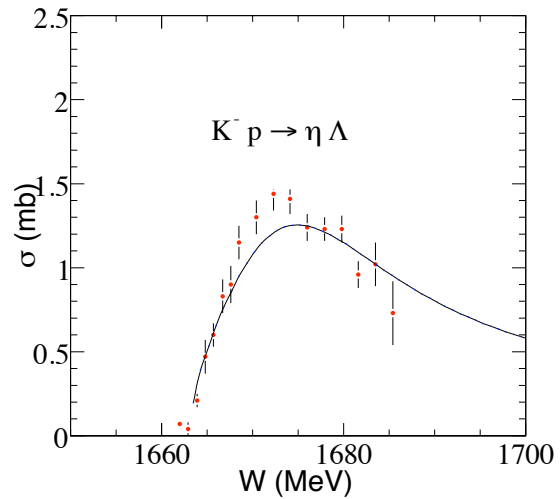}}
\caption{Integrated cross section for $K^-p\rightarrow\eta\Lambda$ compared with the results of our energy-dependent fit. Data are from Starostin 2001 \cite{starostin2001}.}
\label{etalambda}
\end{figure}

 The general qualitative behavior of the partial-wave amplitudes that we wanted to determine is known from earlier partial-wave analyses. Therefore it was convenient to make use of this information in our single-energy fits. In 2007, one of us (J. Tulpan) completed a multichannel fit {\cite{Tulpan2007}} of published partial-wave amplitudes for $\overline{K}N$ scattering to several final states, including $\overline{K}N$, $\overline{K}^*N$, $\overline{K}\Delta$, $\pi \Lambda$, $\pi \Lambda(1520)$, $\pi \Sigma$, and $\pi \Sigma(1385)$. The channels $\sigma\Lambda$, $\sigma\Sigma$, and $\eta\Sigma$ (for $S_{11}$) were included as ``dummy channels",  where $\sigma$ denotes the broad isoscalar S-wave $\pi \pi$ interaction. Also, $\eta\Lambda$ was included for $S_{01}$. Our fit of $S_{01}$ amplitudes included data for $\sigma(K^-p\rightarrow\eta\Lambda)$ up to a c.m.\ energy of 1685 MeV (see Fig.\ \ref{etalambda}). The dummy channels were channels without data and were included to satisfy unitarity in some partial waves.
Tulpan's work resulted in an energy-dependent solution that is consistent with { $S$}-matrix unitarity.
We refer to his solution as the {\it initial global fit}. Within each energy bin, each partial-wave amplitude with a given isospin amplitude was approximated by a first-order Taylor series expansion:
\begin{equation}
       T(W) = T(W_0) + T'(W_0)(W - W_0)
\end{equation}       
        where $W$ is the c.m.\ energy of the data point in the bin and $W_0$ is the central energy of the bin. Here, for simplicity $T(W)$ denotes an isospin amplitude $T_{l\pm}^I$. The complex $T$-matrix amplitude $T(W_0)$ belongs to the parameter set to be optimized at c.m.\ energy $W_0$, and $T'(W_0)$ is called the slope parameter. During fits, the slope parameter was held fixed so that the real and imaginary parts of $T(W_0)$ were our fitting parameters.
During our initial single-energy partial-wave analyses, we calculated the slope parameters $T'(W_0)$ from the initial global fit and kept these parameters constant in our fits.

Because the database is somewhat sparse, additional constraints were introduced in order to determine partial-wave amplitudes with a reasonably smooth variation with energy.
To decrease the number of free parameters to be searched, we also held fixed the very small {$ T$}-matrix amplitudes (those with $|{ T}(W_0)|<0.05$).
This constraint is expected to introduce only a small bias to our final energy-dependent partial-wave solution.

As an additional constraint, we held fixed the ${ D}_{03}$ amplitudes for $\overline{K}N \rightarrow \overline{K}N$ and $\overline{K}N \rightarrow \pi \Sigma$ at the values from the initial global fit in the bin with $W_0=1520$ MeV. This constraint was introduced because of the well-known narrow $\Lambda (1520)$ resonance, which has a width of only about 16 MeV.
Even with this constraint, we ultimately concluded that we could not determine reliable amplitudes in this bin for the  reactions $\overline{K}N \rightarrow \overline{K}N$ and $\overline{K}N \rightarrow \pi \Sigma$.

Finally, in our single-energy fits, we introduced a {\it penalty term} to the $\chi^2$ function that we minimized. This penalty term constrained our fitted amplitudes from differing greatly from the values of the initial global fit. 
For calculating the uncertainties in our single-energy amplitudes, we carried out a {\it zero-iteration} fit in which the initial values of all amplitudes were replaced by the values determined by our $\chi^2$ minimization procedure.
In the zero-iteration fit, all partial-wave amplitudes except ${ G}_{17}$ were treated as free parameters.
The ${ G}_{17}$ amplitude was held fixed in this procedure to remove the ambiguity in determining the global overall phase of our amplitudes. Spin-9/2 waves were not needed in our solution.

Once we had obtained a complete set of amplitudes for $\overline{K}N$, $\pi \Lambda$, and $\pi \Sigma$ reactions from our single-energy analyses, we carried out  global multichannel energy-dependent fits using a procedure similar to that of Tulpan \cite{Tulpan2007}.
The key new ingredient is that our global fit (details of how the partial-wave $S$-matrix was constructed can be found in Ref. \cite{manoj12}) used our own single-energy amplitudes for the $\overline{K}N$, $\pi \Lambda$, and $\pi \Sigma$ channels.
For other final states, 
we used the same input information as Tulpan did from Refs. \cite{Gopal1977, Cameron1978, Cameron1977, Cameron1978_1, Litch1974}. We assumed the same uncertainties used by Tulpan {\cite{Tulpan2007}} for obtaining the inital global fit ($\pm 0.025$ for $\overline{K}N$, $\pm 0.035$ for $\pi \Lambda$ and $\pi \Sigma$, and $\pm 0.050$ otherwise). These uncertainties were necessary because previous published partial-wave amplitudes were without error bars. These uncertainties were estimated by comparing like partial-wave amplitudes from different energy-dependent analyses and estimating the average differences for the real and imaginary parts. The smaller error bars implied the analyses agreed well with each other and the larger error bars implied the analyses agreed less well with each other.

We reduced the number of free amplitudes for a new set of single-energy solutions. At this stage, our free amplitudes included only $ S_{01}, S_{11}, P_{01}, P_{11}, P_{13},$ and $ D_{03}$.
All other amplitudes were held fixed at the values determined from our first new global fit.
In addition, the slope parameters were recalculated based on the new global fit and kept constant during this stage of the single-energy analyses.
We were able to obtain excellent agreement with the observables.
Next, we repeated our global energy-dependent analysis to refit the new set of single-energy amplitudes for $ S_{01}, S_{11}, P_{01}, P_{11}, P_{13},$ and $ D_{03}$.
We then compared our new predictions with the observables in our single-energy fits.
Still the agreement was less than satisfactory, so we carried out yet another round of single-energy analyses.
At this stage, we successfully reduced our free amplitudes to include only $ S_{01}, S_{11}$, and $ P_{01}$.
All other amplitudes were unchanged at the values from our last global fit, and slope parameters were again recalculated from the last global fit, and then held fixed in the single-energy fits.

\section{\emph{\bf  RESULTS AND DISCUSSION }}
The final single-energy fits resulted in an excellent agreement with all observables ($\rm d\sigma/{\rm d\Omega}$, $P$, $P\rm d\sigma/{\rm d\Omega}$, and $\sigma$) yielding a fairly smooth set of partial-wave  amplitudes within the energy range of our analysis. The energy-dependent solutions were finally used to compare with the observable data.
Figures 2 - 7 show representative energy-dependent results for the differential cross section of each $\overline KN$ reaction included in our single-energy fits. 
The cross sections are shown as a function of $\cos\theta$, where $\theta$ is the c.m.\ scattering angle of the meson.
Figure 2 shows the comparison of differential cross section data for $K^-p\rightarrow K^-p$ with our energy-dependent solution at four lab momenta of 514, 935, 1165, and 1483 MeV. Although  the data are from the 1960s and 1070s \cite{Armenteros1970, Conforto1971, Conforto1976, Daum1968} they are in excellent agreement with our solution. For $K^- p \rightarrow \overline K^0 n$ (Fig.\ 3) the Crystal Ball data with smaller error bars at $ P_{\rm Lab}$ = 514 MeV and 714 MeV are well described by our solution in the forward and backward angles with a slight under fitting in the intermediate angles. The other data \cite{Jones1975, Griselin1975} at $P_{\rm Lab}$ = 936 MeV and 1434 MeV with larger error bars are in good agreement with our energy-dependent solution. Similarly, Fig.\ 4 shows the excellent agreement between our solution and differential cross section data at $P_{\rm Lab}$ = 514, 750, 1153, and 1465 MeV for $K^-p\rightarrow \pi^0\Lambda$. Figure 5 shows a comparison of data from Refs. \cite{Armenteros1970, Armenteros1968, Conforto1976, Berthon1970} with our solution for $K^-p\rightarrow \pi^+\Sigma^-$. Except for some under representation of data at $P_{\rm Lab}$ = 1245 MeV we have an excellent agreement with the data. Figures 6 and 7 show an excellent agreement of our energy-dependent solution with the differential cross section data at various lab momenta of kaons for $K^-p\rightarrow \pi^0\Sigma^0$ and $K^-p\rightarrow \pi^-\Sigma^+$, respectively.
\begin{figure}[htpb]
\vspace{-15mm}
\scalebox{0.35}{\includegraphics{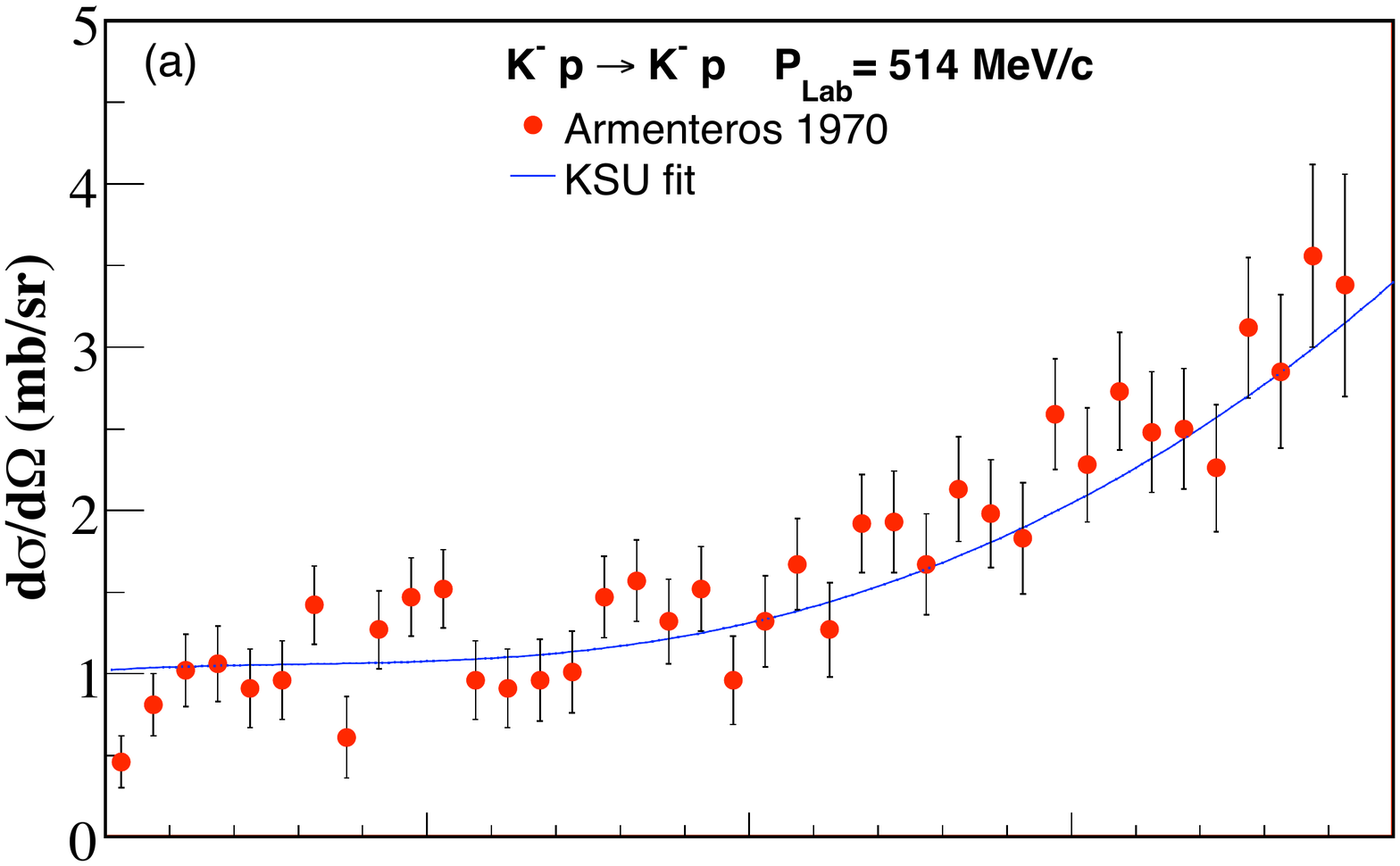}}
\vspace{-1mm}
\vspace{-25mm}
\scalebox{0.35}{\includegraphics{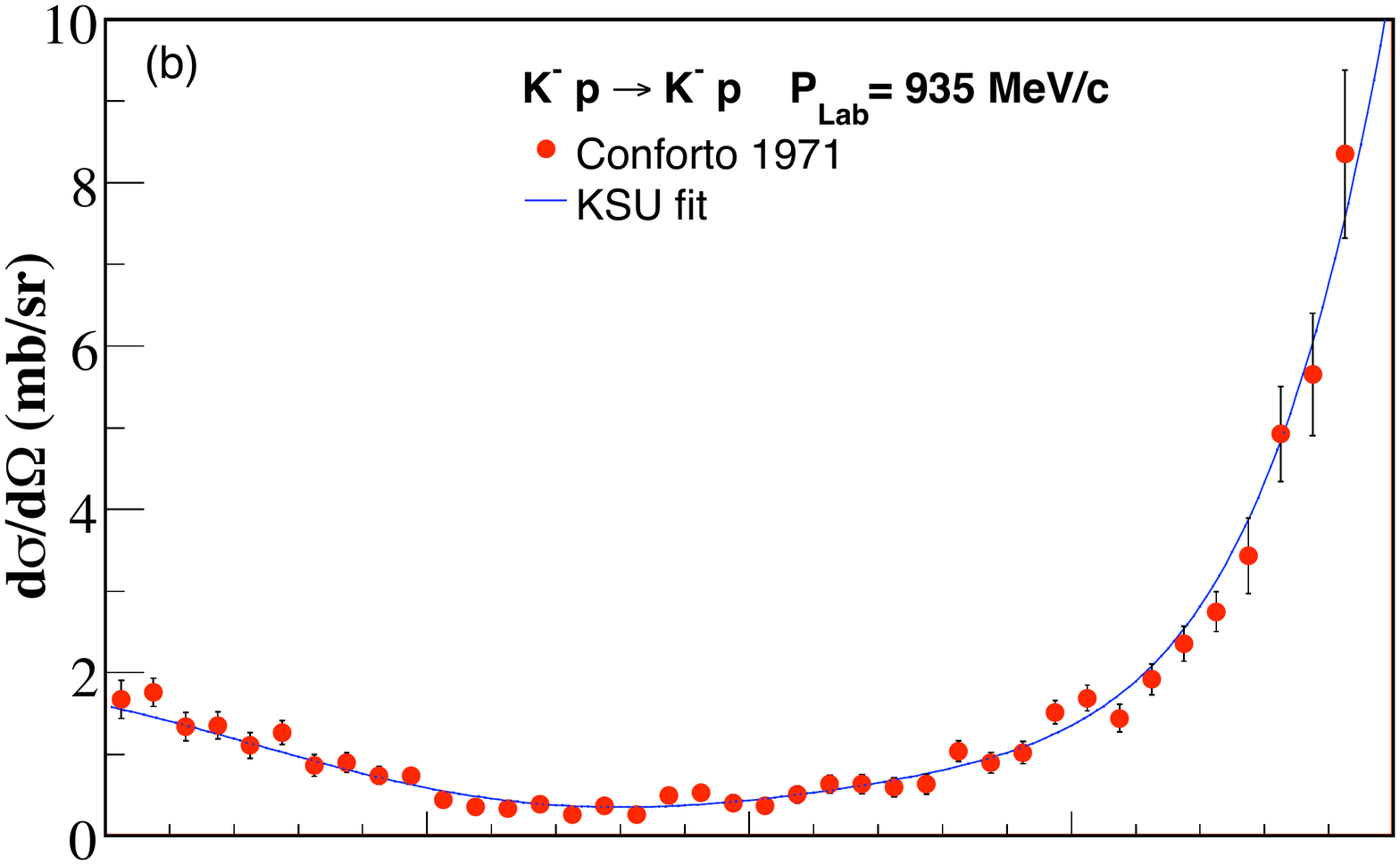}}
\vspace{-25.5mm}
\vspace{-1mm}
\scalebox{0.35}{\includegraphics{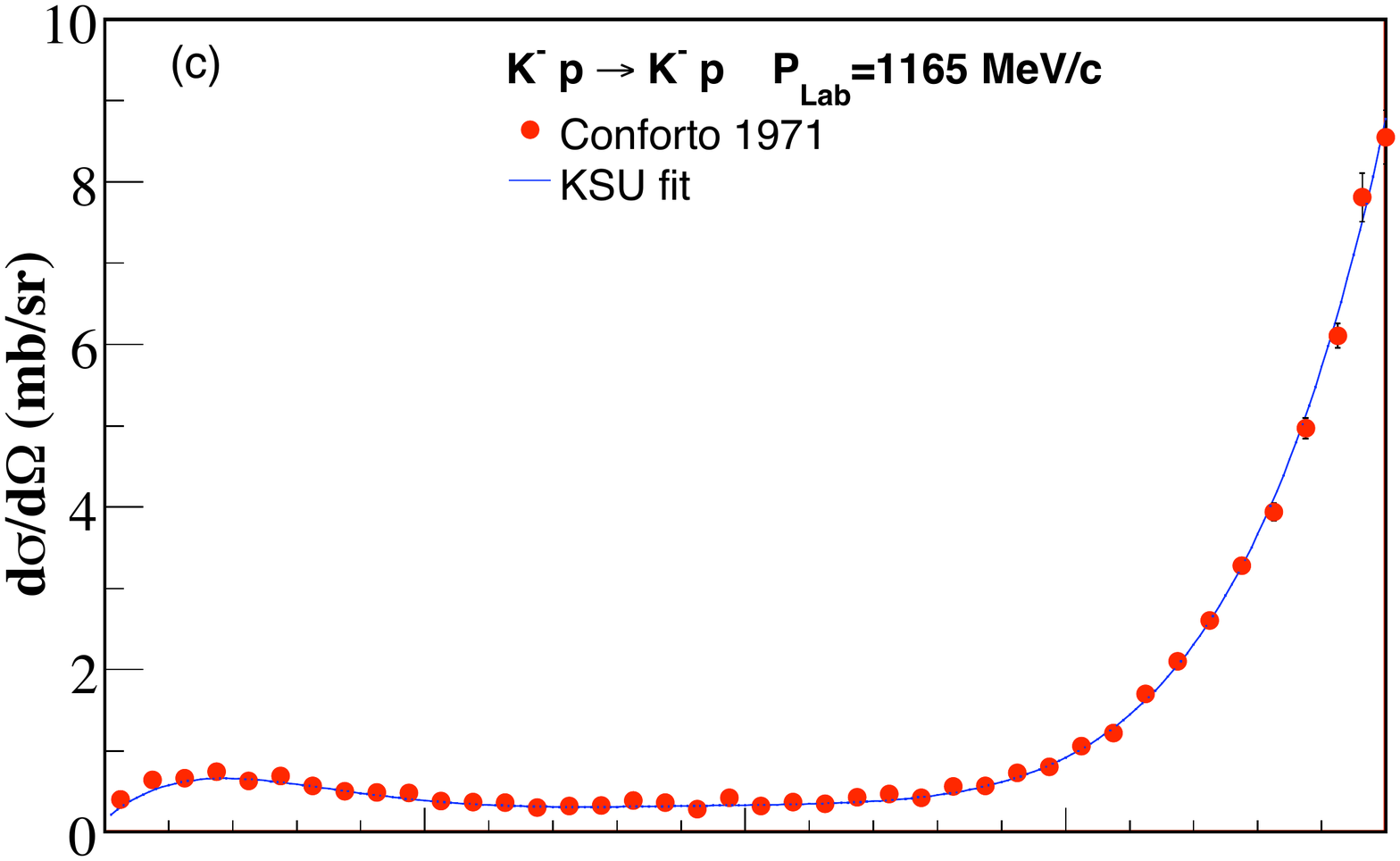}} 
\vspace{-20mm}
\vspace{-1mm}
\scalebox{0.35}{\includegraphics{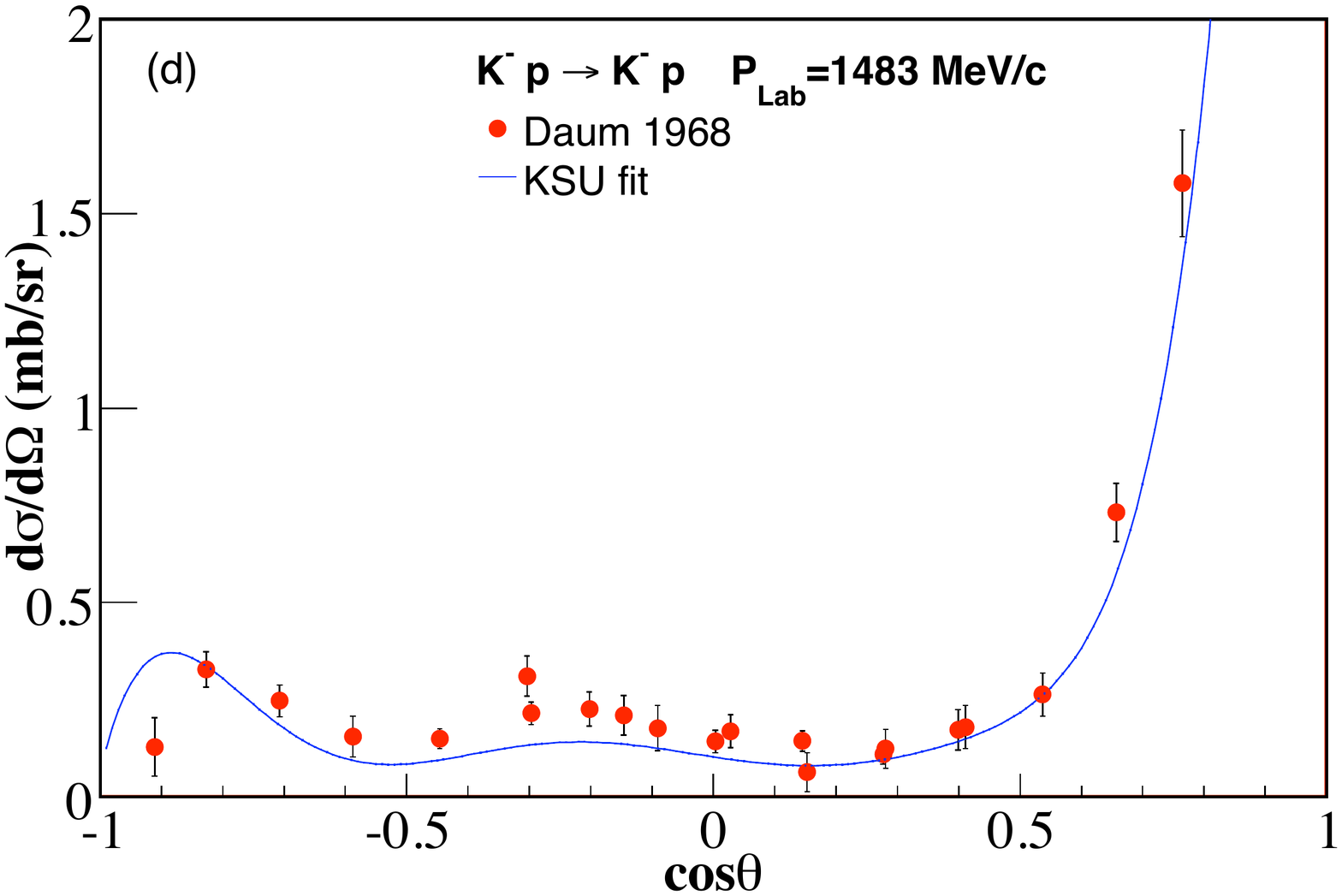}}
\vspace{-5mm}
\caption{(Color online) Representative results of our energy-dependent fit for the $K^- p \rightarrow K^- p$ differential cross section. Data are from Armenteros 1970 \cite{Armenteros1970}, Conforto 1971 \cite{Conforto1971}, and Duam 1968 \cite{Daum1968}.}
\label{fig:dSigma_11_New}
\end{figure}

\begin{figure}[htpb]
\vspace{-15mm}
\scalebox{0.35}{\includegraphics{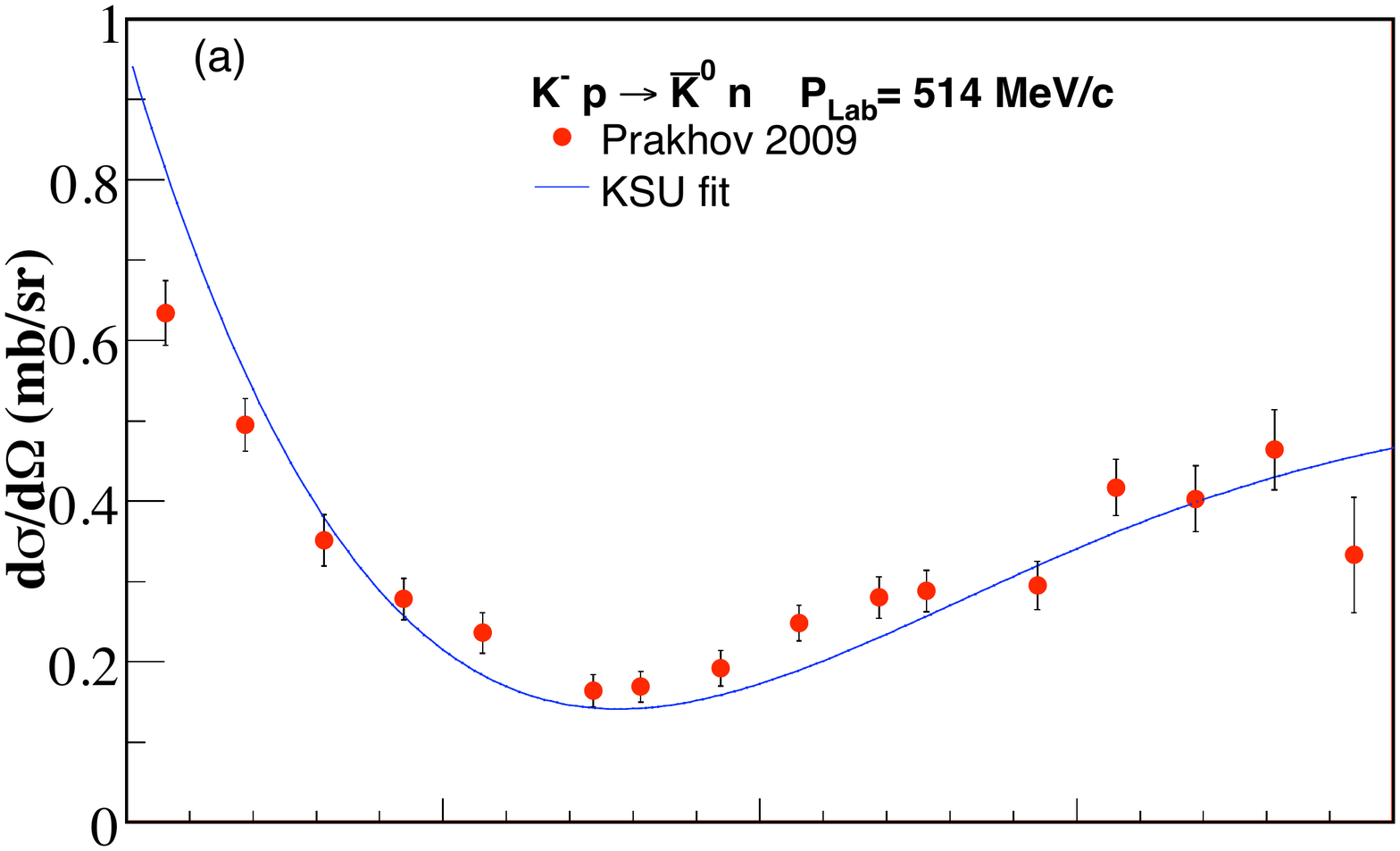}}
\vspace{-1mm}
\vspace{-25mm}
\scalebox{0.35}{\includegraphics{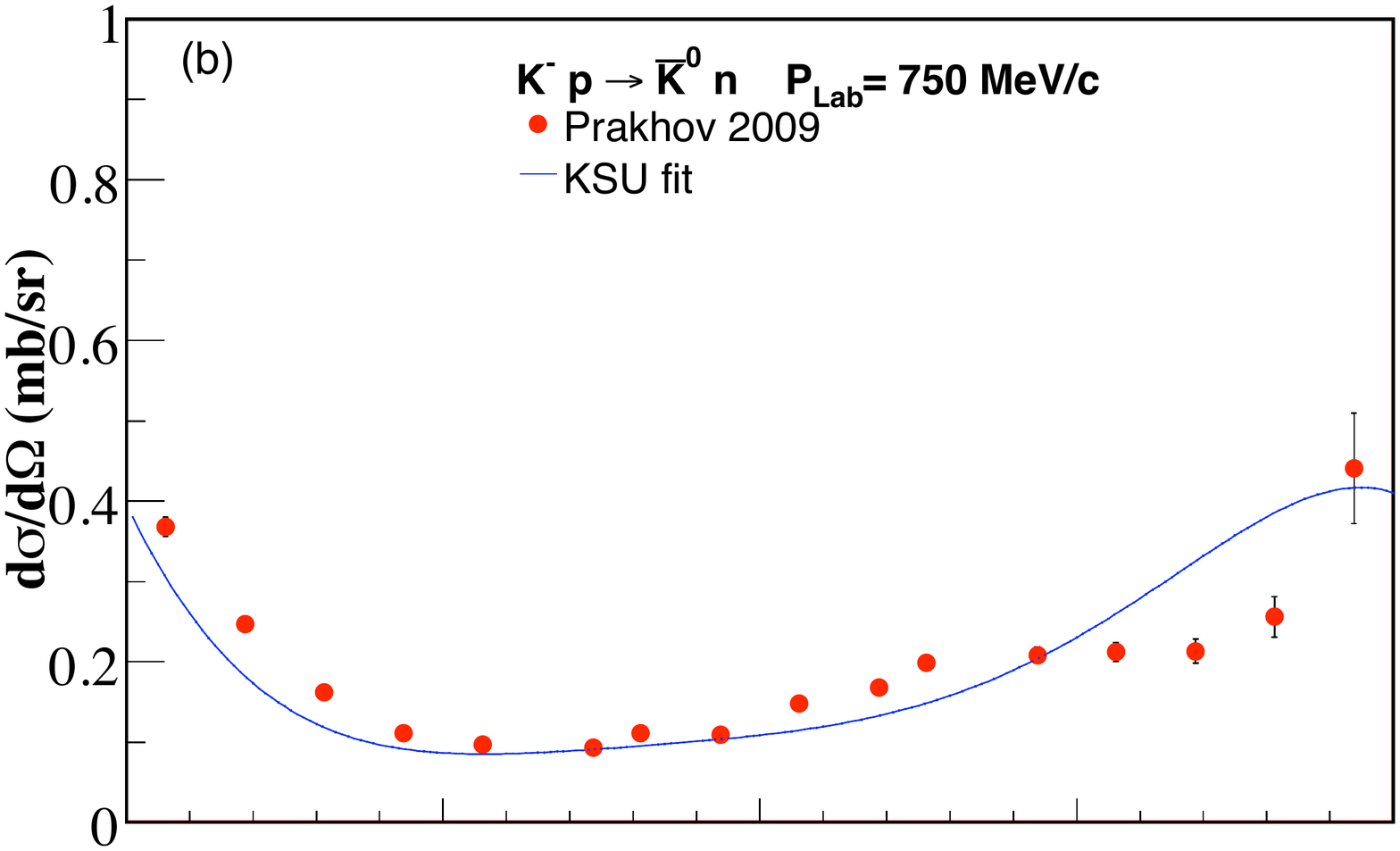}}
\vspace{-25.5mm}
\vspace{-1mm}
\scalebox{0.35}{\includegraphics{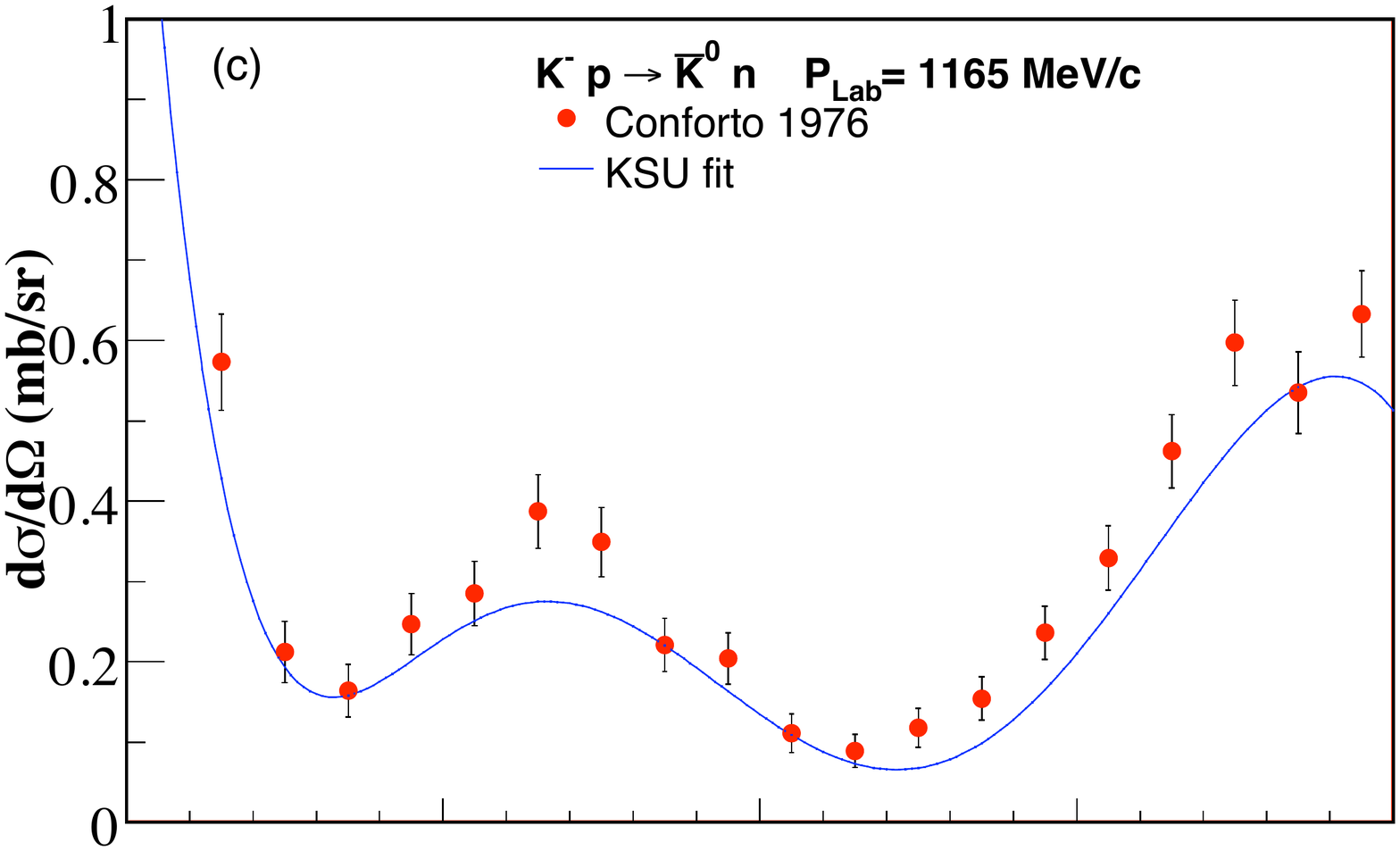}} 
\vspace{-20mm}
\vspace{-1mm}
\scalebox{0.35}{\includegraphics{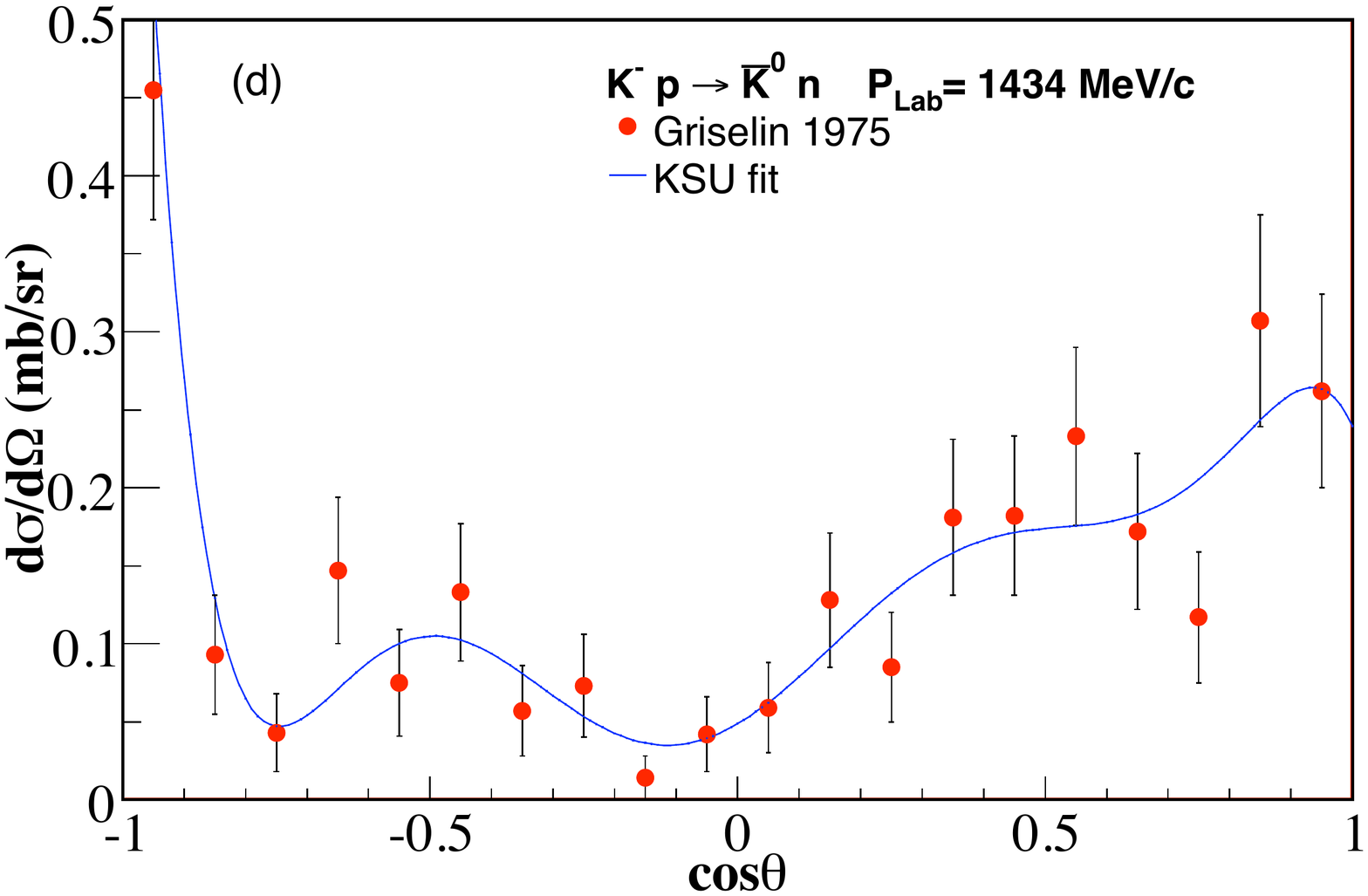}}
\vspace{-5mm}
\caption{(Color online) Representative results of our energy-dependent fit for the $K^- p \rightarrow \overline K^0 n$ differential cross section. Data are from Prakhov 2009 \cite{CrystalBall2005}, Conforto 1976 \cite{Conforto1976}, and Griselin 1975 \cite{Griselin1975}.}
\label{fig:dSigma_11_New}
\end{figure}

\begin{figure}[htpb]
\vspace{-15mm}
\scalebox{0.35}{\includegraphics{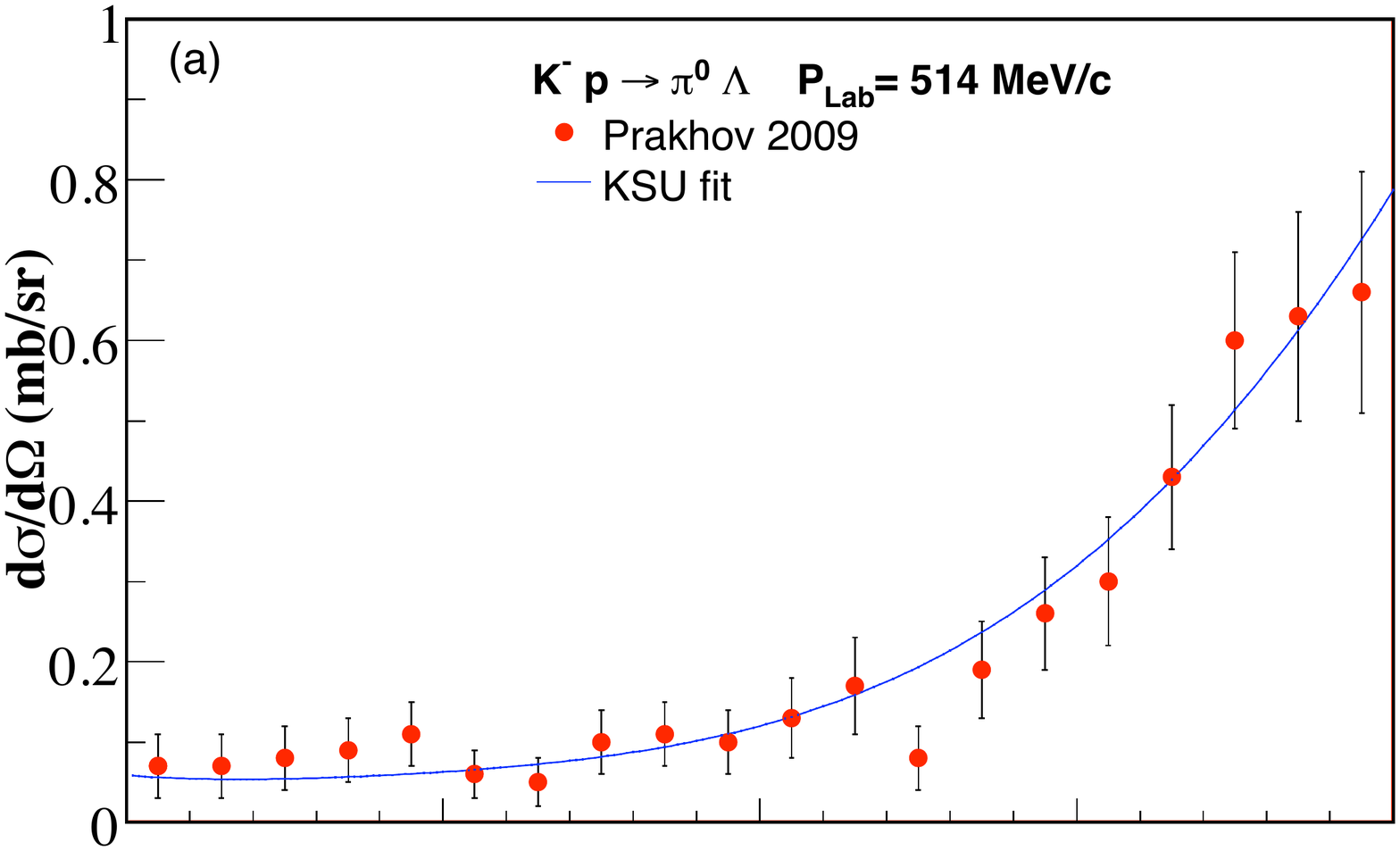}}
\vspace{-1mm}
\vspace{-25mm}
\scalebox{0.35}{\includegraphics{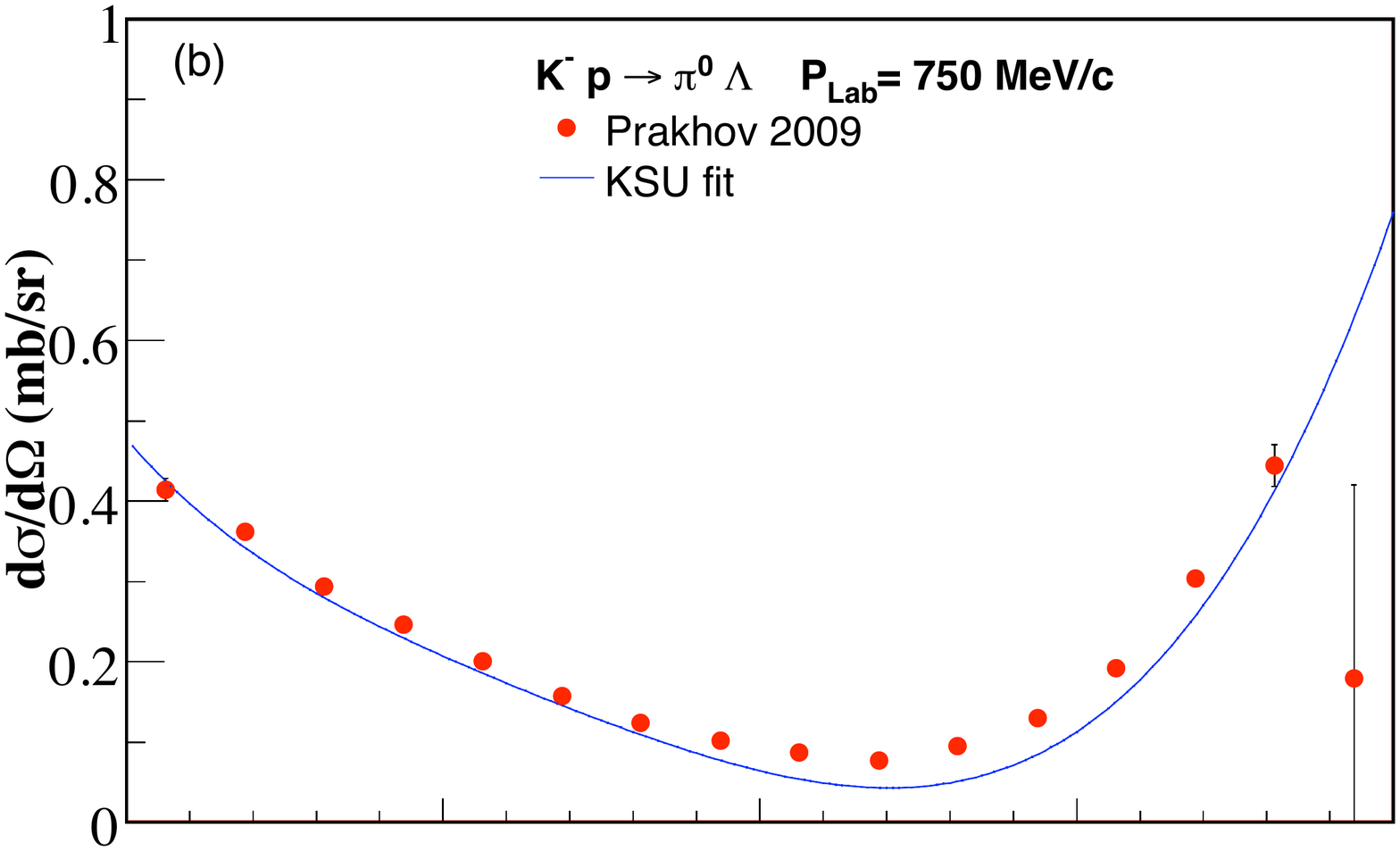}}
\vspace{-25mm}
\vspace{-1mm}
\scalebox{0.35}{\includegraphics{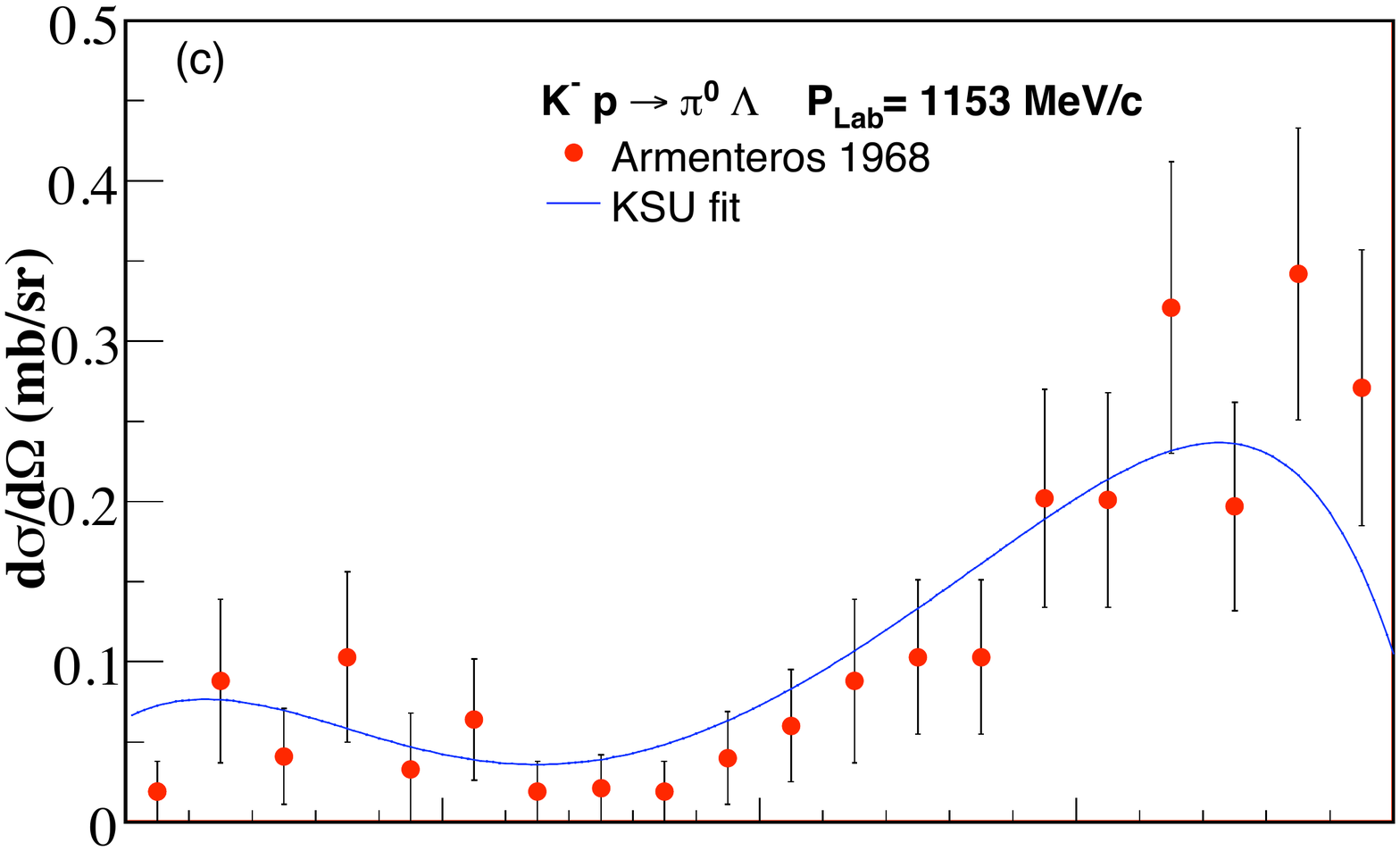}} 
\vspace{-20mm}
\vspace{-1mm}
\scalebox{0.35}{\includegraphics{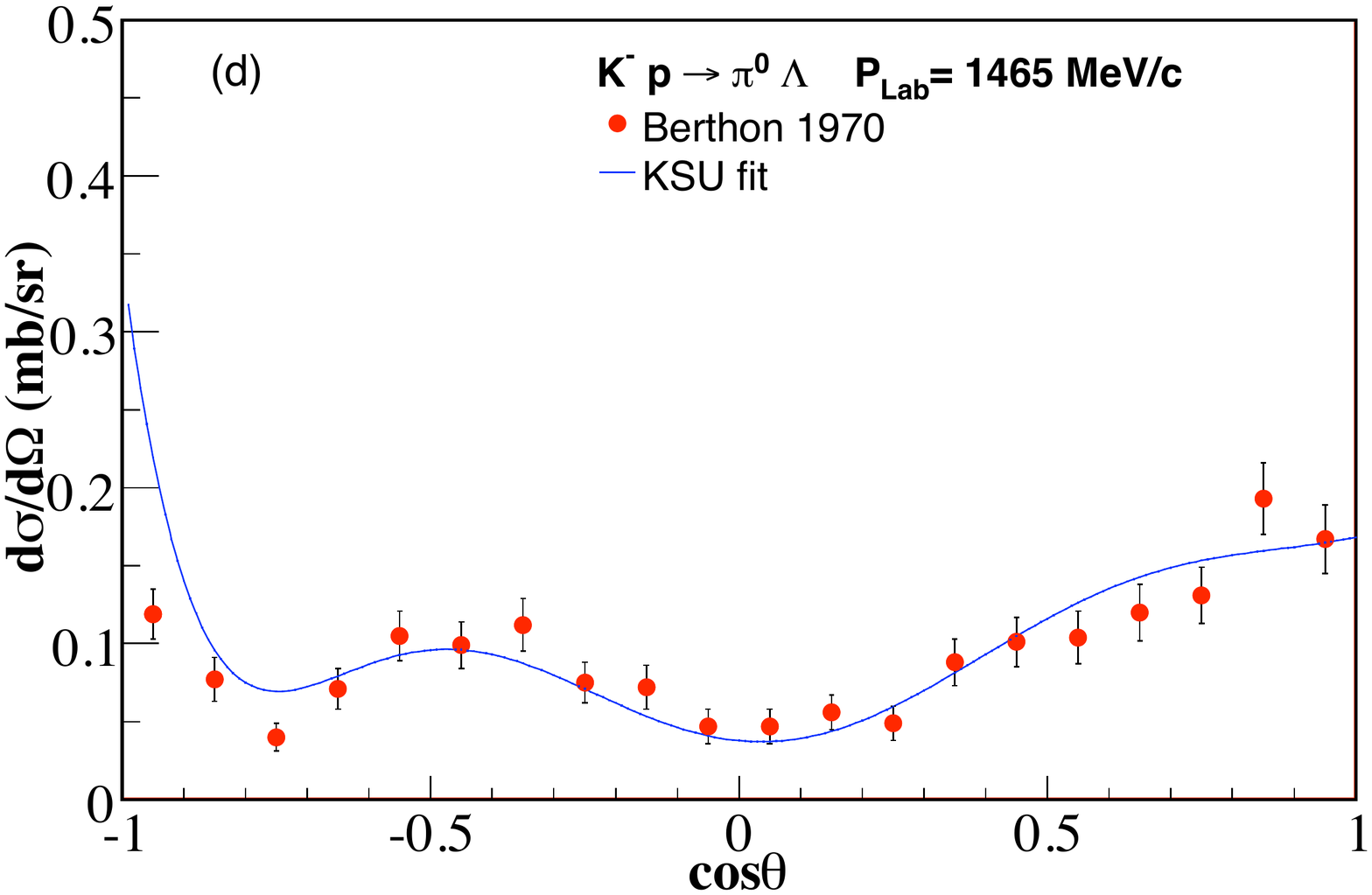}}
\vspace{-5mm}
\caption{(Color online) Representative results of our energy-dependent fit for the $K^- p \rightarrow \pi^0\Lambda$ differential cross section. Data are from Prakhov 2009 \cite{CrystalBall2005}, Armenteros 1968 \cite{Armenteros1968}, and Berthon 1970 \cite{Berthon1970}.}
\label{fig:dSigma_11_New}
\end{figure}

\begin{figure}[htpb]
\vspace{-15mm}
\scalebox{0.35}{\includegraphics{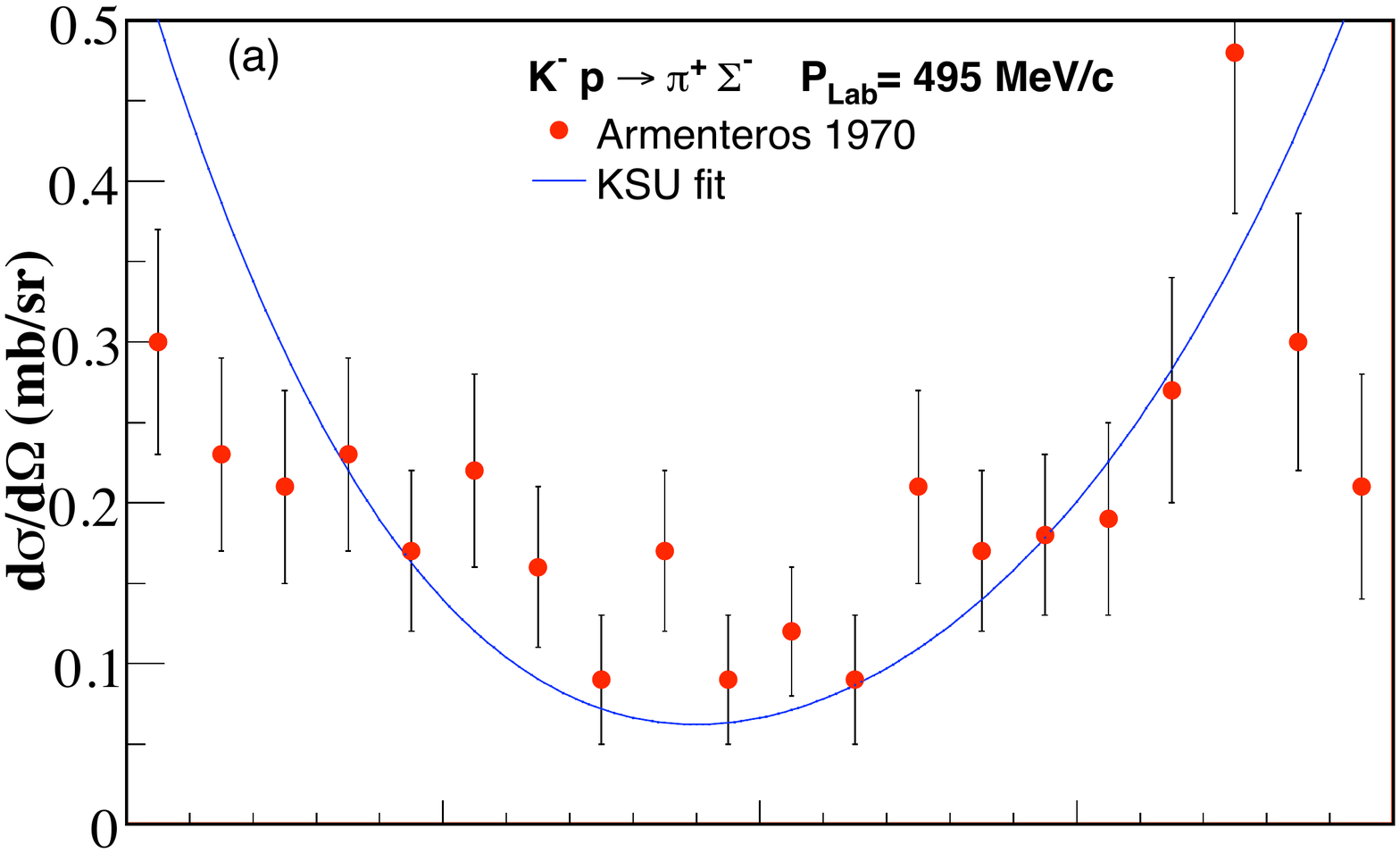}}
\vspace{-1mm}
\vspace{-25mm}
\scalebox{0.35}{\includegraphics{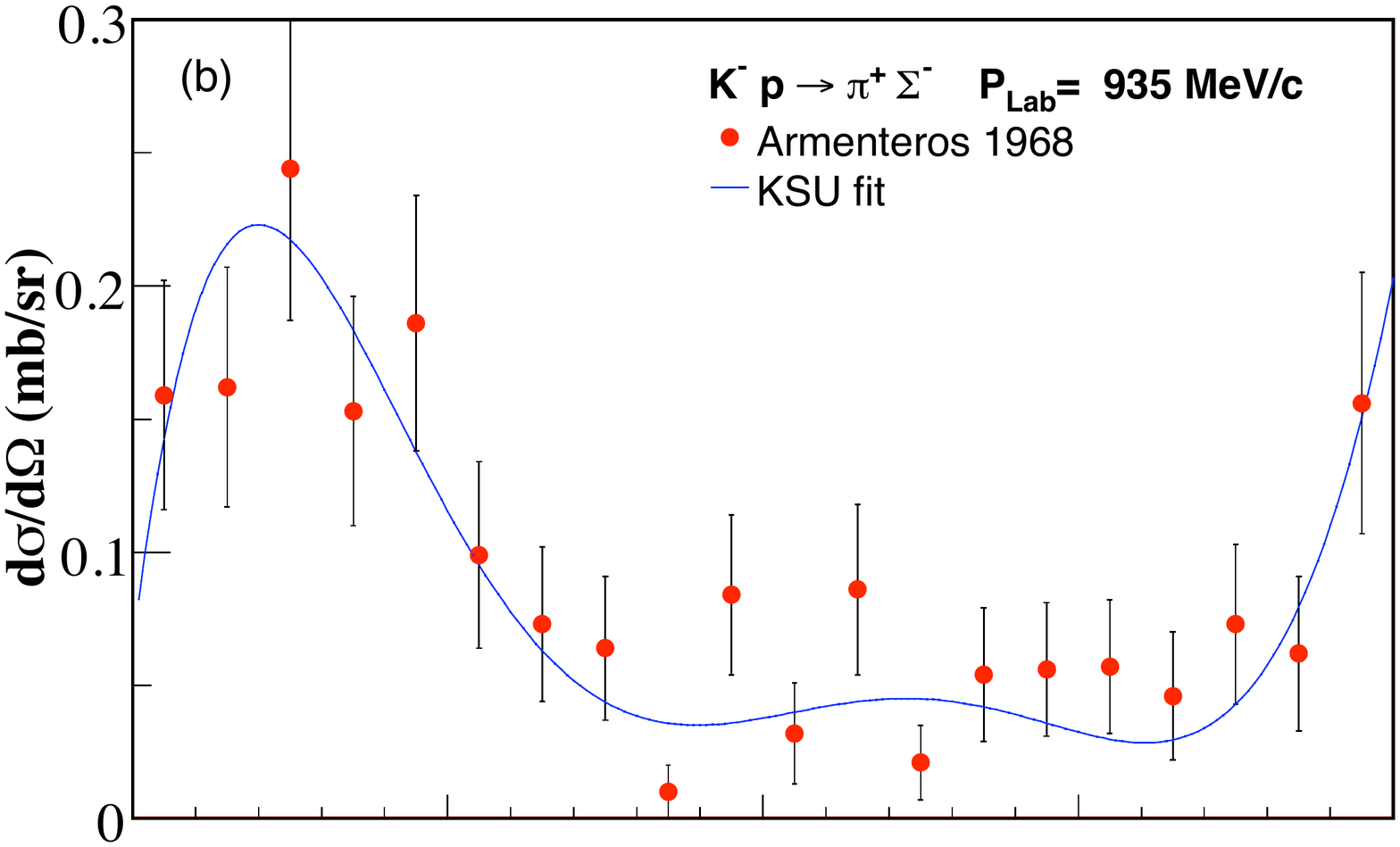}}
\vspace{-25mm}
\vspace{-1mm}
\scalebox{0.35}{\includegraphics{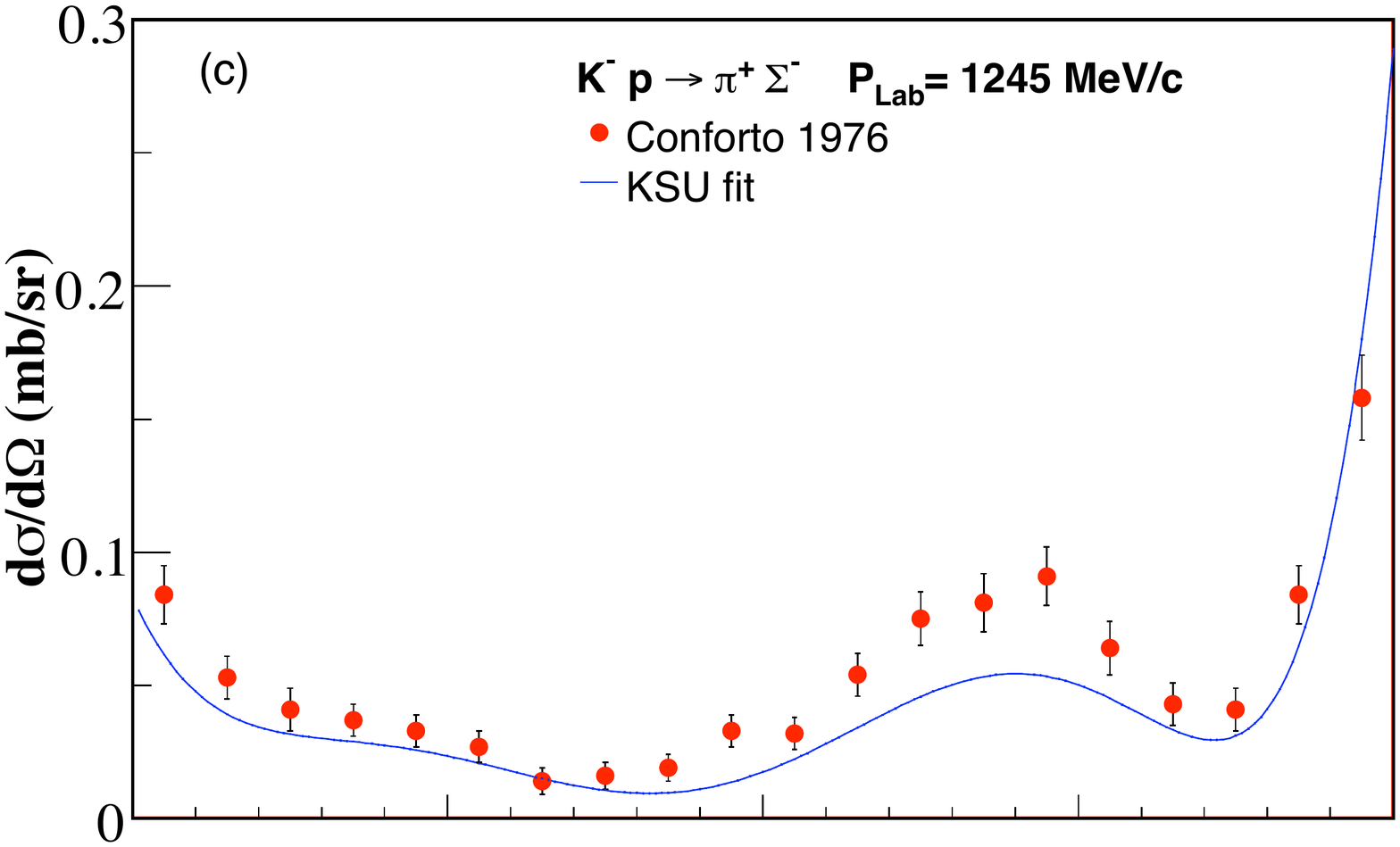}} 
\vspace{-20mm}
\vspace{-1mm}
\scalebox{0.35}{\includegraphics{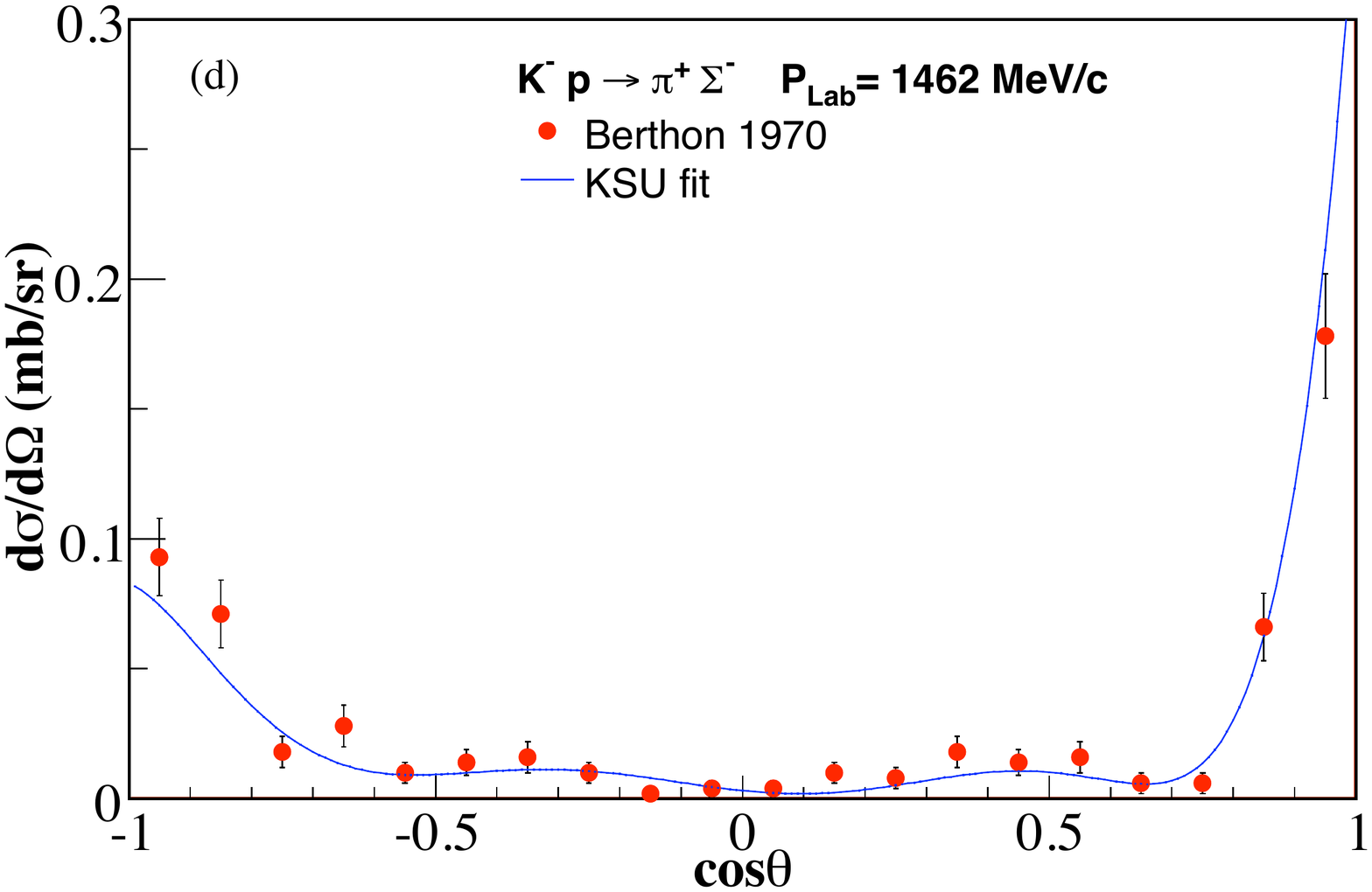}}
\vspace{-5mm}
\caption{(Color online) Representative results of our energy-dependent fit for the $K^- p \rightarrow \pi^+\Sigma^-$ differential cross section. Data are from Armenteros 1970 \cite{Armenteros1970}, Armenteros 1968 \cite{Armenteros1968}, Conforto 1976 \cite{Conforto1976}, and Berthon 1970 \cite{Berthon1970}.}
\label{fig:dSigma_11_New}
\end{figure}

\begin{figure}[htpb]
\vspace{-15mm}
\scalebox{0.35}{\includegraphics{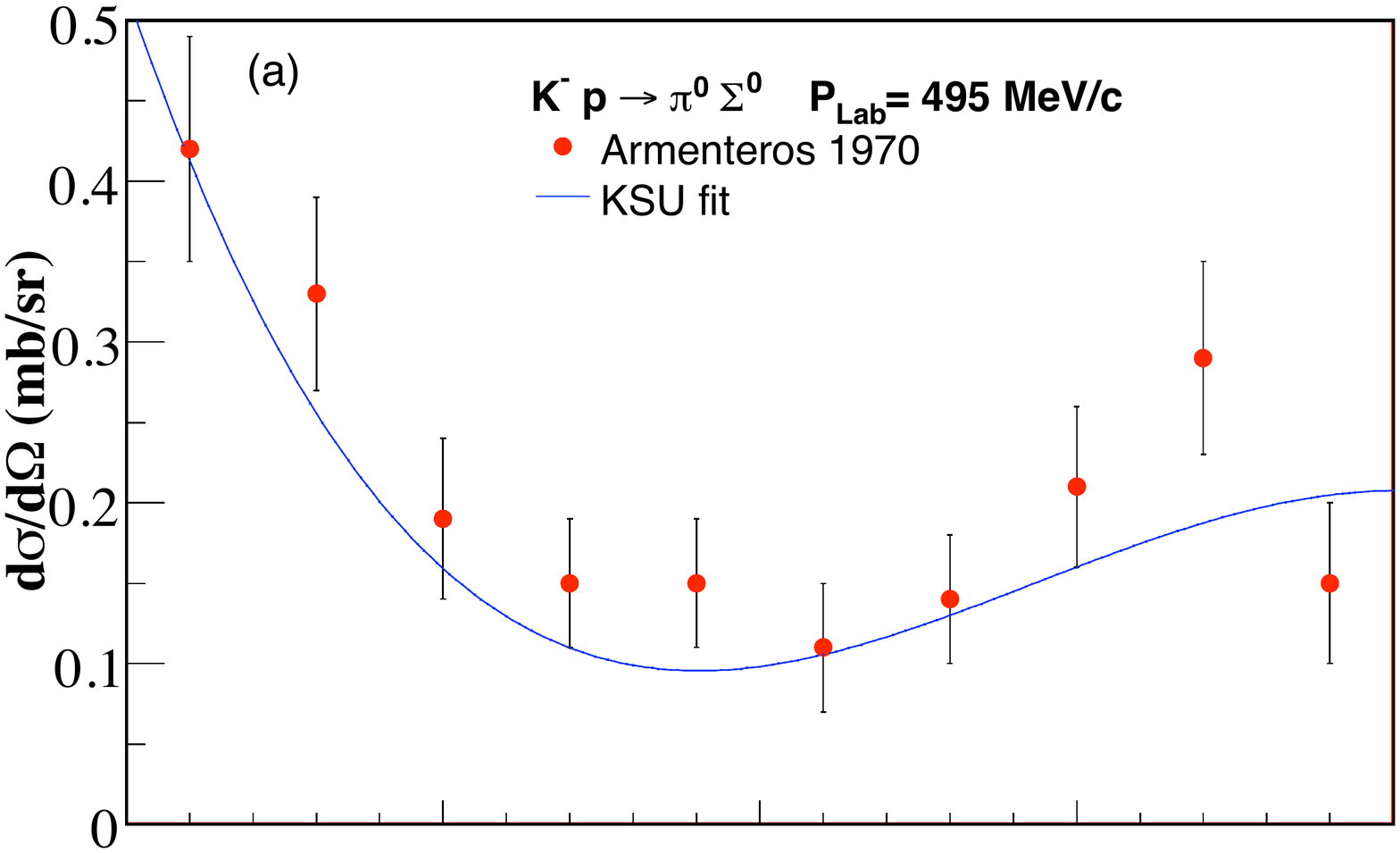}}
\vspace{-1mm}
\vspace{-25mm}
\scalebox{0.35}{\includegraphics{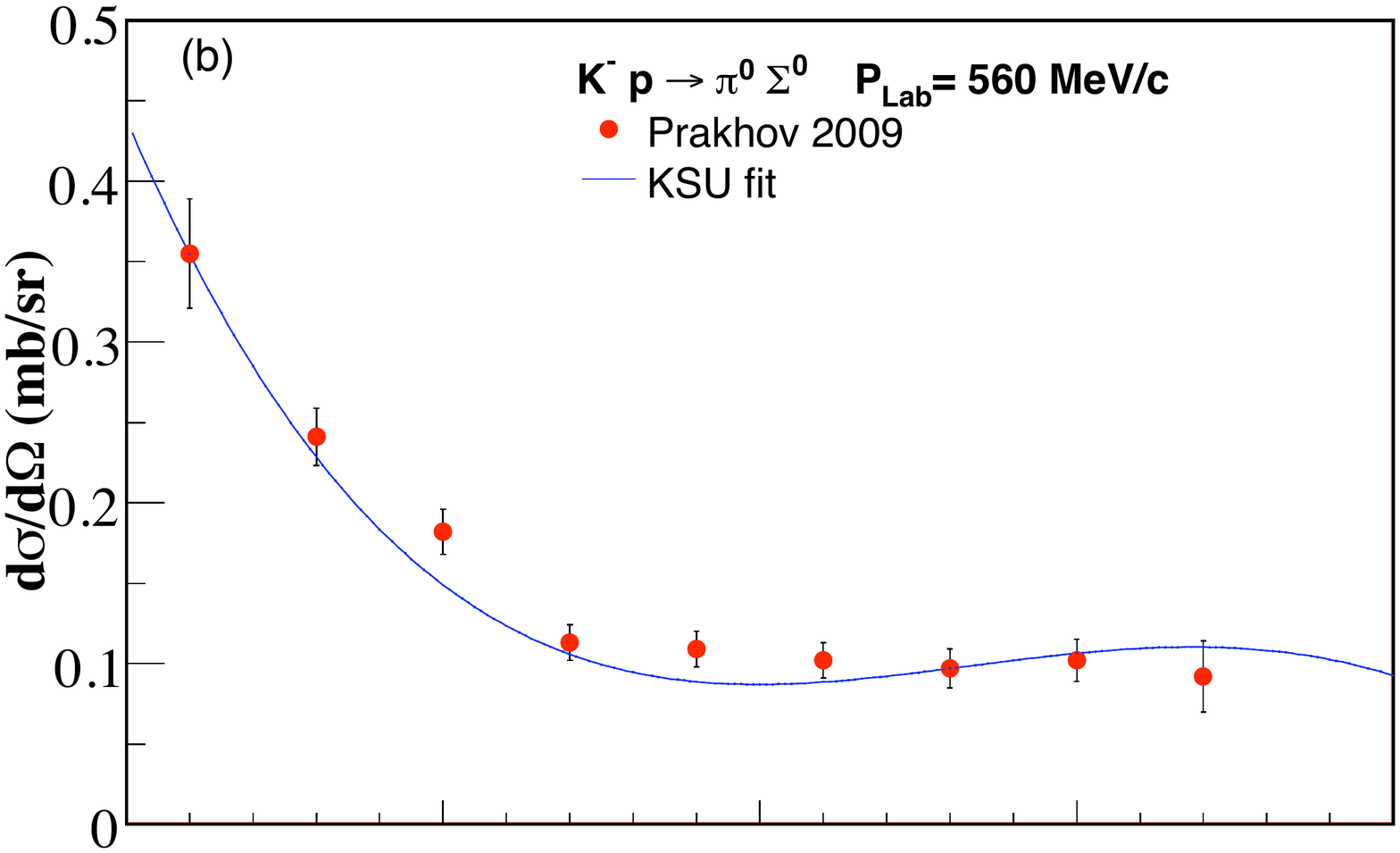}}
\vspace{-25mm}
\vspace{-1mm}
\scalebox{0.35}{\includegraphics{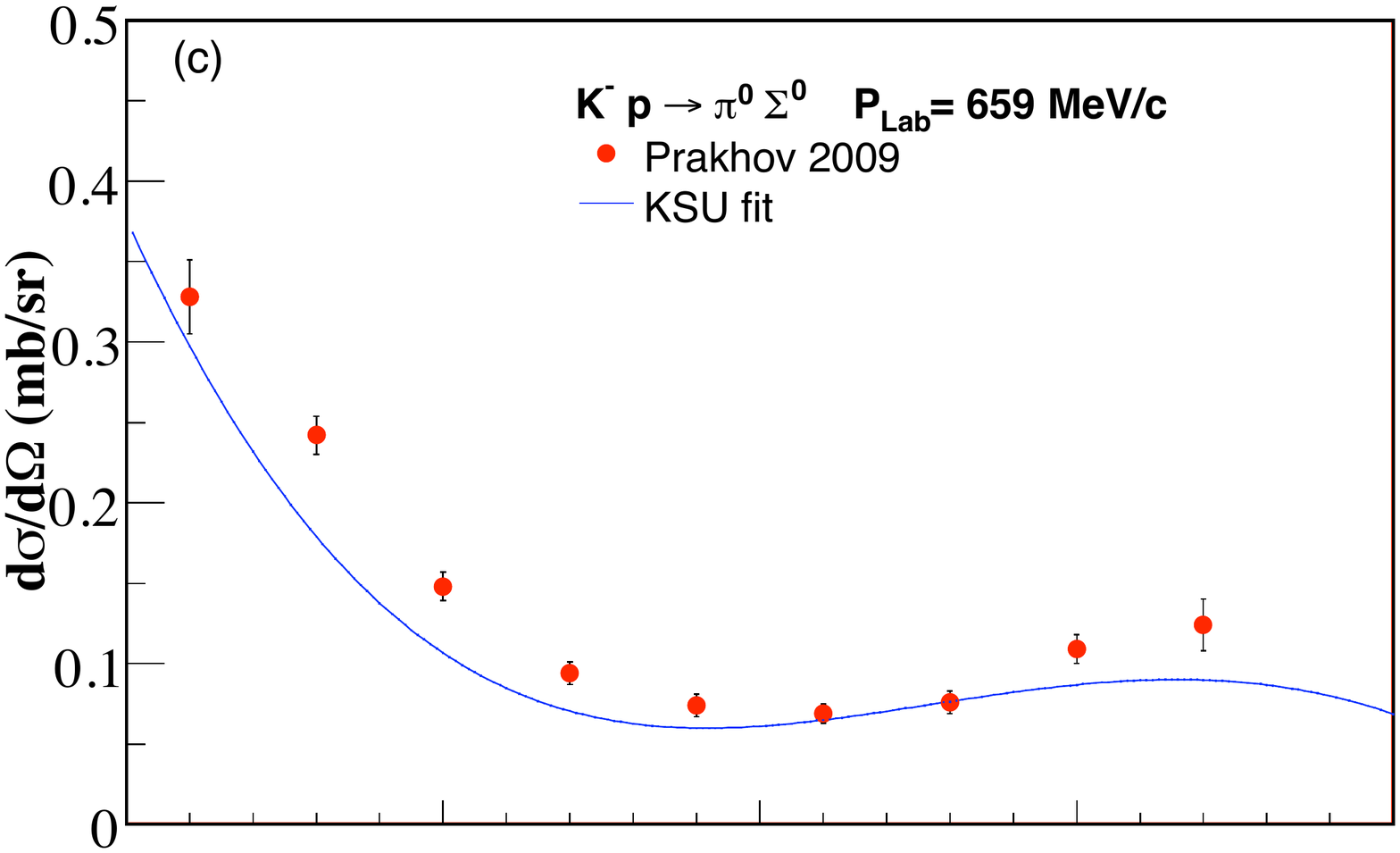}} 
\vspace{-20mm}
\vspace{-1mm}
\scalebox{0.35}{\includegraphics{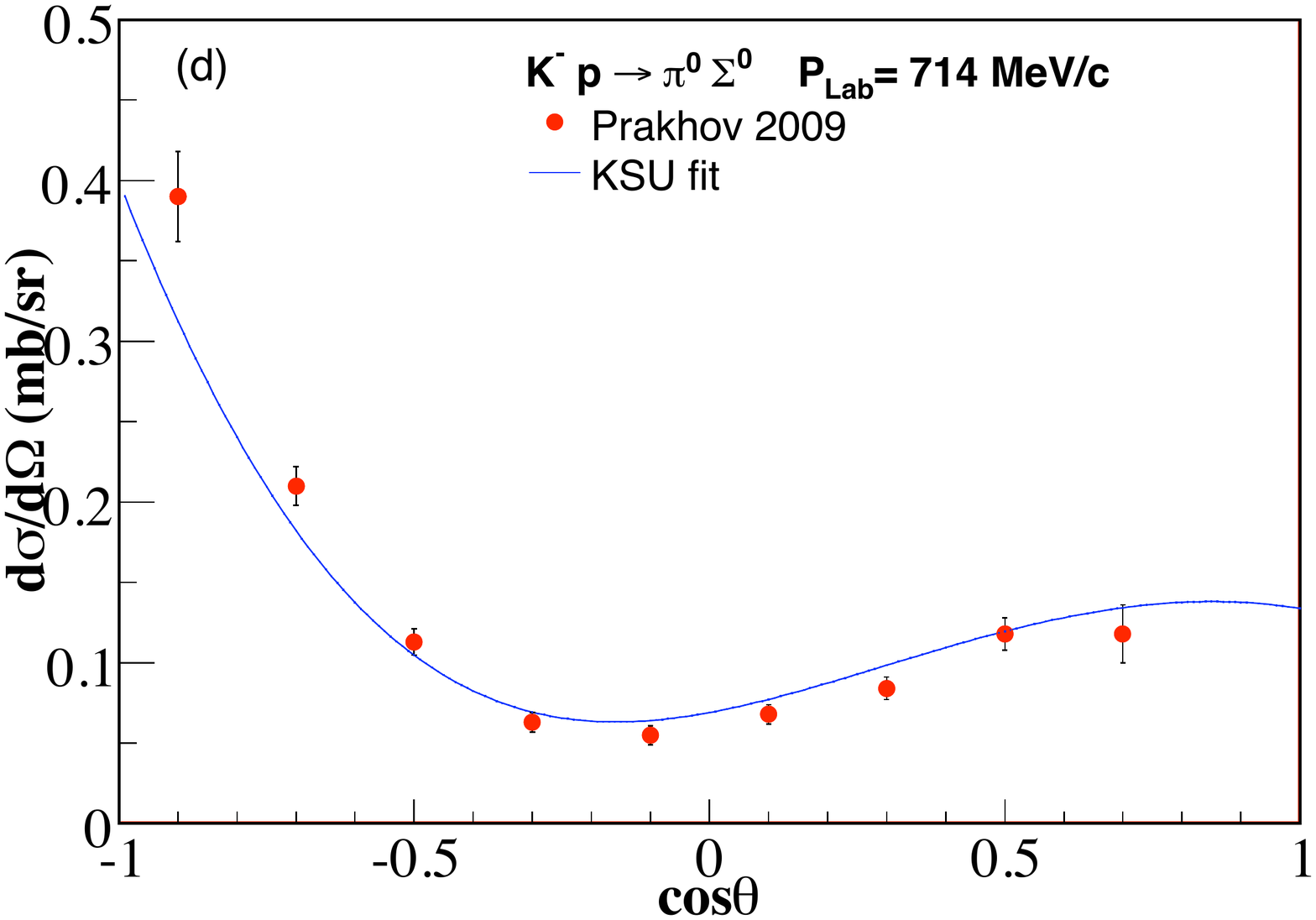}}
\vspace{-5mm}
\caption{(Color online) Representative results of our energy-dependent fit for the $K^- p \rightarrow \pi^0\Sigma^0$ differential cross section. Data are from Armenteros 1970 \cite{Armenteros1970} and Prakhov 2009 \cite{CrystalBall2005}.}
\label{fig:dSigma_11_New}
\end{figure}


\begin{figure}[htpb]
\vspace{-15mm}
\scalebox{0.35}{\includegraphics{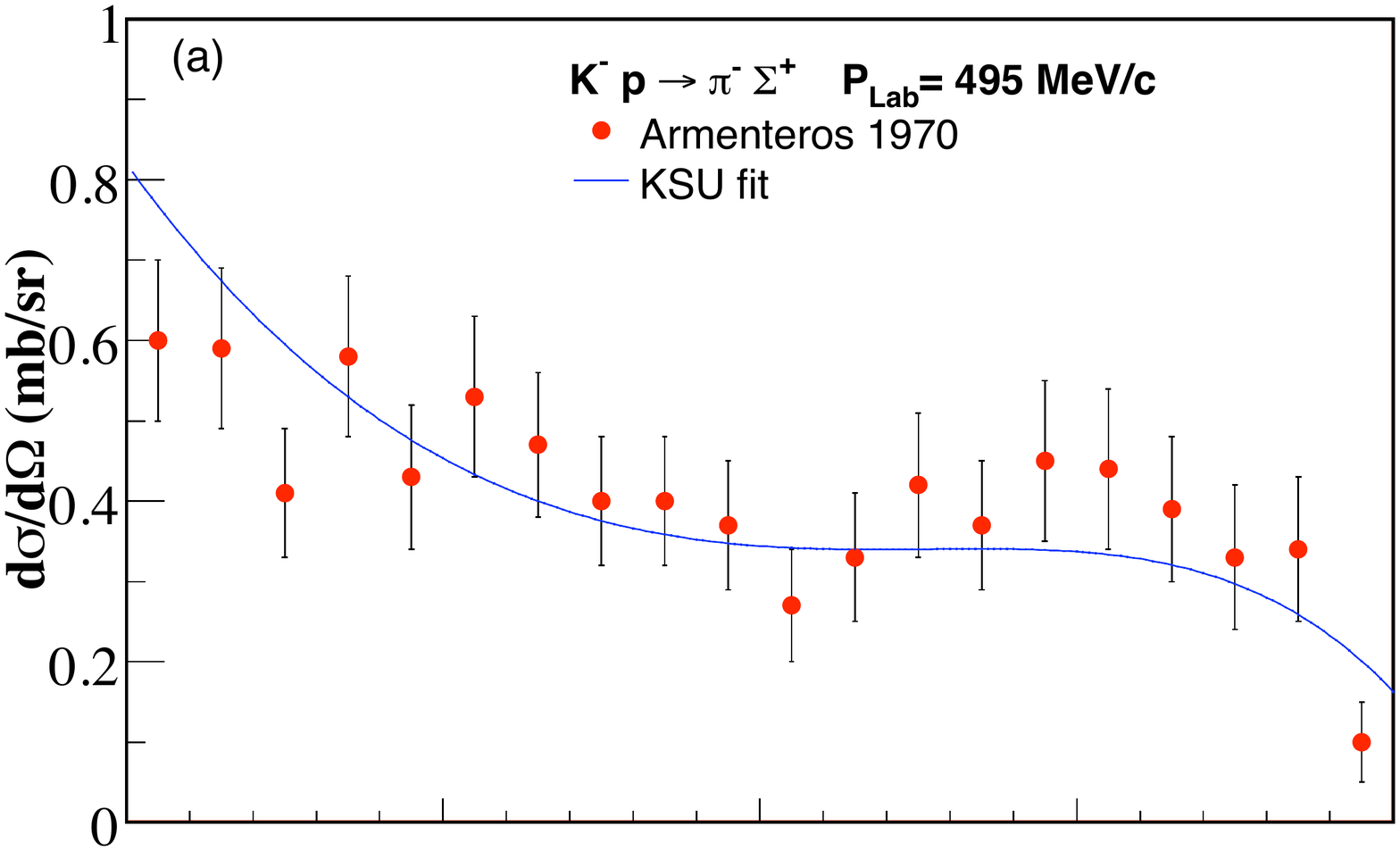}}
\vspace{-1mm}
\vspace{-25mm}
\scalebox{0.35}{\includegraphics{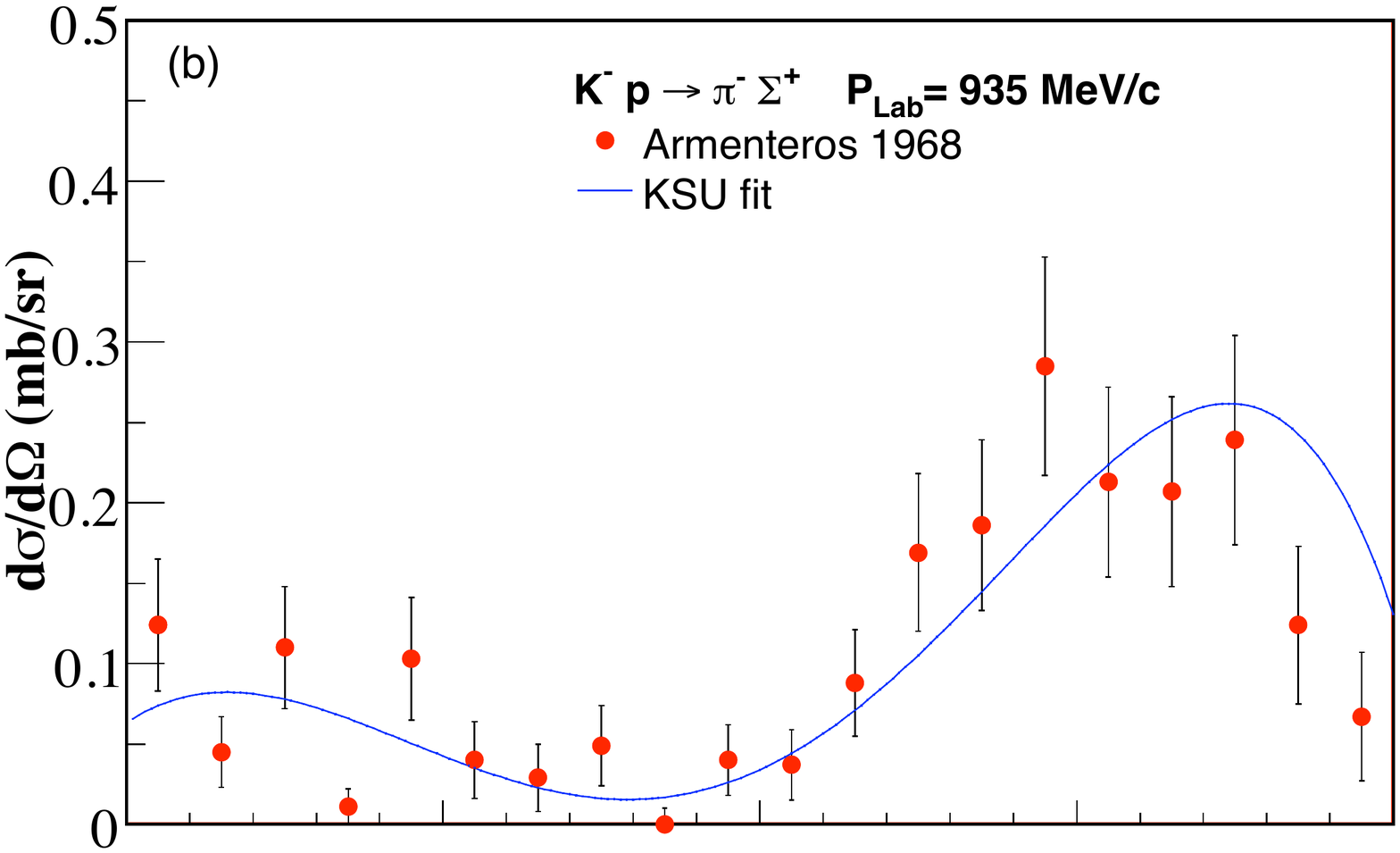}}
\vspace{-25mm}
\vspace{-1mm}
\scalebox{0.35}{\includegraphics{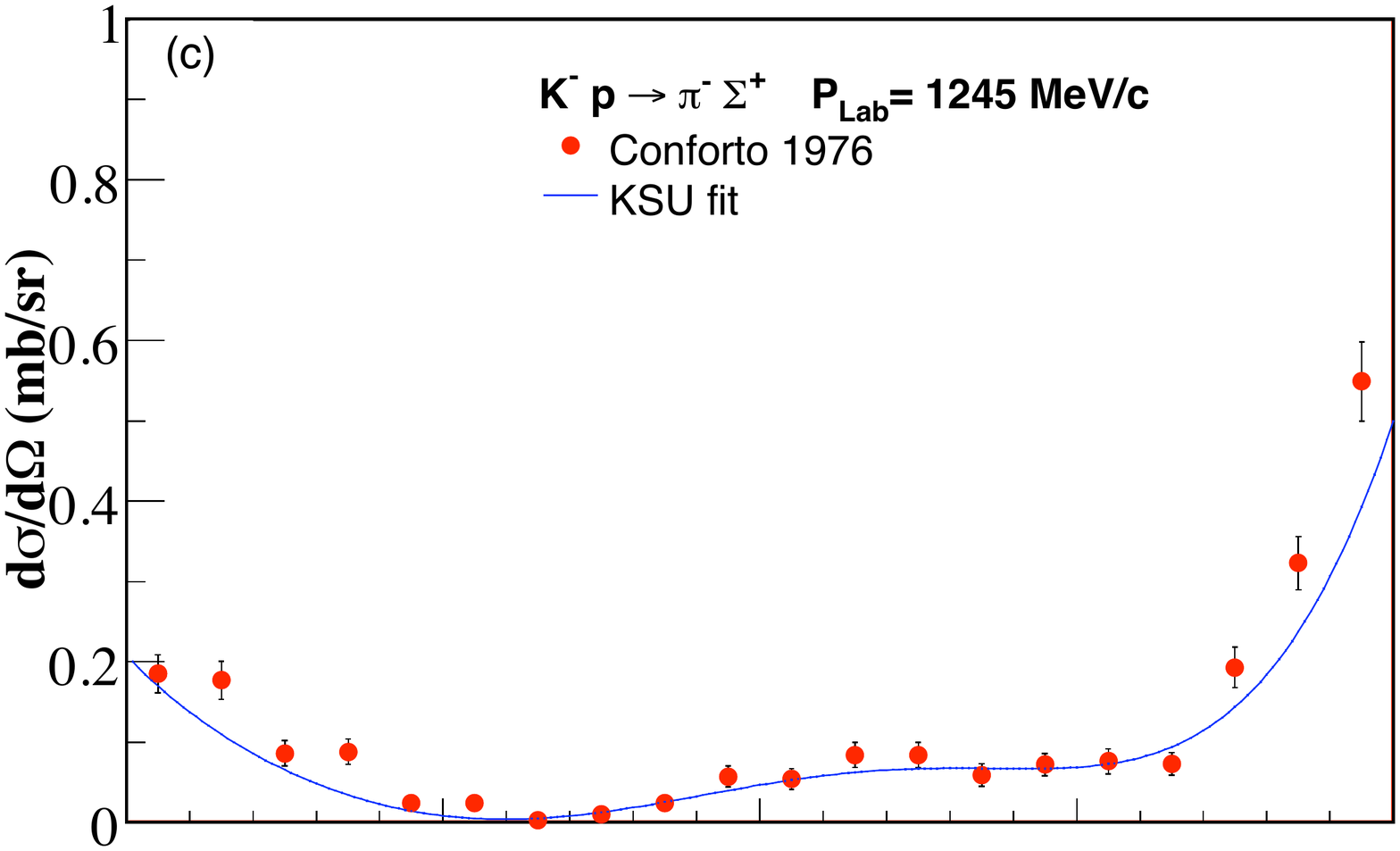}} 
\vspace{-25mm}
\vspace{3mm}
\scalebox{0.35}{\includegraphics{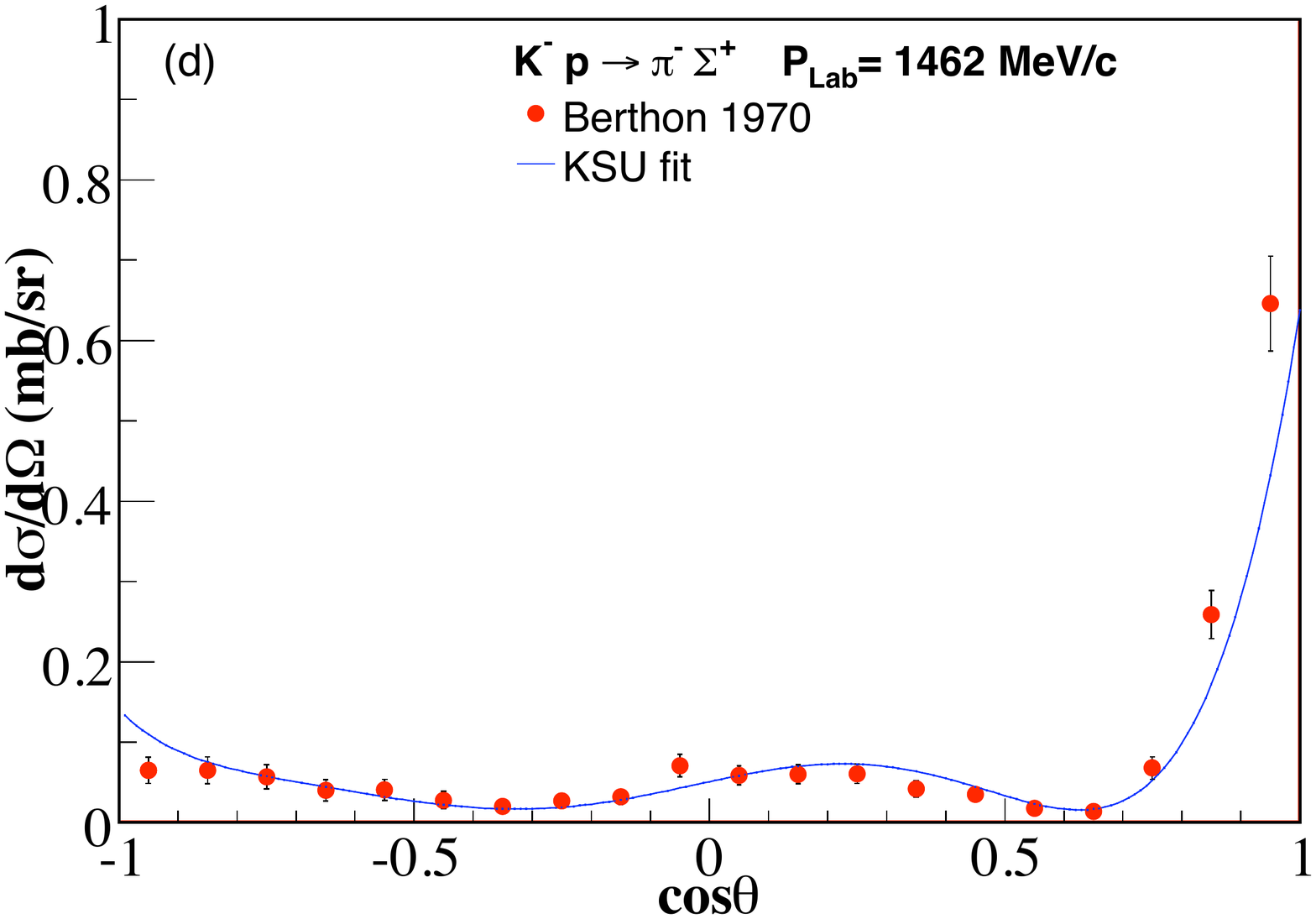}}
\vspace{-2mm}
\caption{(Color online) Representative results of our energy-dependent fit for the $K^- p \rightarrow \pi^-\Sigma^+$ differential cross section. Data are from Armenteros 1970 \cite{Armenteros1970}, Armenteros 1968 \cite{Armenteros1968}, Conforto 1976 \cite{Conforto1976}, and Berthon 1970 \cite{Berthon1970}.}
\label{fig:dSigma_11_New}
\end{figure}

Figures 8, 9, 10, and 11 show representative energy-dependent fit results for the polarization in reactions $K^- p \rightarrow K^- p$, $K^- p \rightarrow \pi\Lambda$, $K^- p \rightarrow \pi^0\Sigma^0 $, and $K^- p \rightarrow \pi^- \Sigma^+$, respectively. The polarizations are shown as a function of $\cos\theta$, where $\theta$ is the c.m.\ scattering angle of the meson. Figure 8 shows the excellent agreement of our energy-dependent solution with the $K^-p\rightarrow K^-p$ polarization at $P_{\rm Lab}$ = 1383, 1483, 1584, 1684 MeV from Ref. \cite{Daum1968}. For $K^-p\rightarrow\pi^0\Lambda$ (Fig.\ 9) our solution is in good agreement with the polarization data at $P_{\rm Lab}$ = 514, 936, and 1165 MeV. Our solution also agrees well the Crystal Ball data at $P_{\rm Lab}$ = 714 MeV at forward angles but there is a slight under representation of data at backward angles. Similarly, Fig.\ 10 shows a very good agreement between our solution and the $K^-p\rightarrow \pi^0\Sigma^0$ polarization data within the given uncertainties at $P_{\rm Lab}$ = 514, 581, 687, and 750 MeV, all from Crystal Ball Collaboration. Finally, Fig.\ 11 shows a comparison of our solution with the $K^-p\rightarrow \pi^-\Sigma^+$ polarization data at  $P_{\rm Lab}$ = 862, 936, 1001, and 1125 MeV. Except for small forward angles at $P_{\rm Lab}$ = 1001 MeV, we have good agreement with the data.

\begin{figure}[htpb]
\vspace{-15mm}
\scalebox{0.35}{\includegraphics{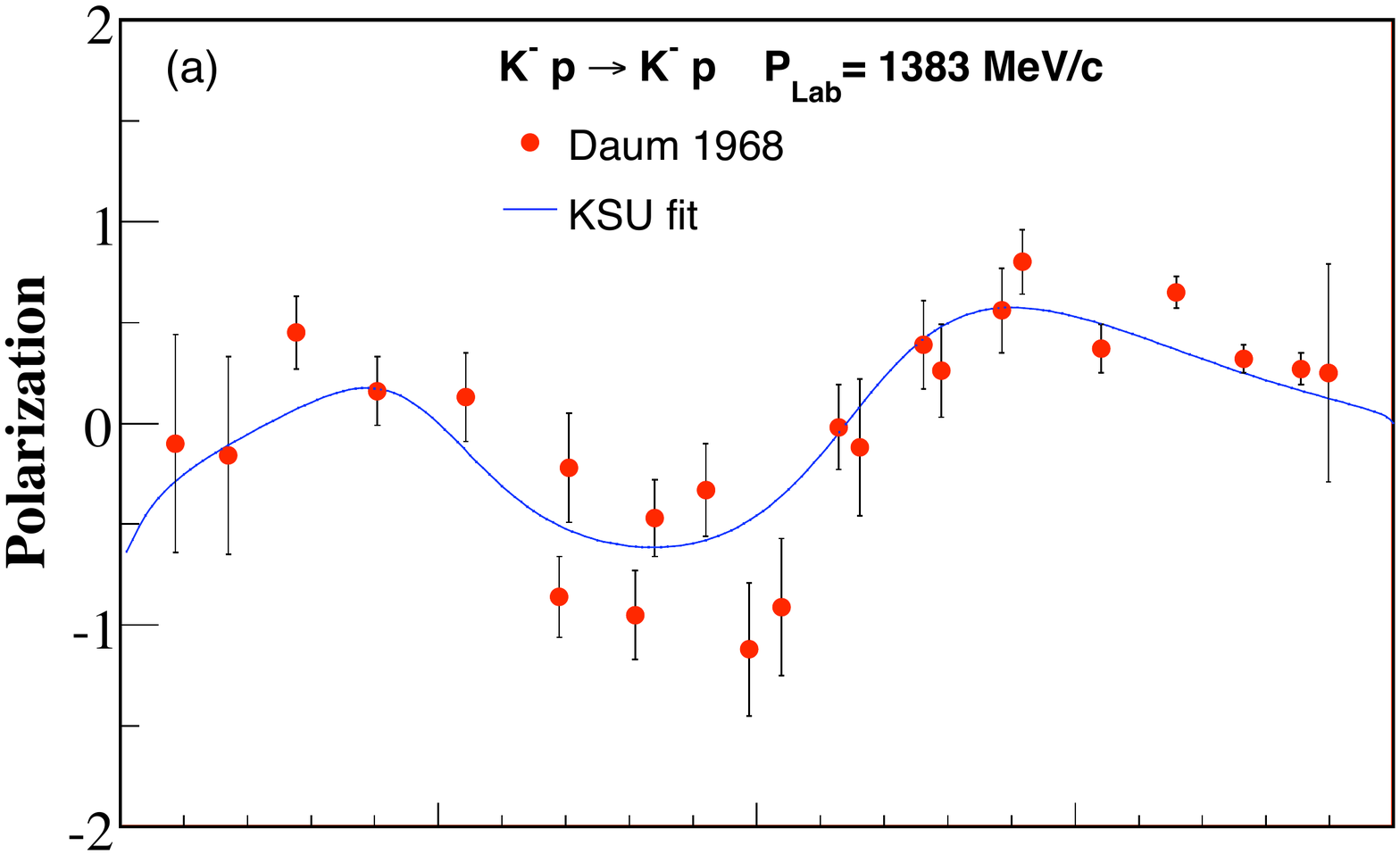}}
\vspace{-1mm}
\vspace{-25mm}
\scalebox{0.35}{\includegraphics{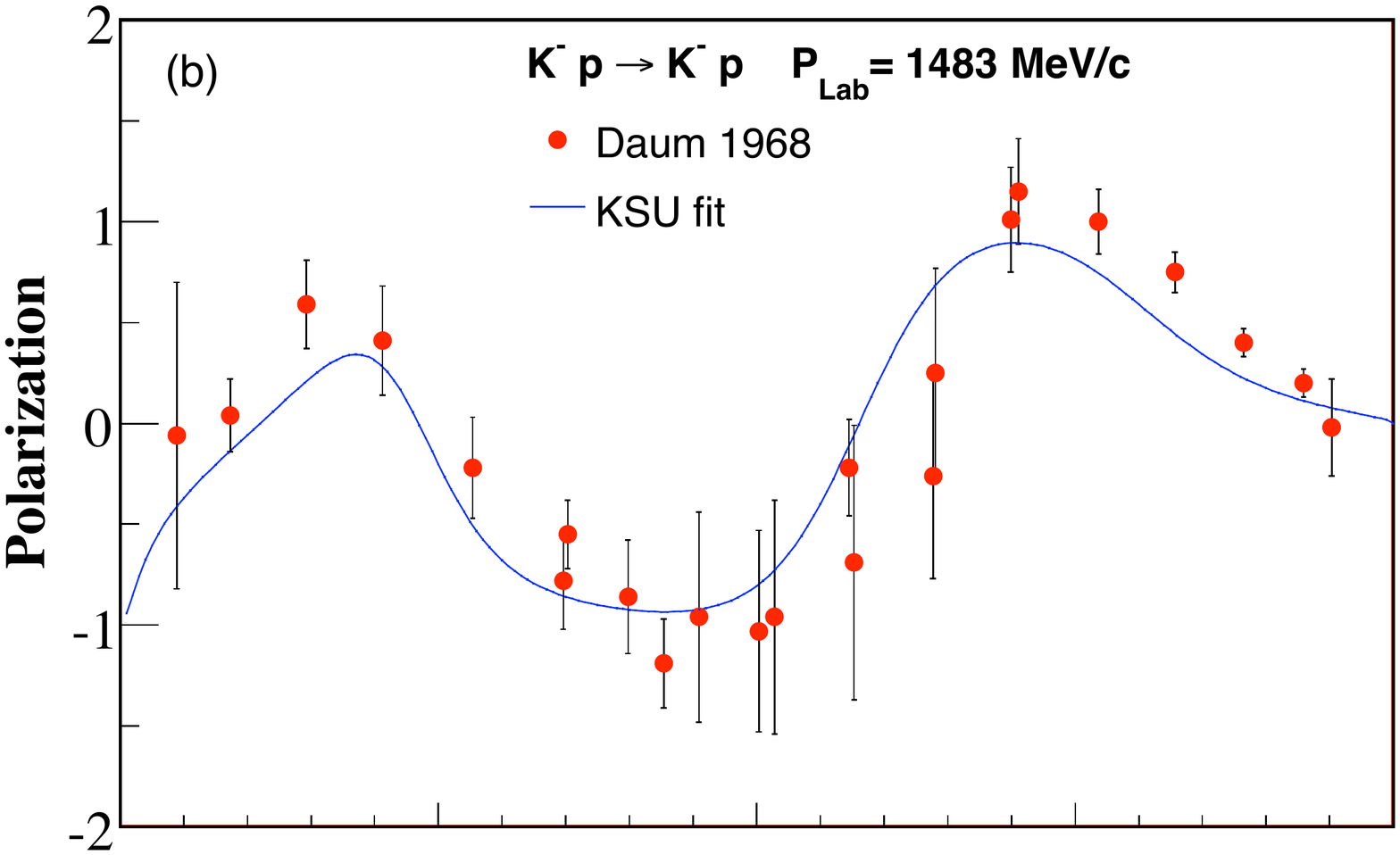}}
\vspace{-25mm}
\vspace{-1mm}
\scalebox{0.35}{\includegraphics{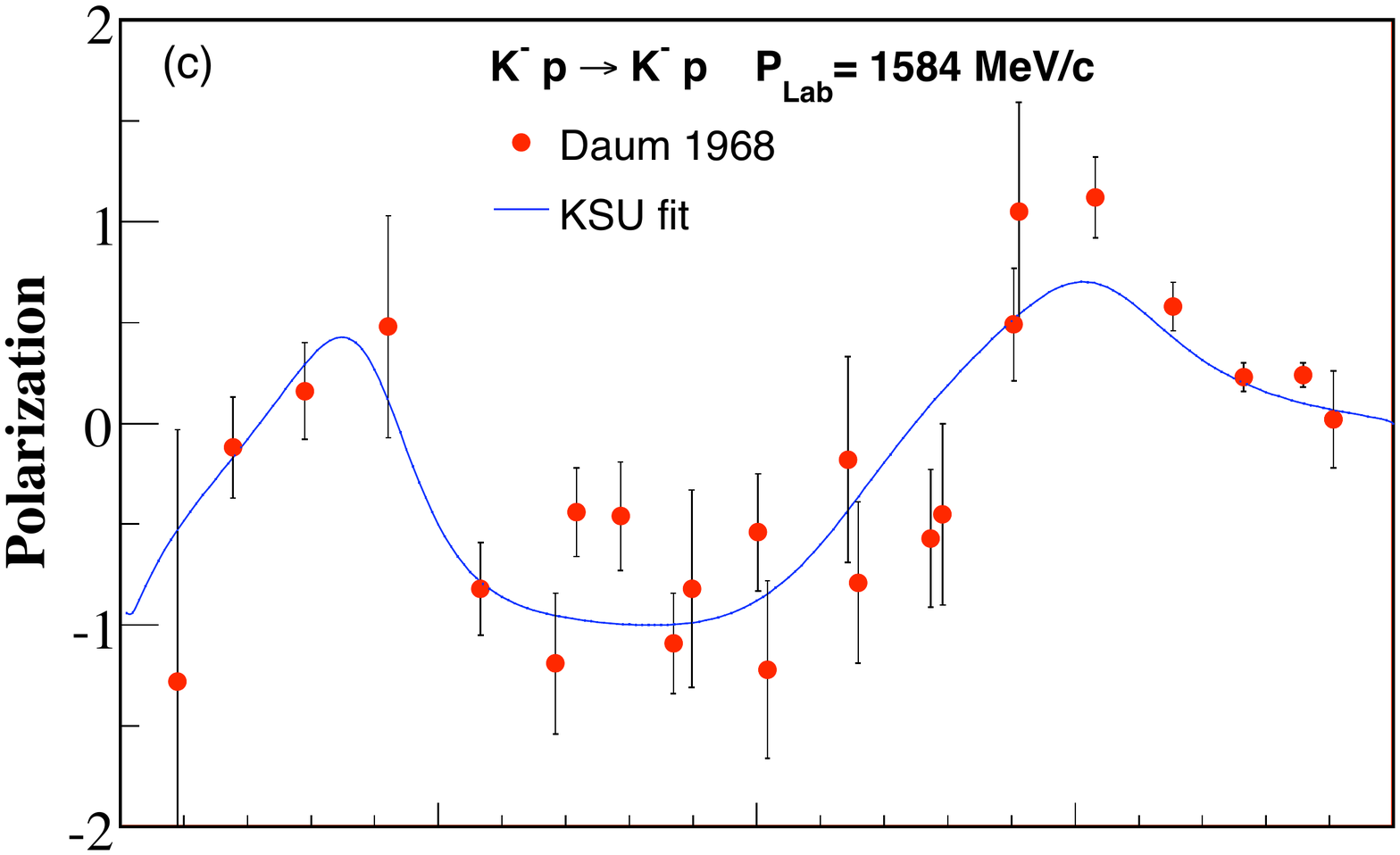}} 
\vspace{-25mm}
\vspace{3mm}
\scalebox{0.35}{\includegraphics{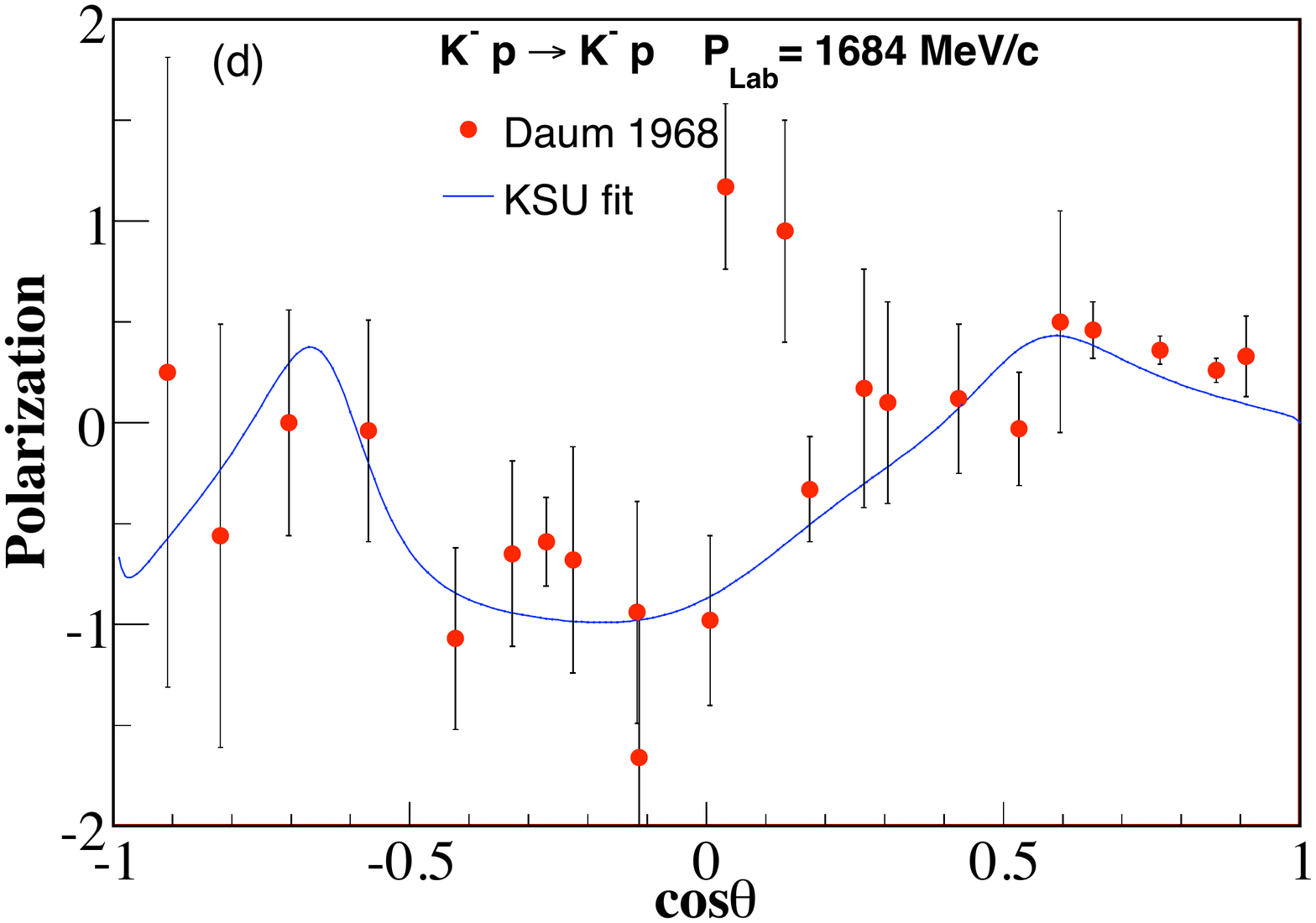}}
\vspace{-5mm}
\caption{(Color online) Representative results of our energy-dependent fit for the $K^- p \rightarrow K^-p$ differential cross section. Data are from Daum 1968 \cite{Daum1968}.} 
\label{fig:dSigma_11_New}
\end{figure}


\begin{figure}[htpb]
\vspace{-15mm}
\scalebox{0.35}{\includegraphics{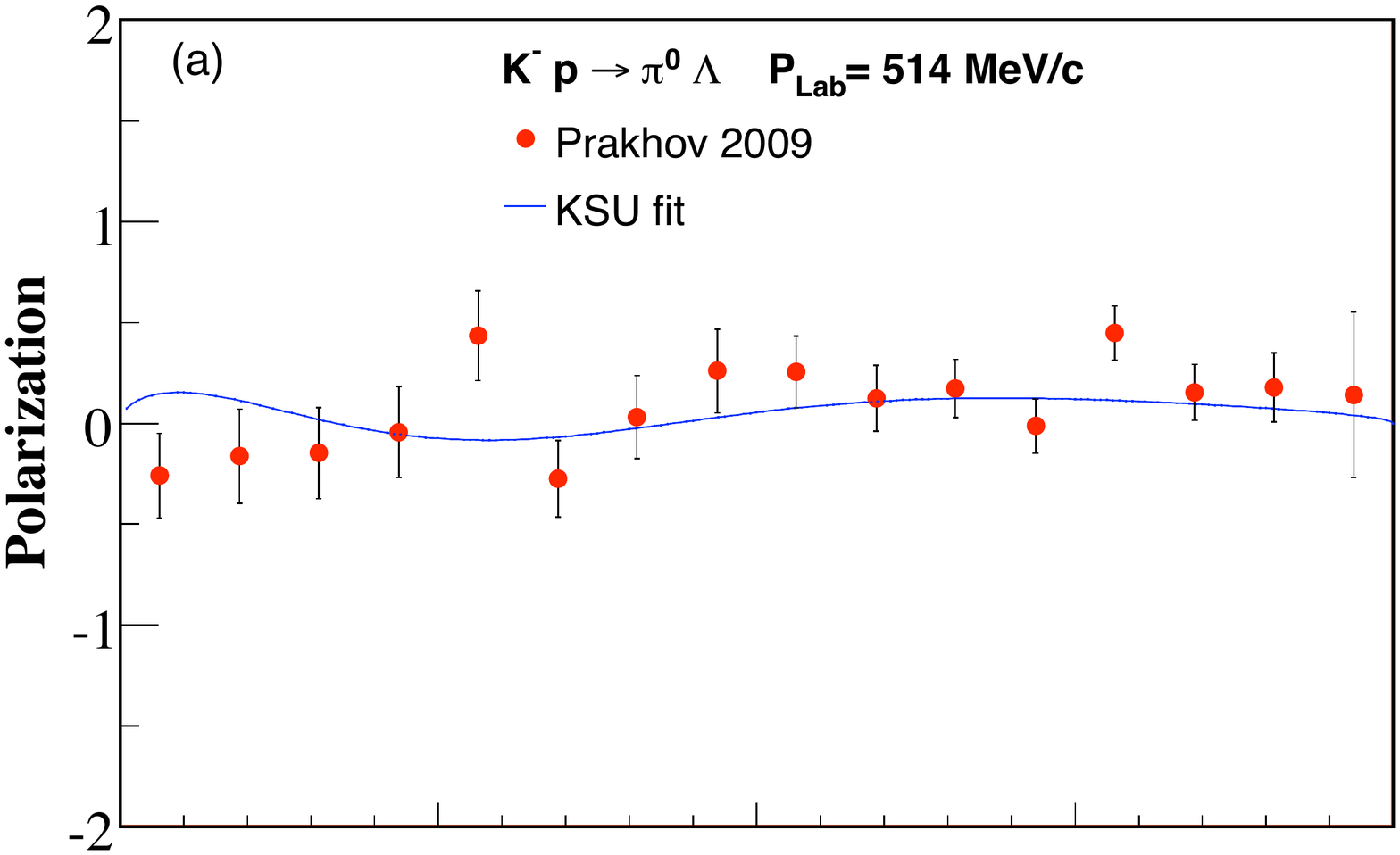}}
\vspace{-1mm}
\vspace{-25mm}
\scalebox{0.35}{\includegraphics{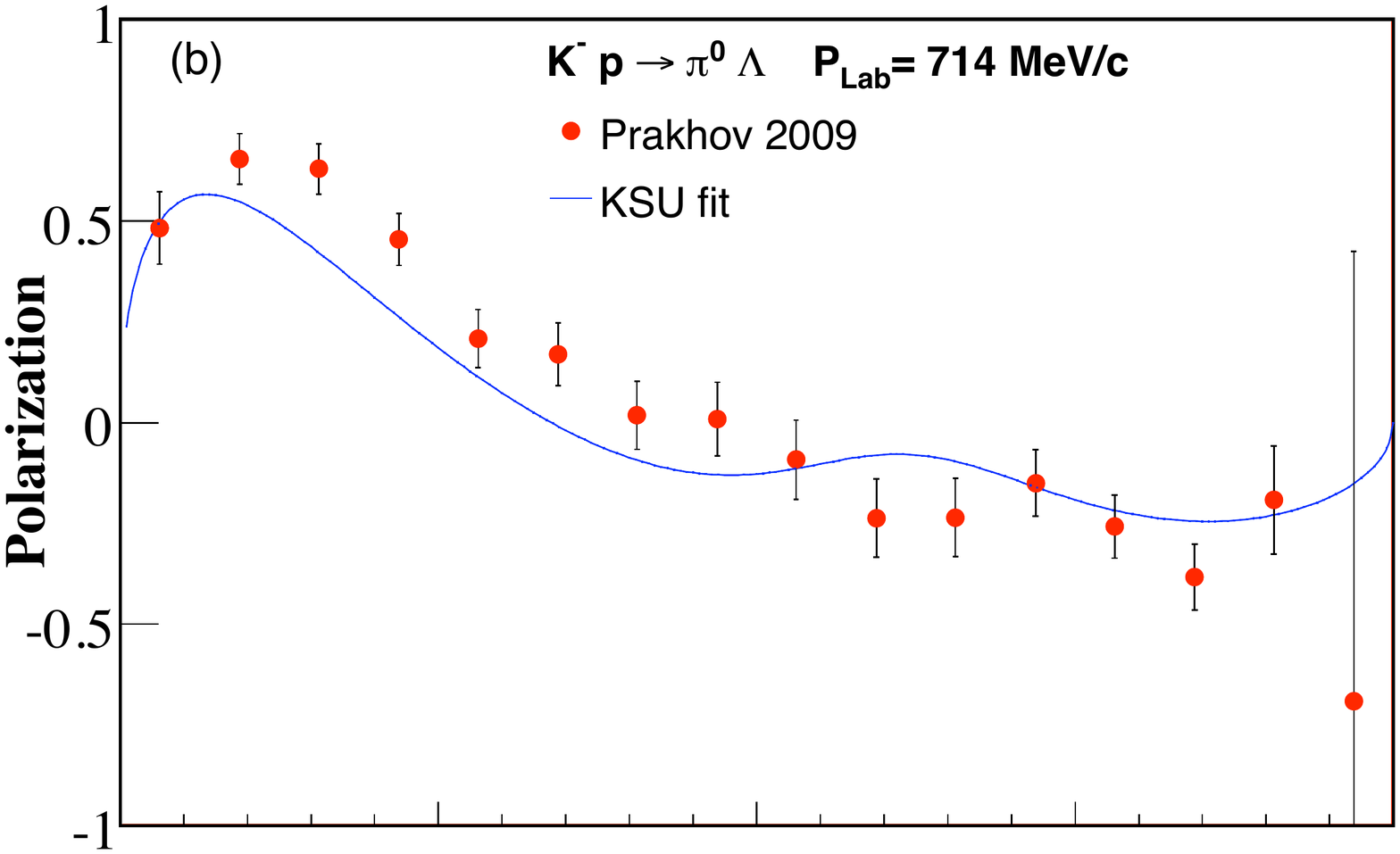}}
\vspace{-25mm}
\vspace{-1mm}
\scalebox{0.35}{\includegraphics{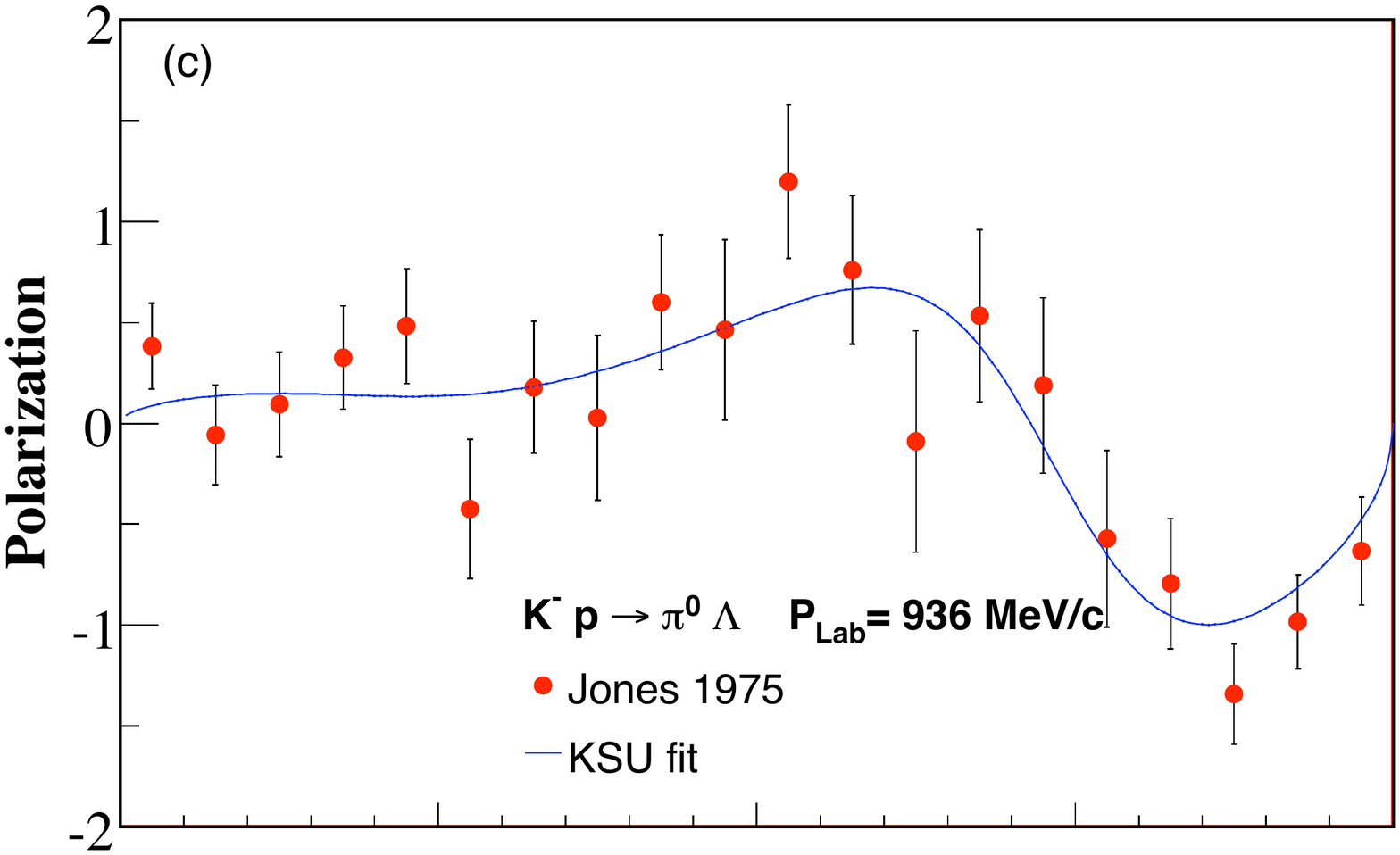}} 
\vspace{-25mm}
\vspace{-1mm}
\scalebox{0.35}{\includegraphics{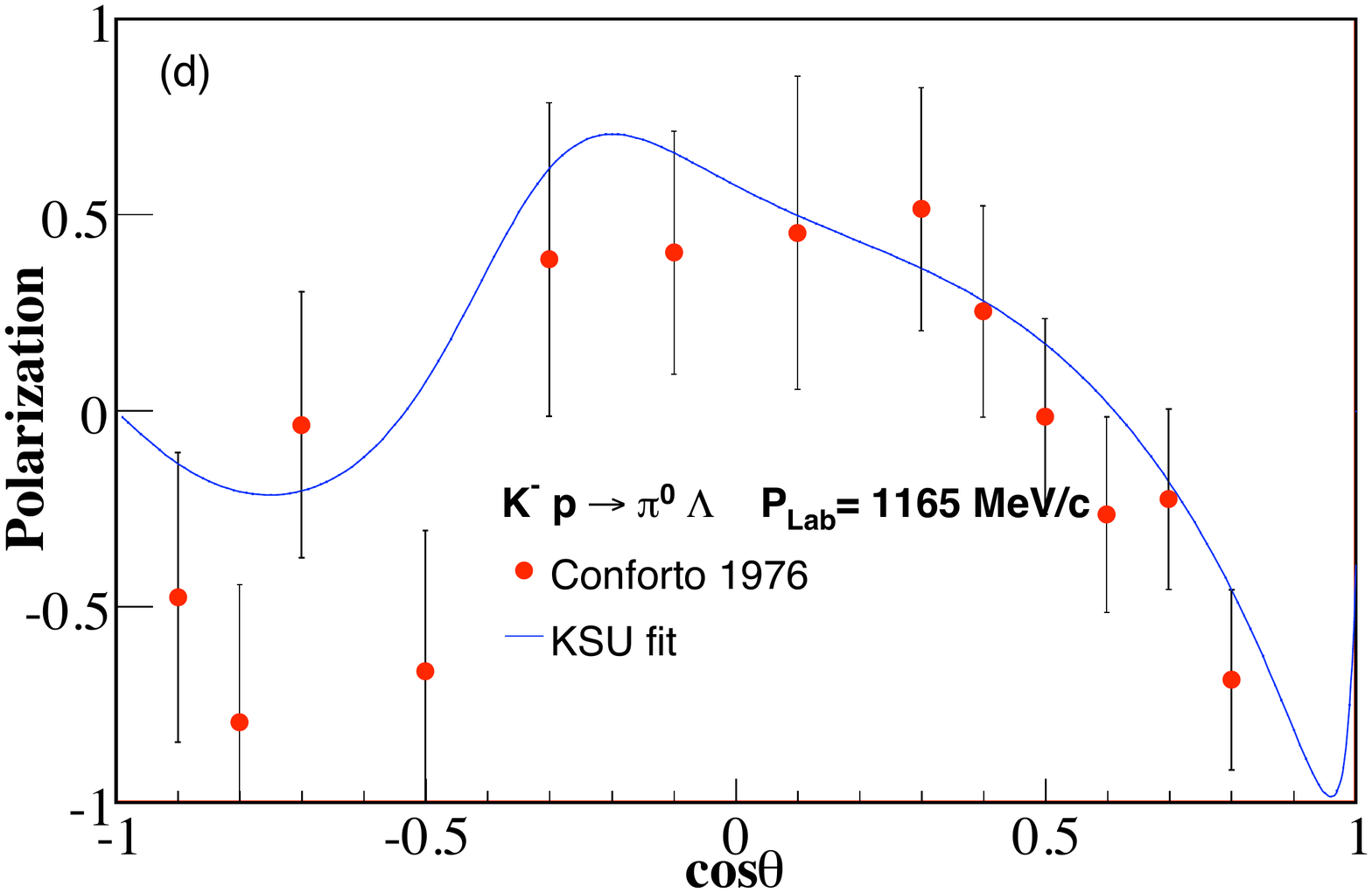}}
\vspace{-5mm}
\caption{(Color online) Representative results of our energy-dependent fit for the $K^- p \rightarrow \pi^0\Lambda$ differential cross section. Data are from Armenteros 1970 \cite{Armenteros1970}, Armenteros 1968 \cite{Armenteros1968}, Conforto 1976 \cite{Conforto1976}, and Berthon 1970 \cite{Berthon1970}.}
\label{fig:dSigma_11_New}
\end{figure}

\begin{figure}[htpb]
\vspace{-15mm}
\scalebox{0.35}{\includegraphics{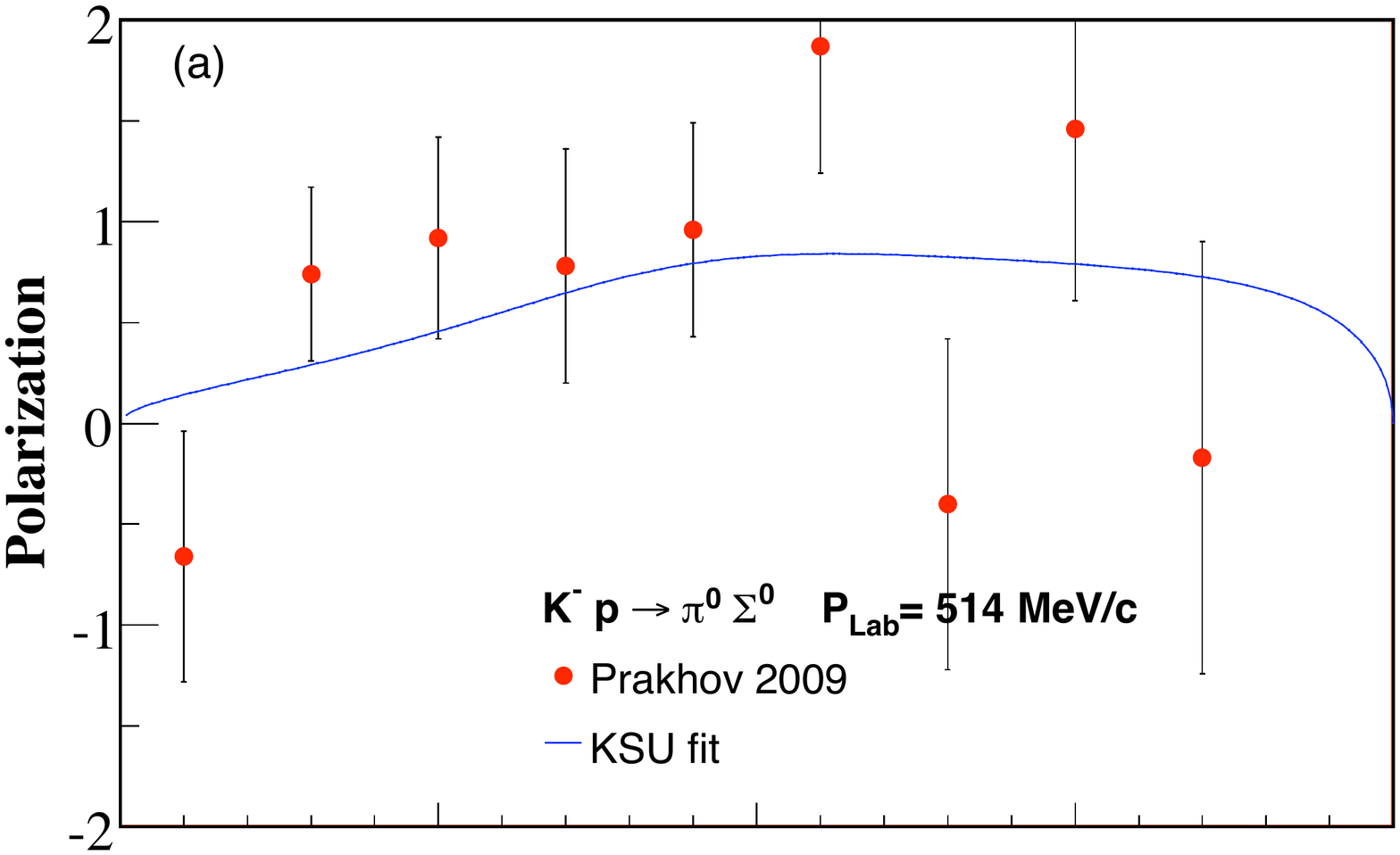}}
\vspace{-1mm}
\vspace{-25mm}
\scalebox{0.35}{\includegraphics{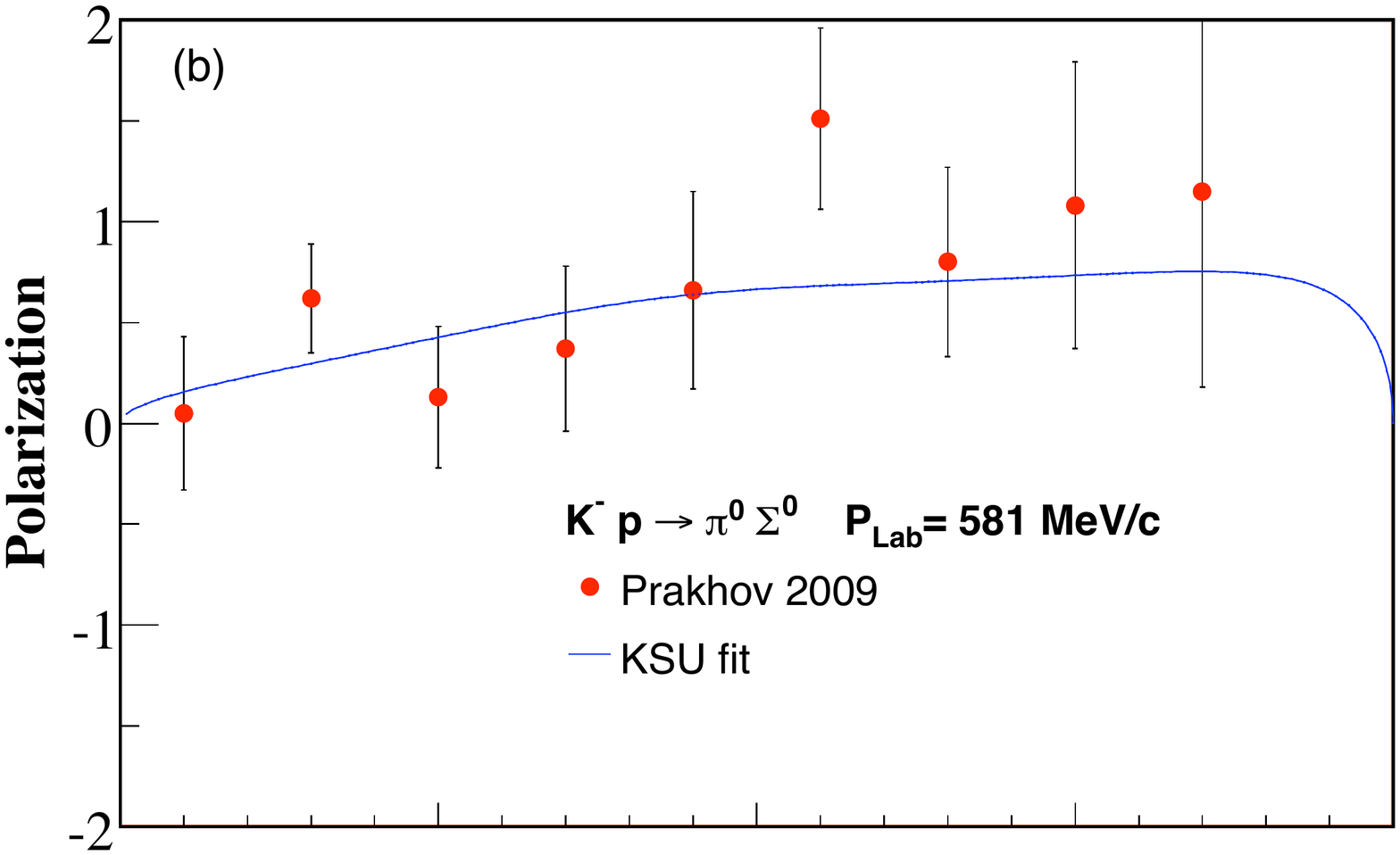}}
\vspace{-25mm}
\vspace{-1mm}
\scalebox{0.35}{\includegraphics{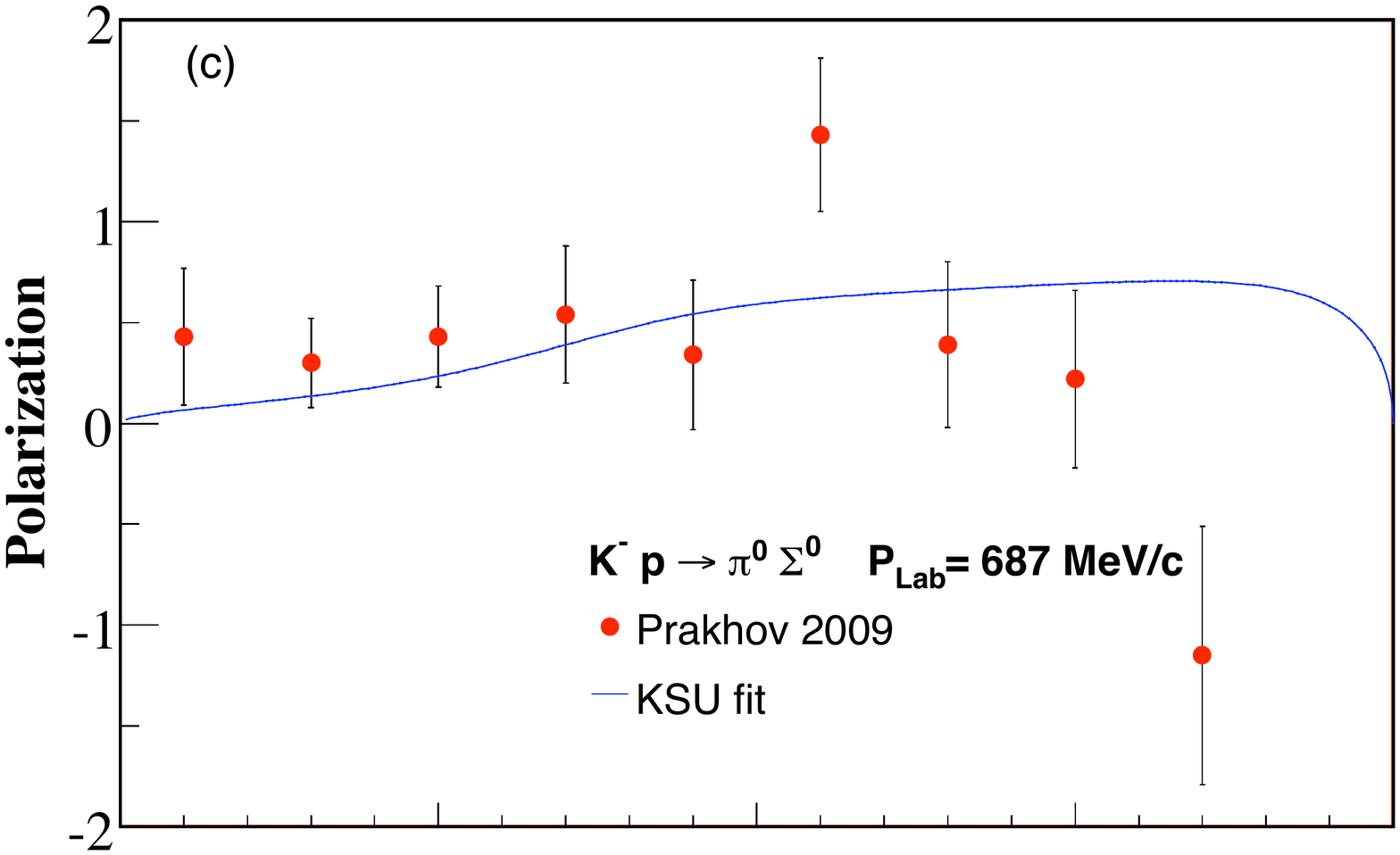}} 
\vspace{-25mm}
\vspace{-1mm}
\scalebox{0.35}{\includegraphics{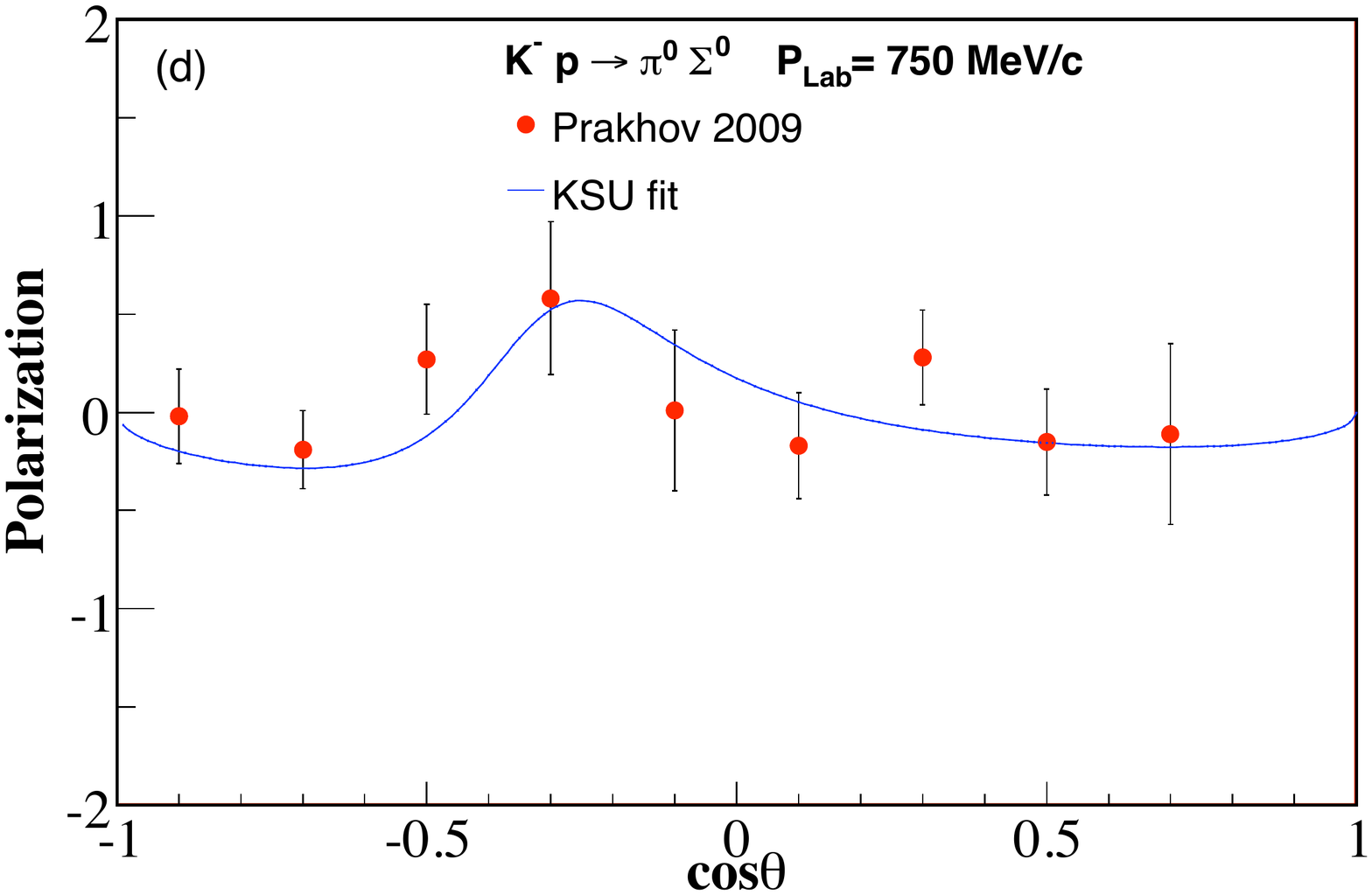}}
\vspace{-5mm}
\caption{(Color online) Representative results of our energy-dependent fit for the $K^- p \rightarrow \pi^-\Sigma^+$ differential cross section. Data are from Armenteros 1970 \cite{Armenteros1970}, Armenteros 1968 \cite{Armenteros1968}, Conforto 1976 \cite{Conforto1976}, and Berthon 1970 \cite{Berthon1970}.}
\label{fig:dSigma_11_New}
\end{figure}

\begin{figure}[htpb]
\vspace{-15mm}
\scalebox{0.35}{\includegraphics{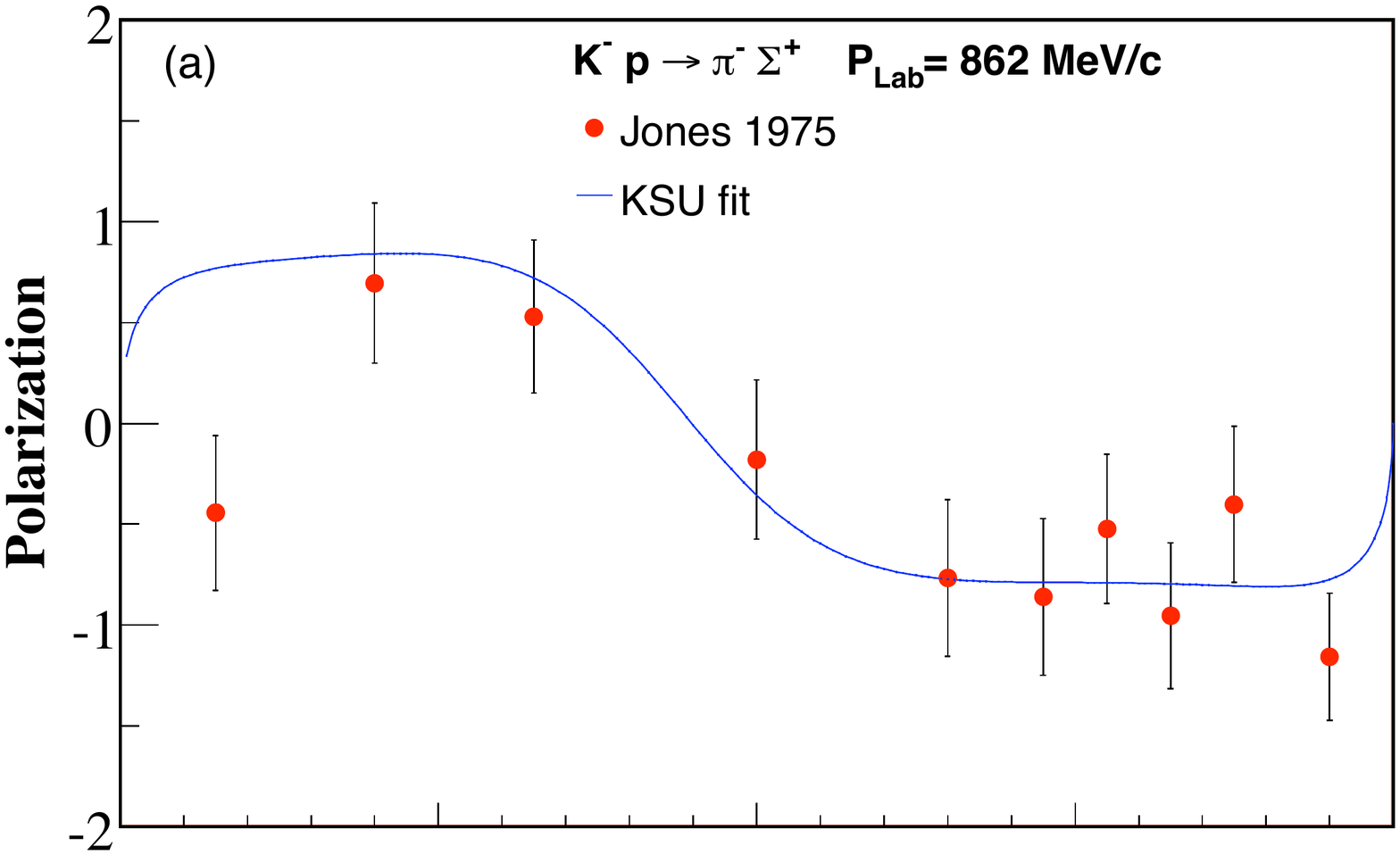}}
\vspace{-1mm}
\vspace{-25mm}
\scalebox{0.35}{\includegraphics{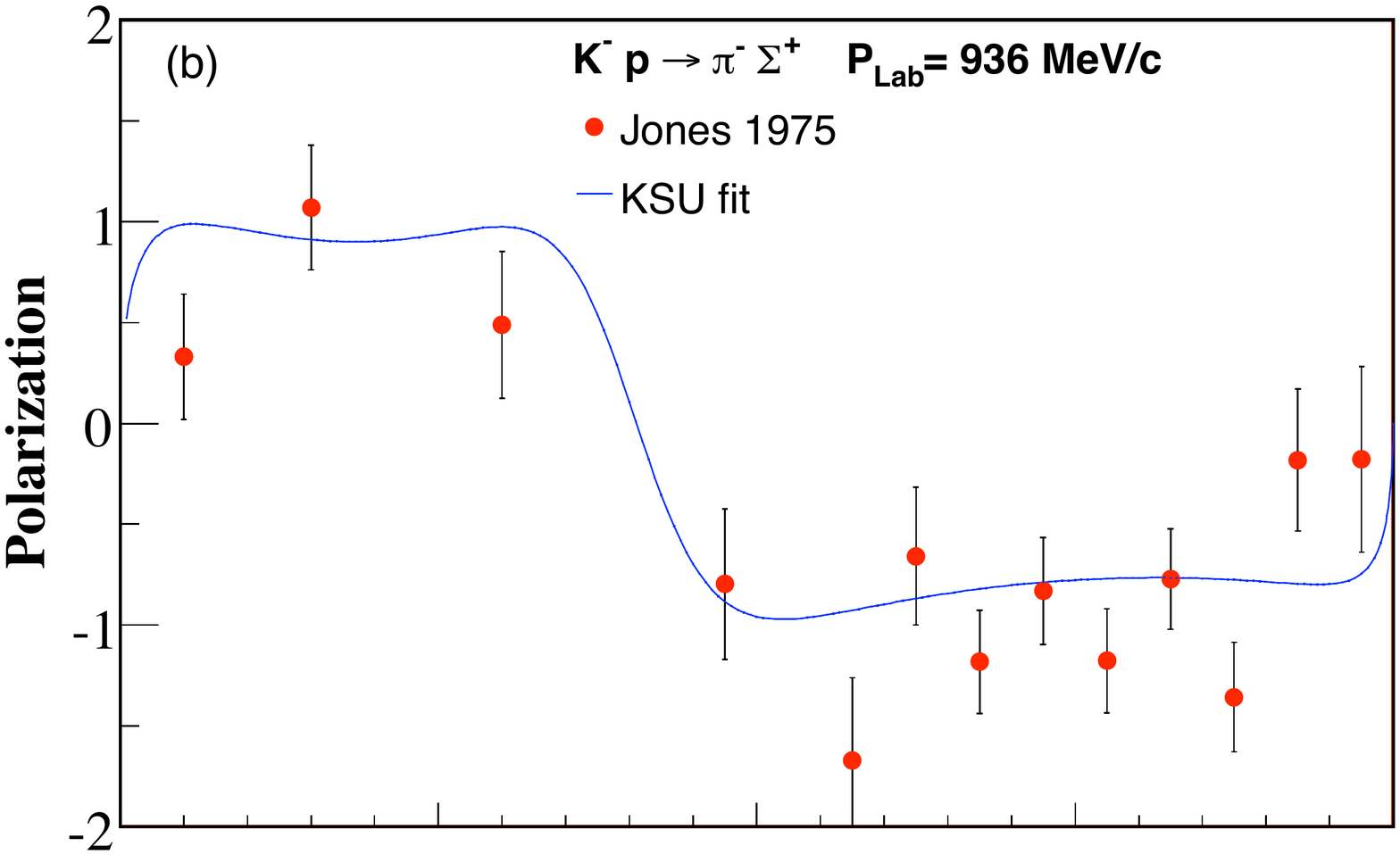}}
\vspace{-25mm}
\vspace{-1mm}
\scalebox{0.35}{\includegraphics{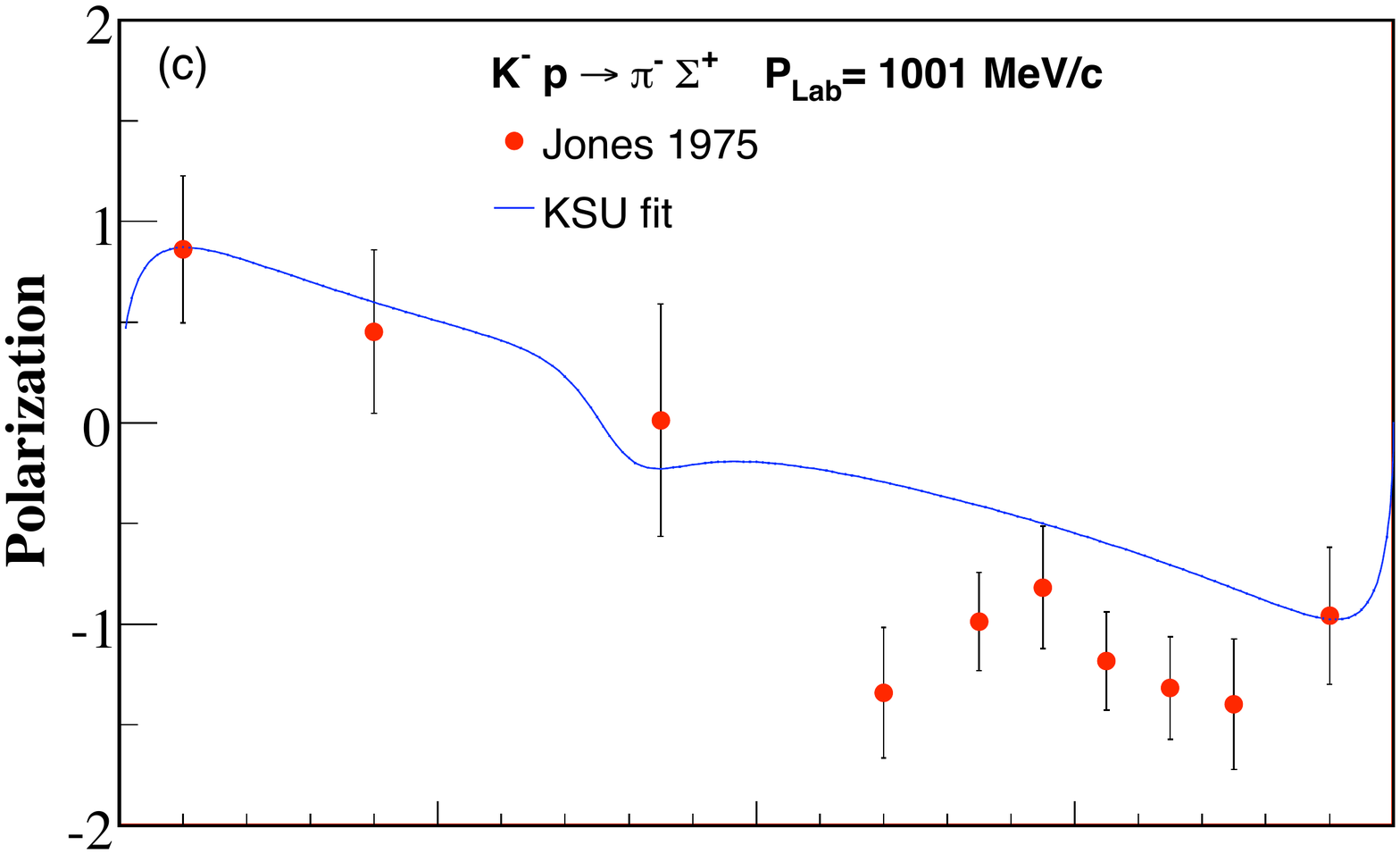}} 
\vspace{-25mm}
\vspace{-1mm}
\scalebox{0.35}{\includegraphics{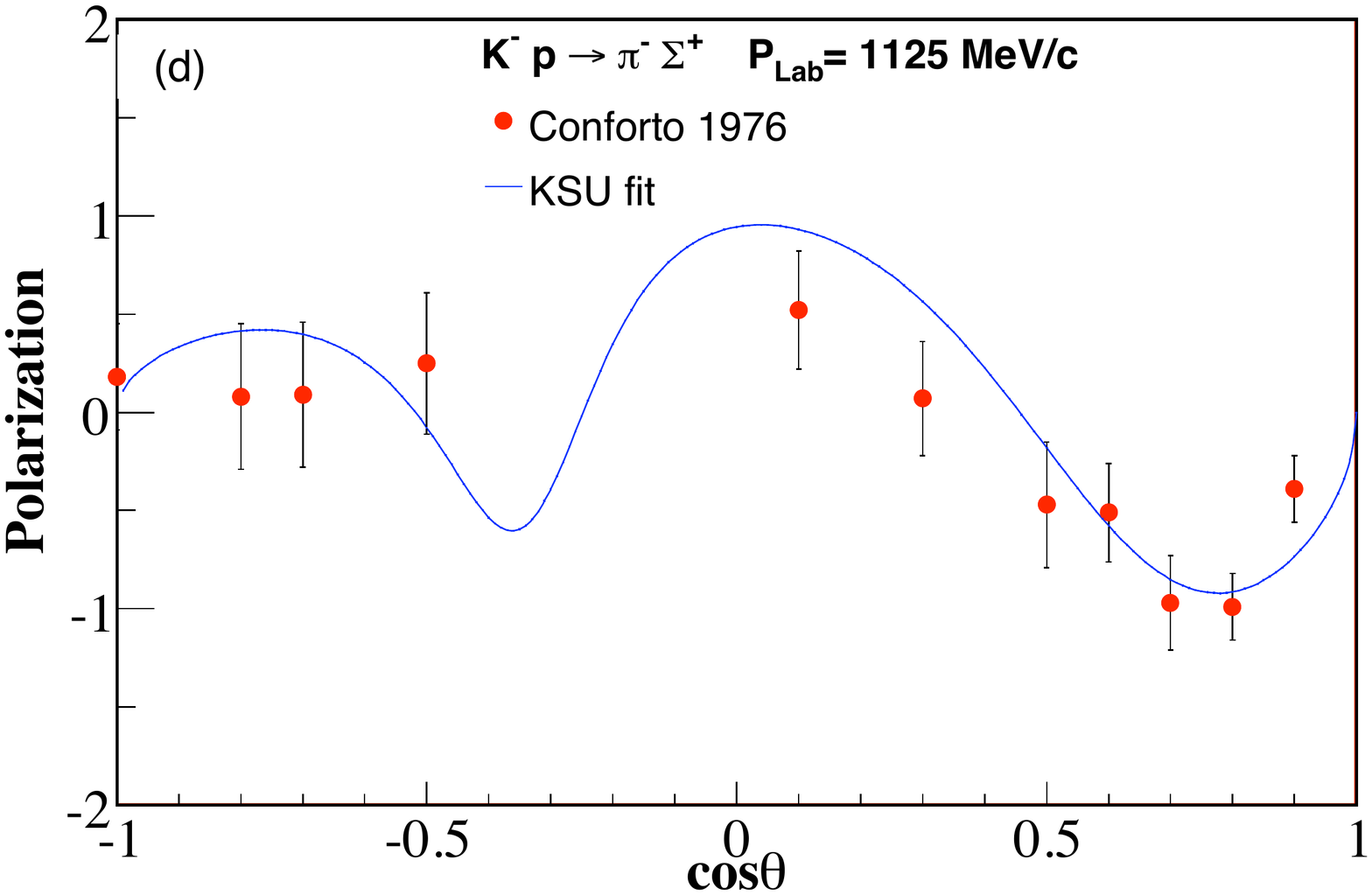}}
\vspace{-5mm}
\caption{(Color online) Representative results of our energy-dependent fit for the $K^- p \rightarrow \pi^-\Sigma^+$ differential cross section. Data are from Armenteros 1970 \cite{Armenteros1970}, Armenteros 1968 \cite{Armenteros1968}, Conforto 1976 \cite{Conforto1976}, and Berthon 1970 \cite{Berthon1970}.}
\label{fig:dSigma_11_New}
\end{figure}

Figures 12, 13, 14, and 15 show representative energy-dependent fit results for the polarized cross section in reactions $K^- p \rightarrow K^- p$, $K^- p \rightarrow \pi\Lambda$, $K^- p \rightarrow \pi^0\Sigma^0 $, and $K^- p \rightarrow \pi^- \Sigma^+$, respectively. The polarized cross sections are shown as a function of $\cos\theta$, where $\theta$ is the c.m.\ scattering angle of the meson.
Within the uncertainties associated with the polarized cross section data our results are in good agreement with the data for  these reactions.

\begin{figure}[htpb]
\vspace{-15mm}
\scalebox{0.35}{\includegraphics{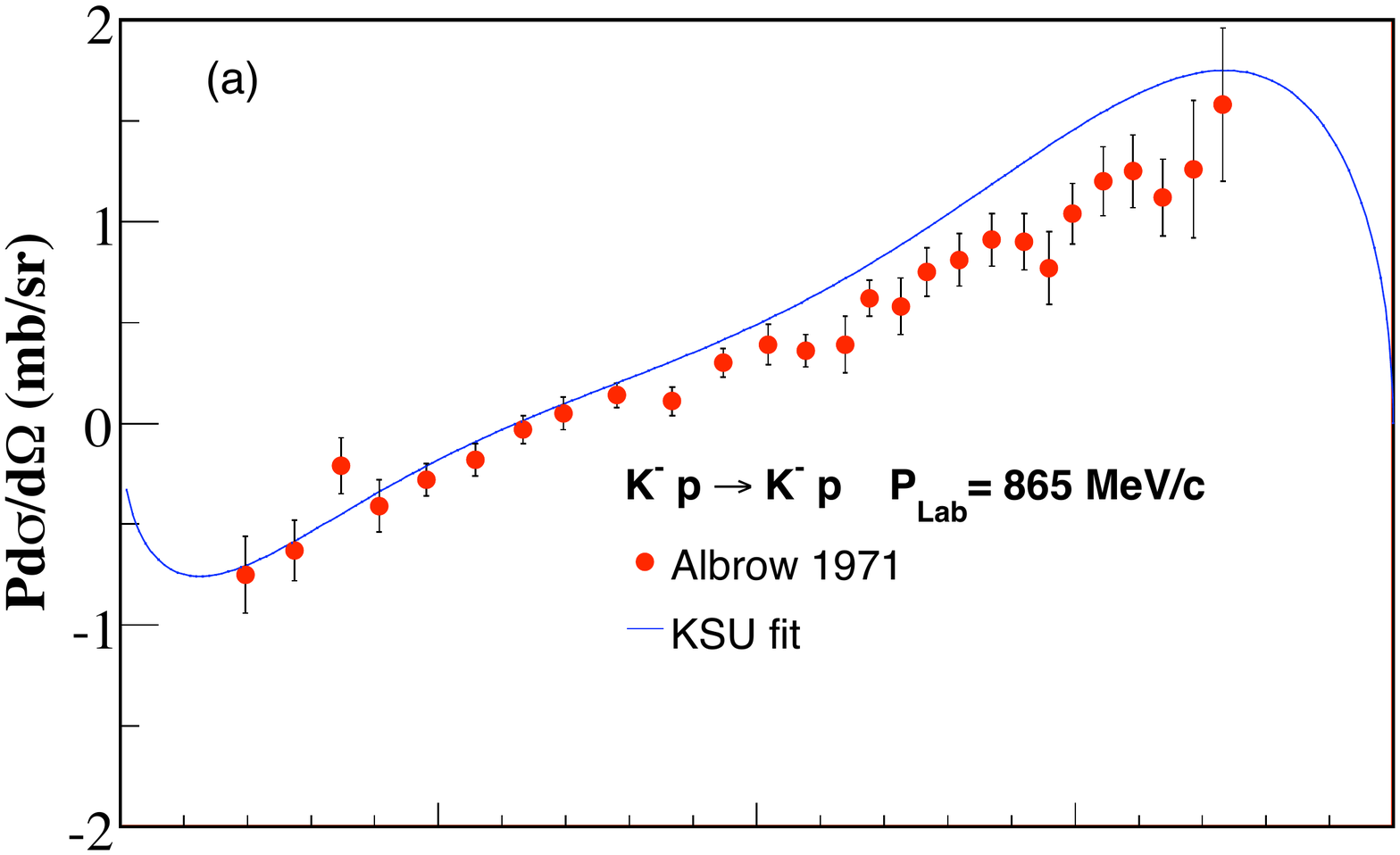}}
\vspace{-1mm}
\vspace{-25mm}
\scalebox{0.35}{\includegraphics{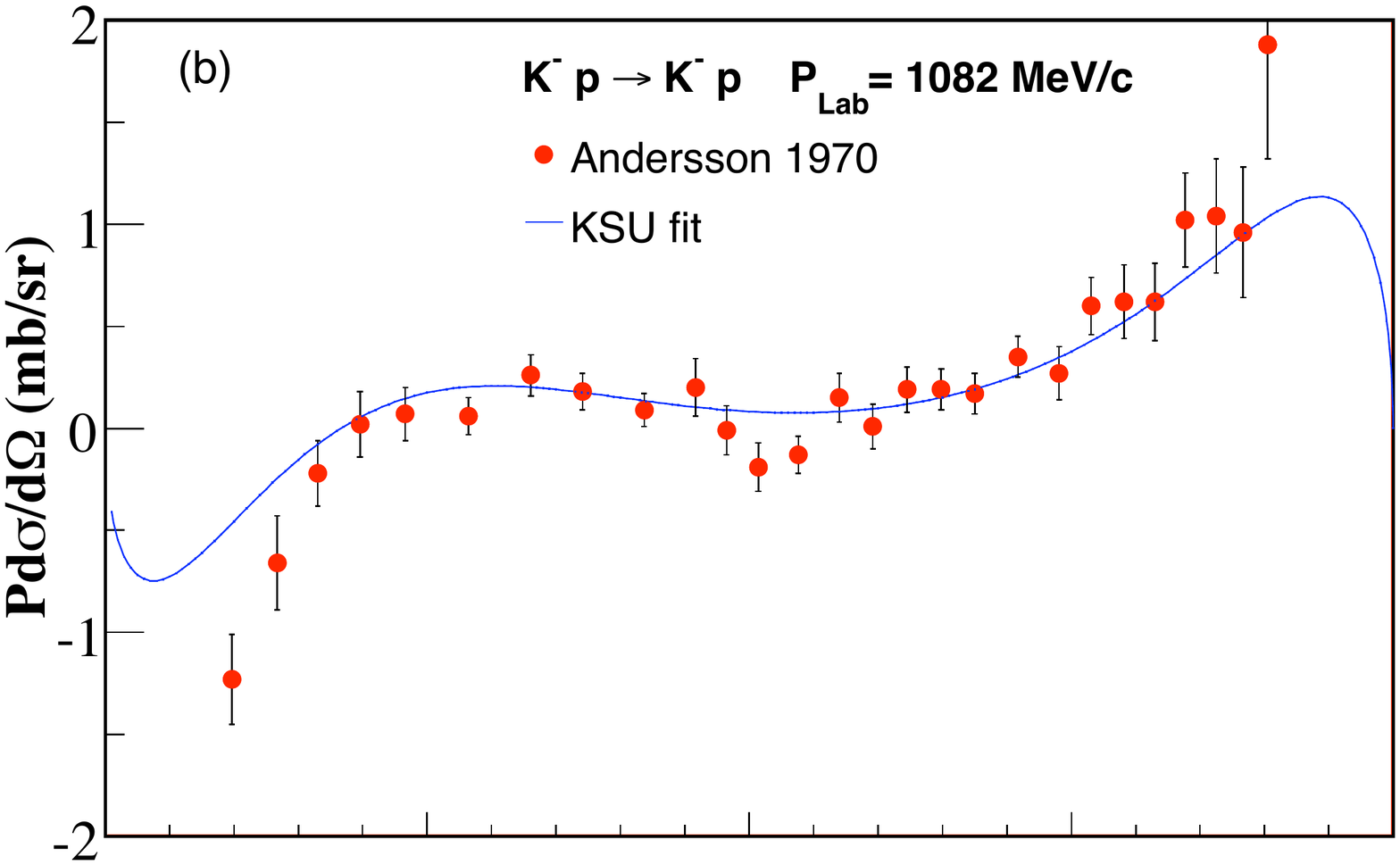}}
\vspace{-25mm}
\vspace{-1mm}
\scalebox{0.35}{\includegraphics{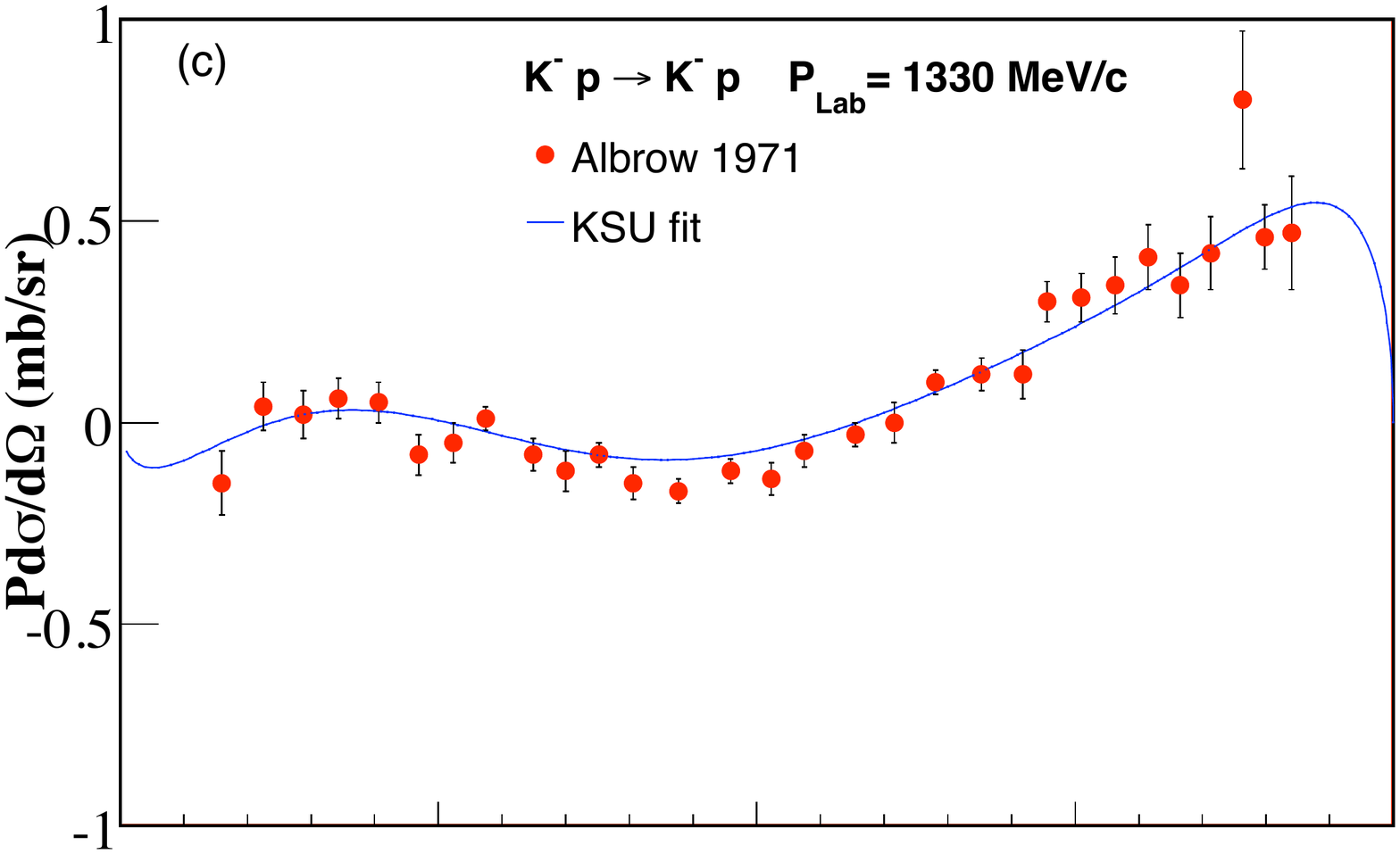}} 
\vspace{-25mm}
\vspace{-1mm}
\scalebox{0.35}{\includegraphics{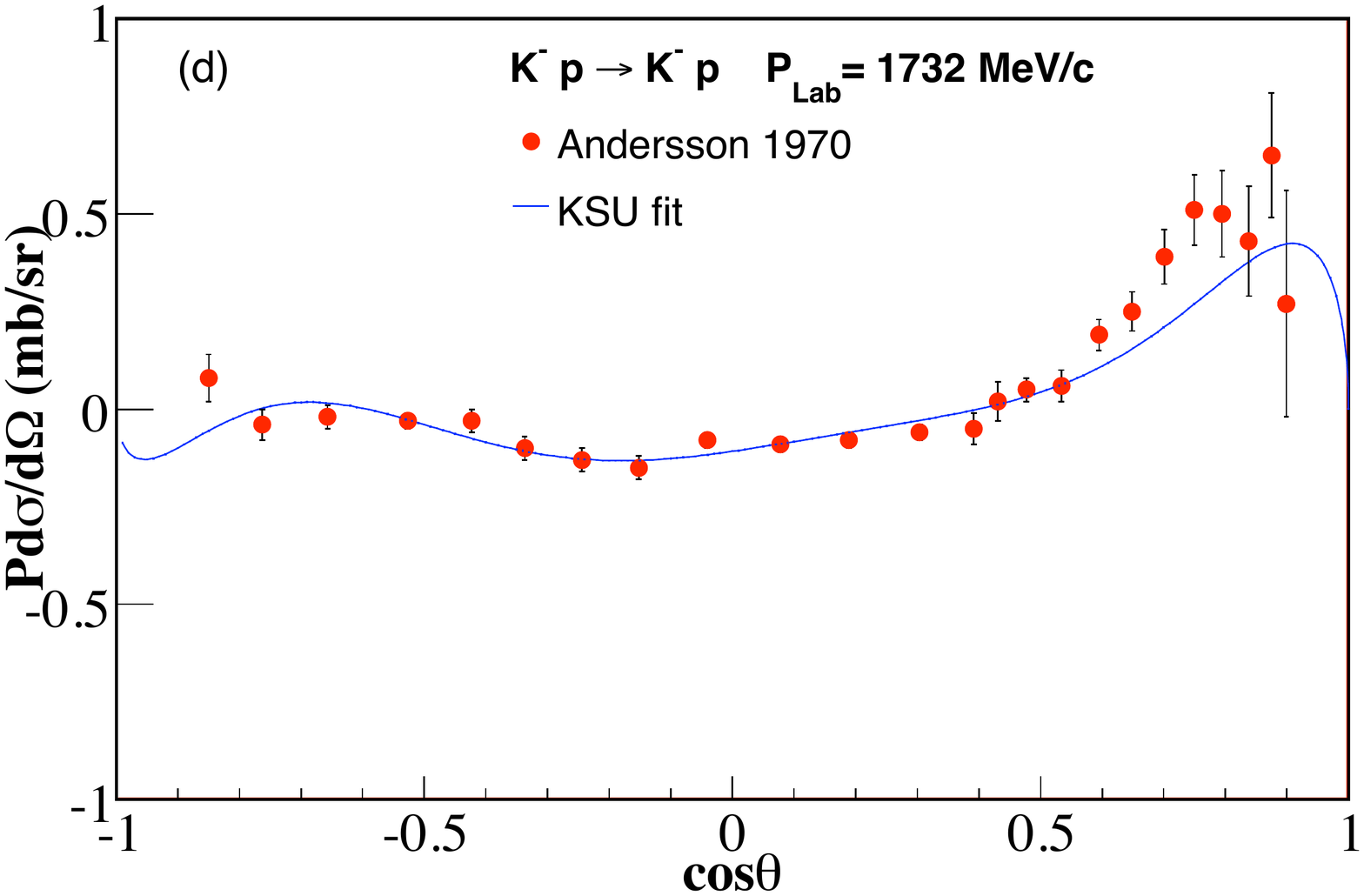}}
\vspace{-5mm}
\caption{(Color online) Representative results of our energy-dependent fit for the $K^- p \rightarrow K^- p$ polarized cross section. Data are from Albrow 1971 \cite{Albrow1971}, Andersson 1970 \cite{Andersson1970}, and Armenteros 1970 \cite{Armenteros1970}.}
\label{fig:dSigma_11_New}
\end{figure}

\begin{figure}[htpb]
\vspace{-10mm}
\includegraphics[width=0.51\textwidth]{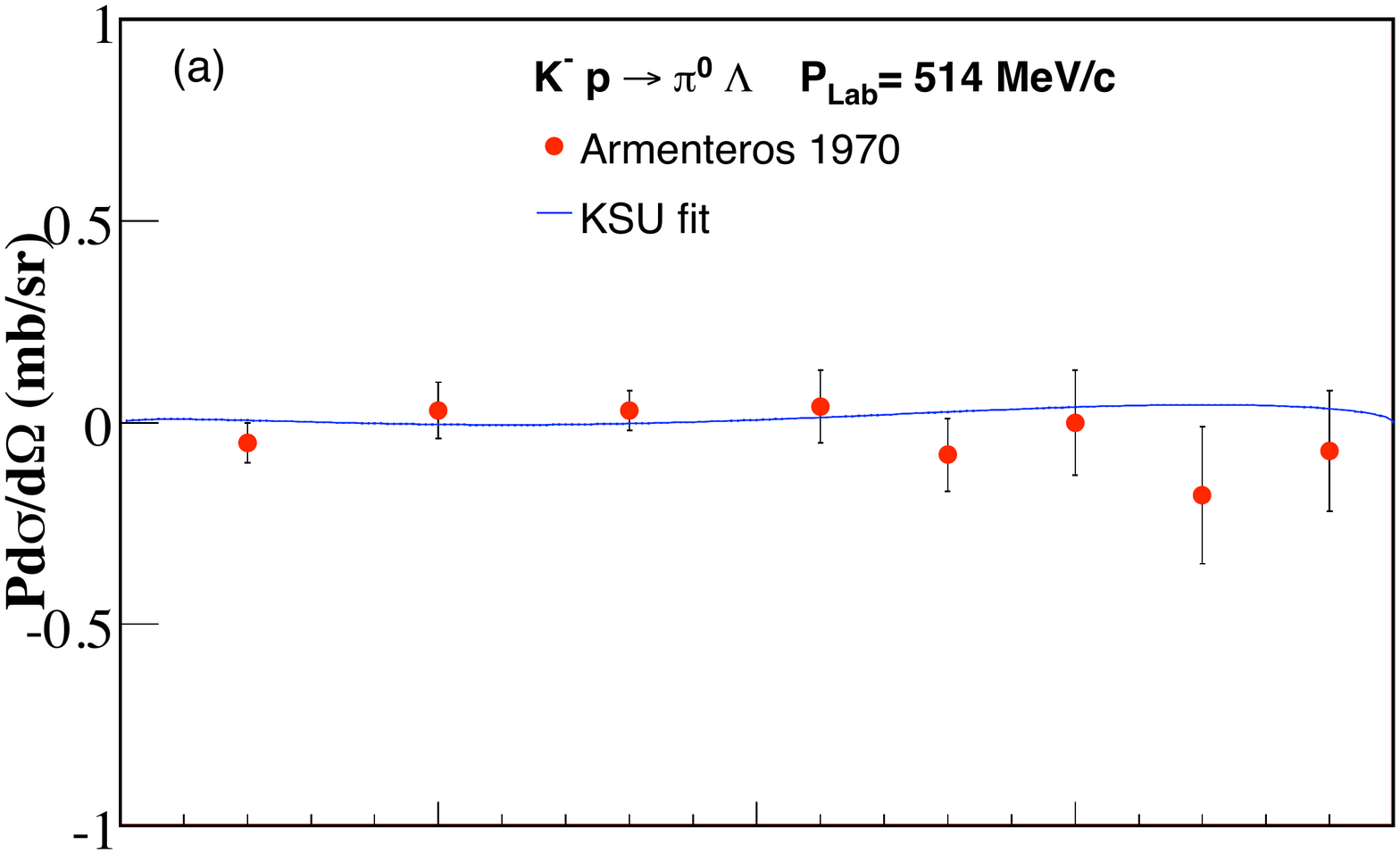}
\vspace{-1mm}
\vspace{-20mm}
\includegraphics[width=0.51\textwidth]{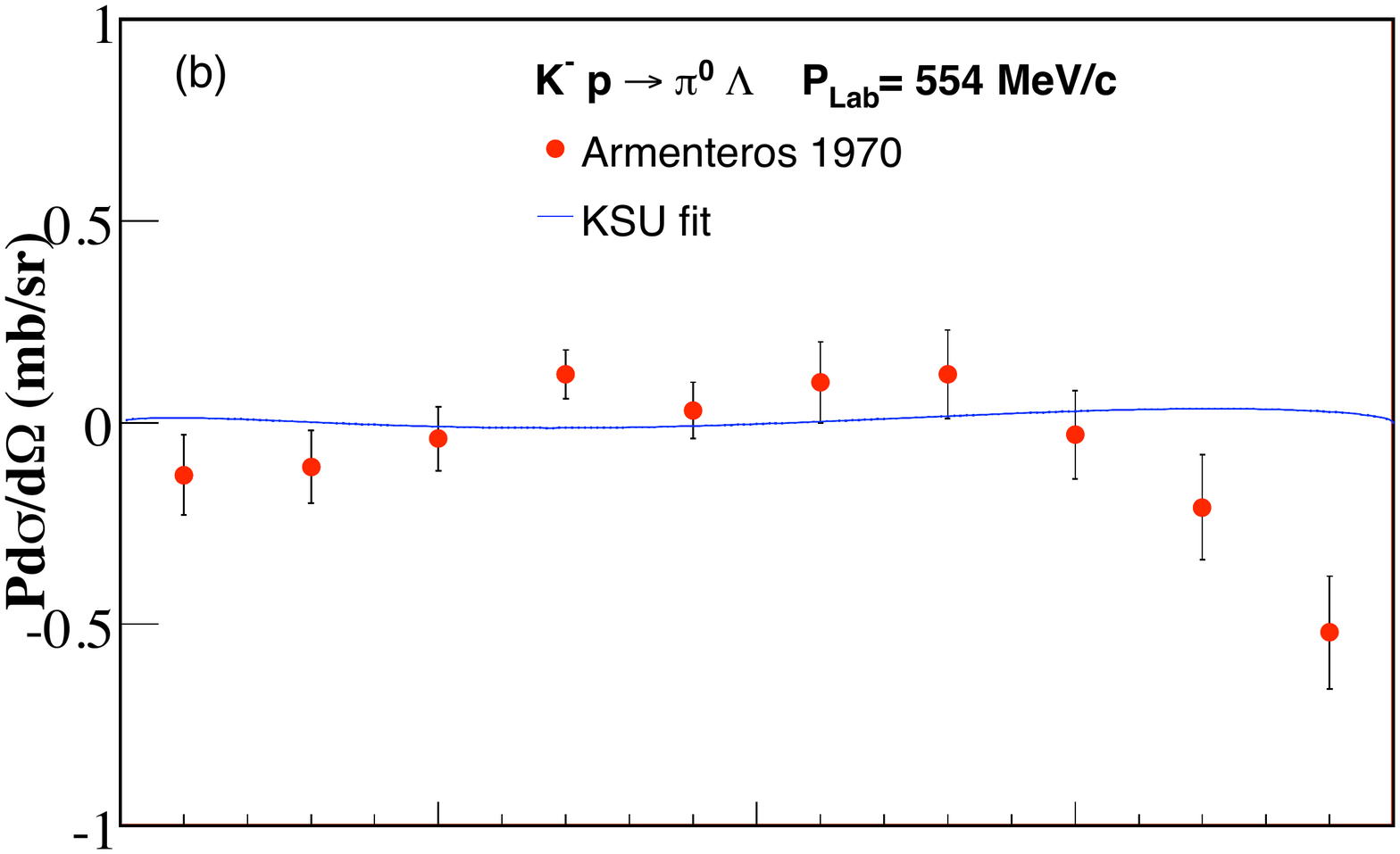}
\vspace{-1mm}
\vspace{-20mm}
\includegraphics[width=0.51\textwidth]{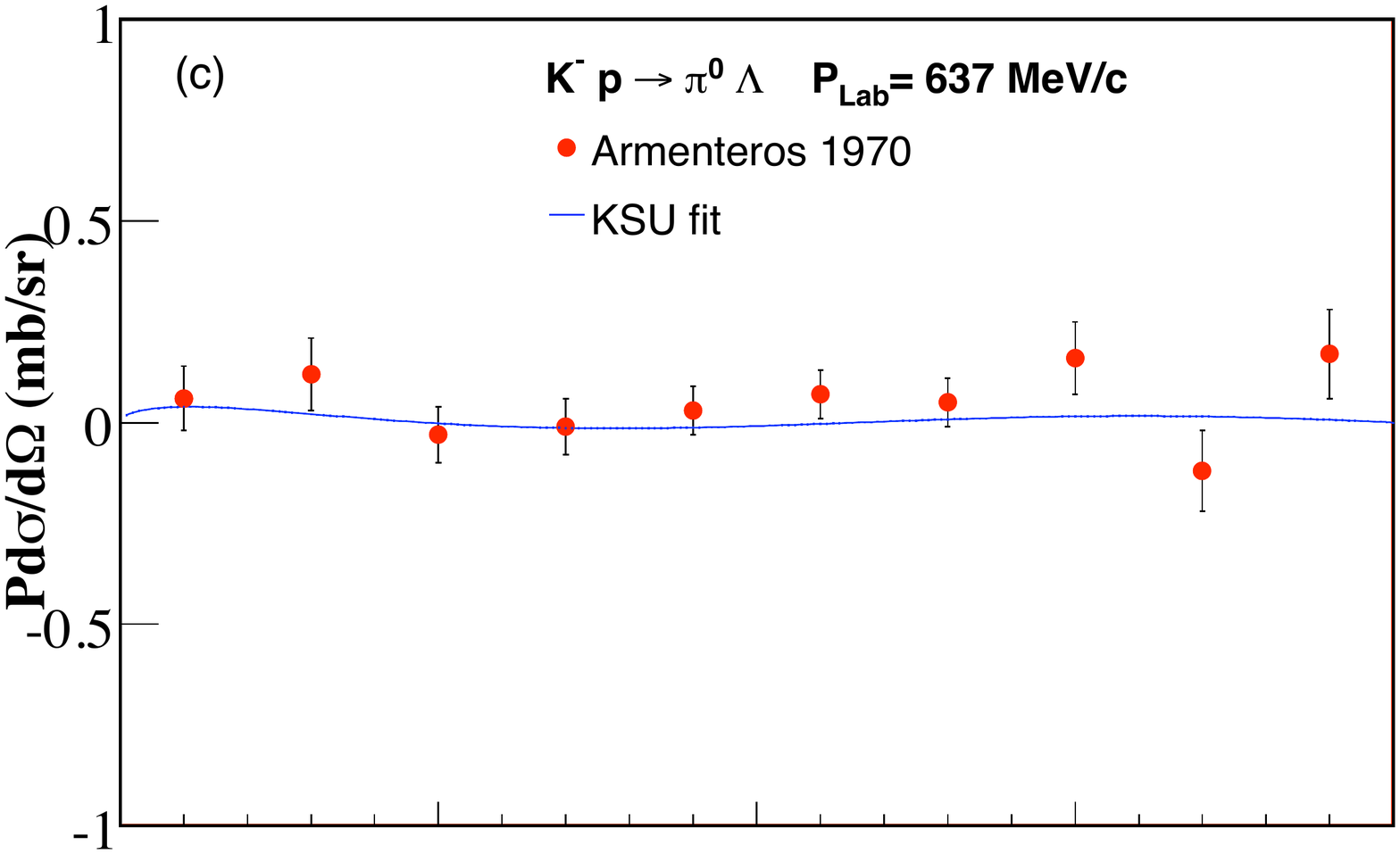}
\vspace{-1mm}
\vspace{-20mm}
\includegraphics[width=0.51\textwidth]{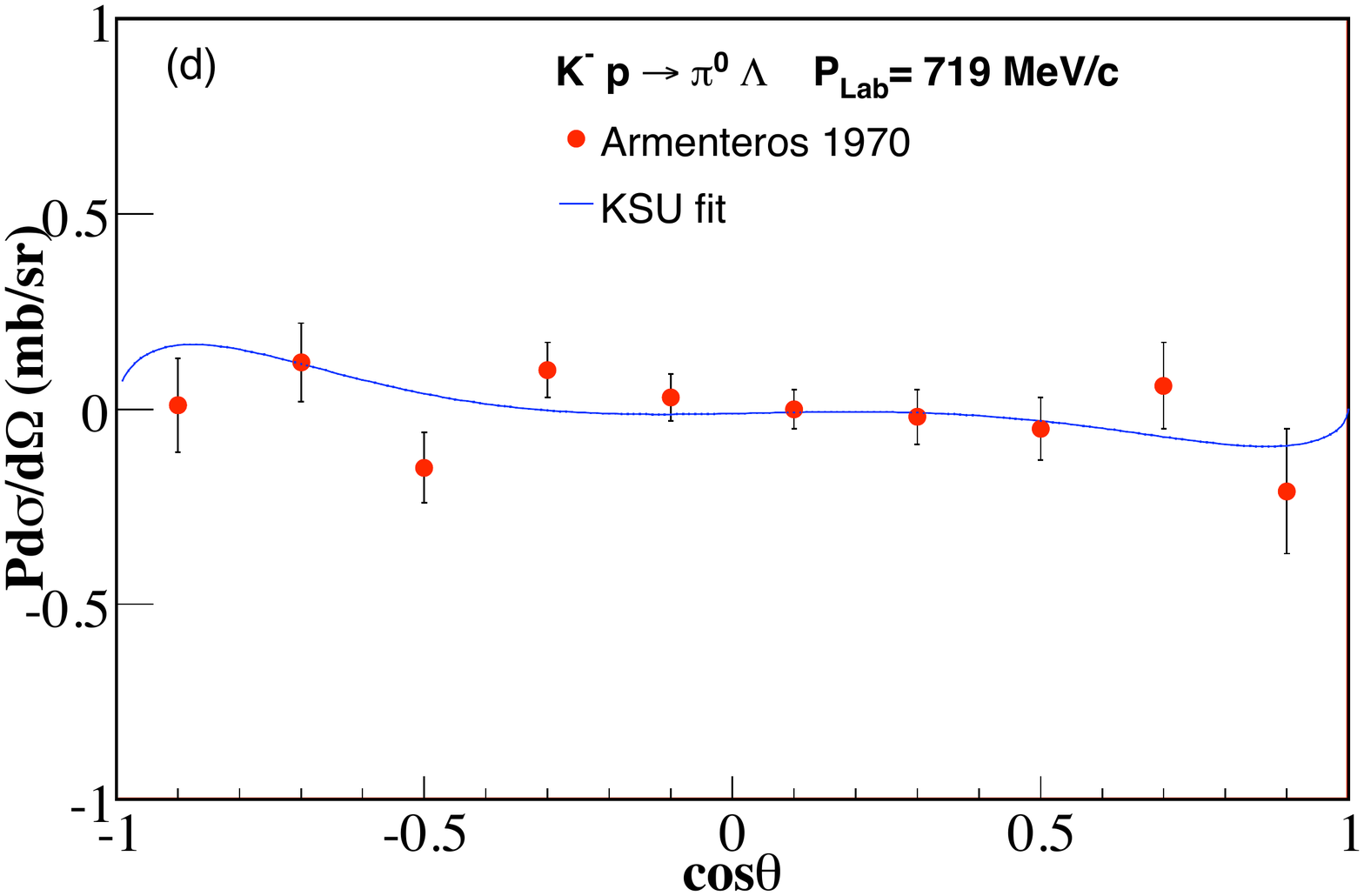}
\caption{(Color online) Representative results of our energy-dependent fit for the $K^- p \rightarrow \pi^0\Lambda$ polarized cross section. Data are from Armenteros 1970 \cite{Armenteros1970}}
\label{fig:dSigma_11_New}
\end{figure}

\begin{figure}[H]
\vspace{-10mm}
\includegraphics[width=0.51\textwidth]{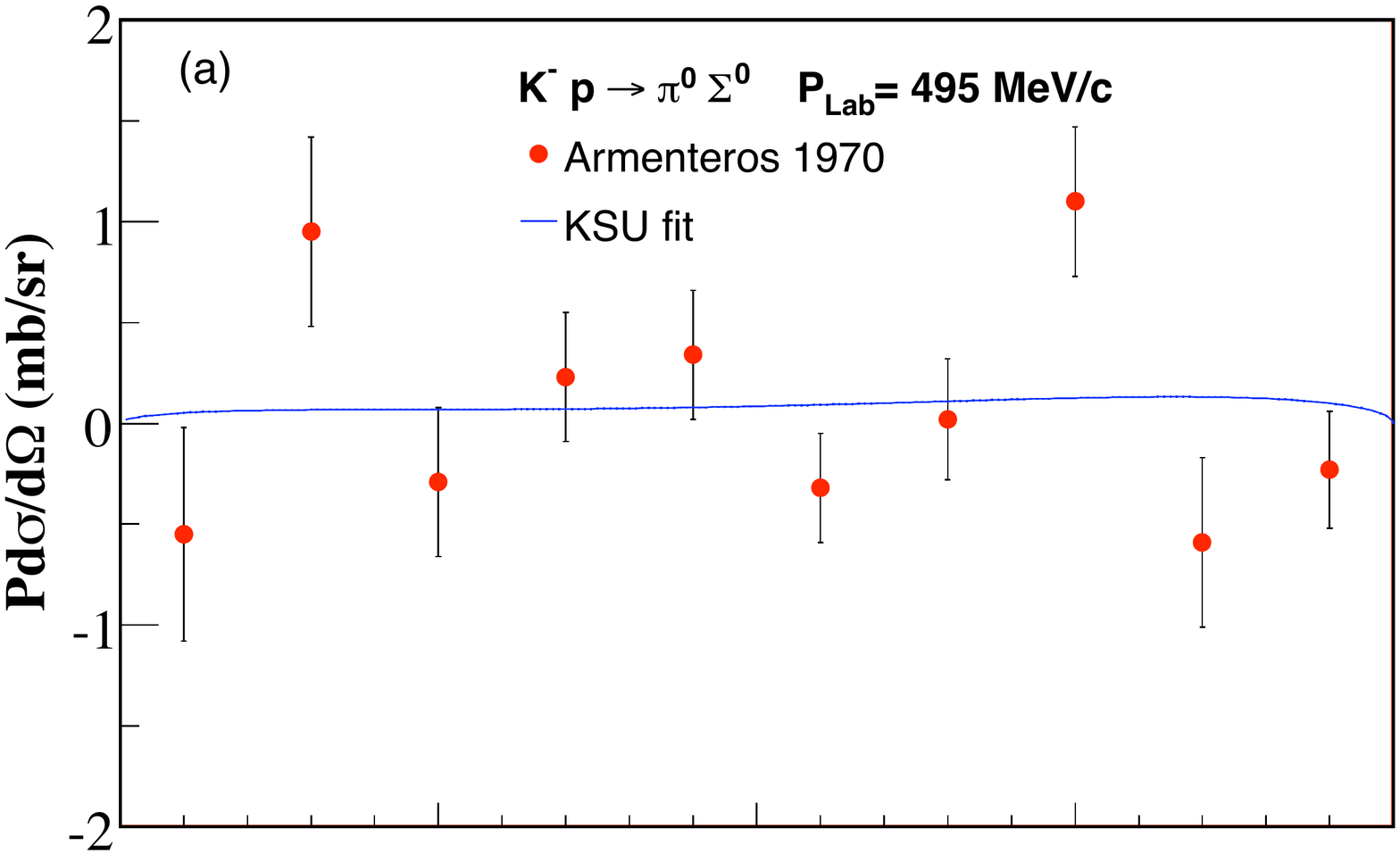}
\vspace{-1mm}
\vspace{-20mm}
\includegraphics[width=0.51\textwidth]{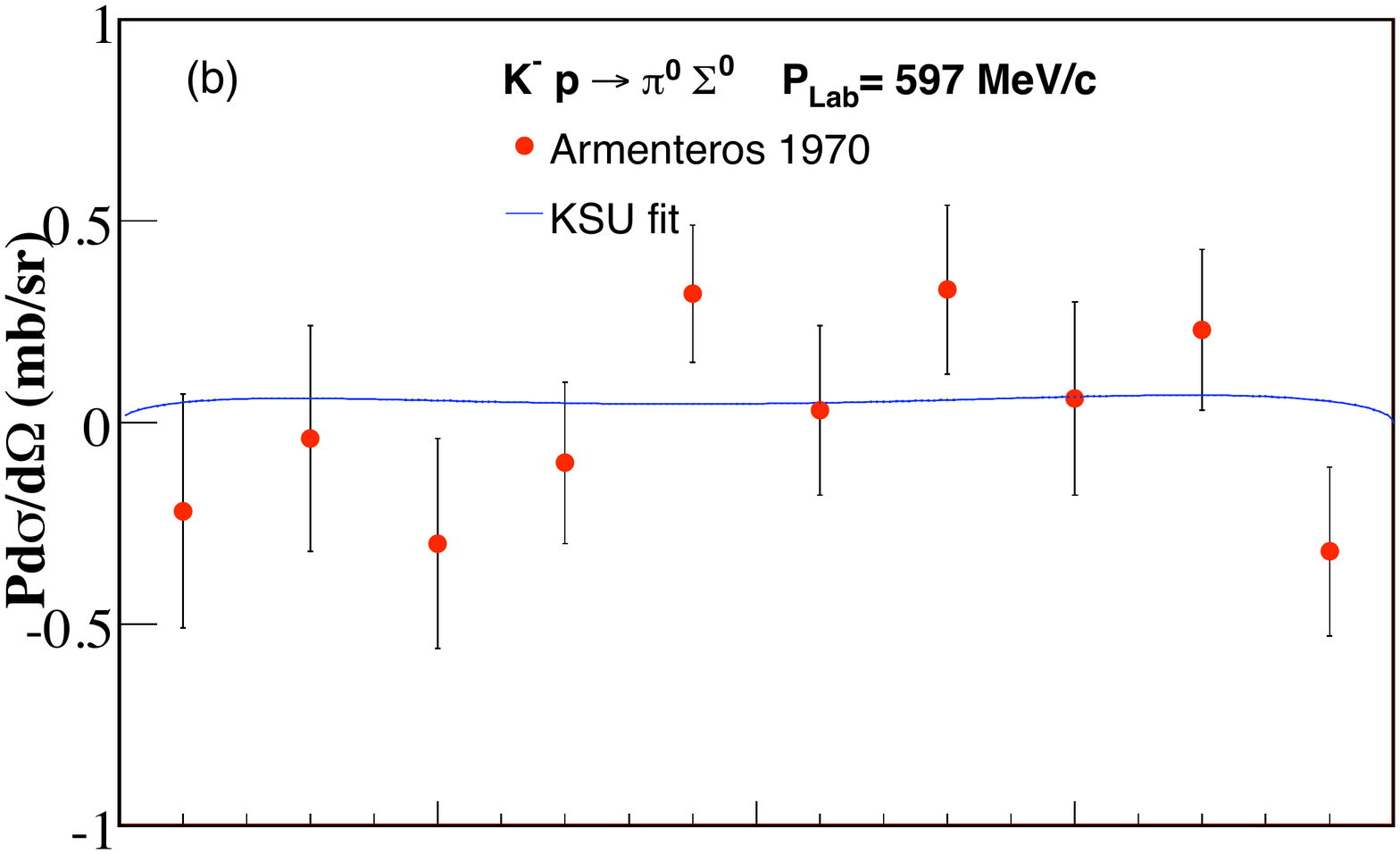}
\vspace{-1mm}
\vspace{-20mm}
\includegraphics[width=0.51\textwidth]{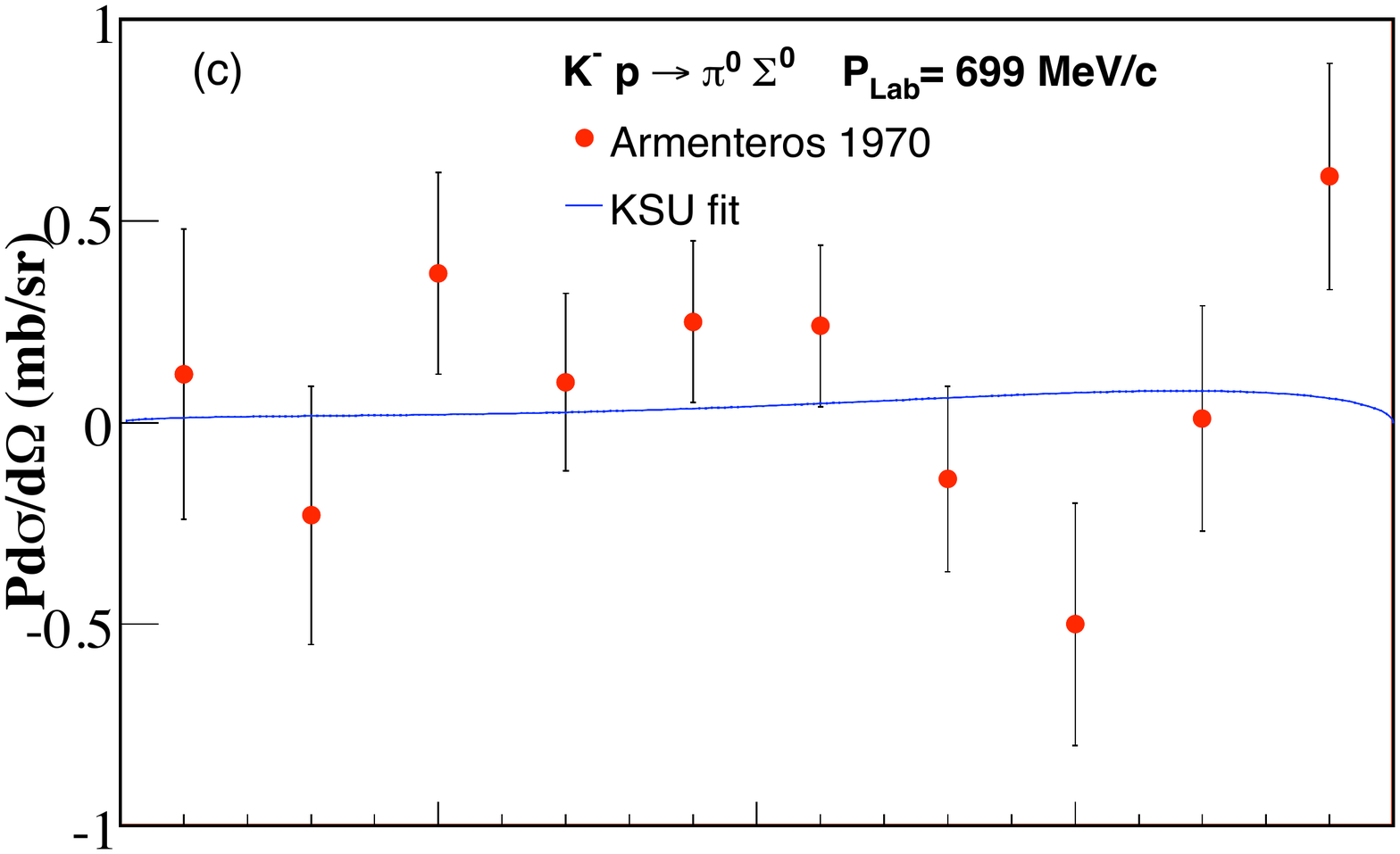}
\vspace{-1mm}
\vspace{-20mm}
\includegraphics[width=0.51\textwidth]{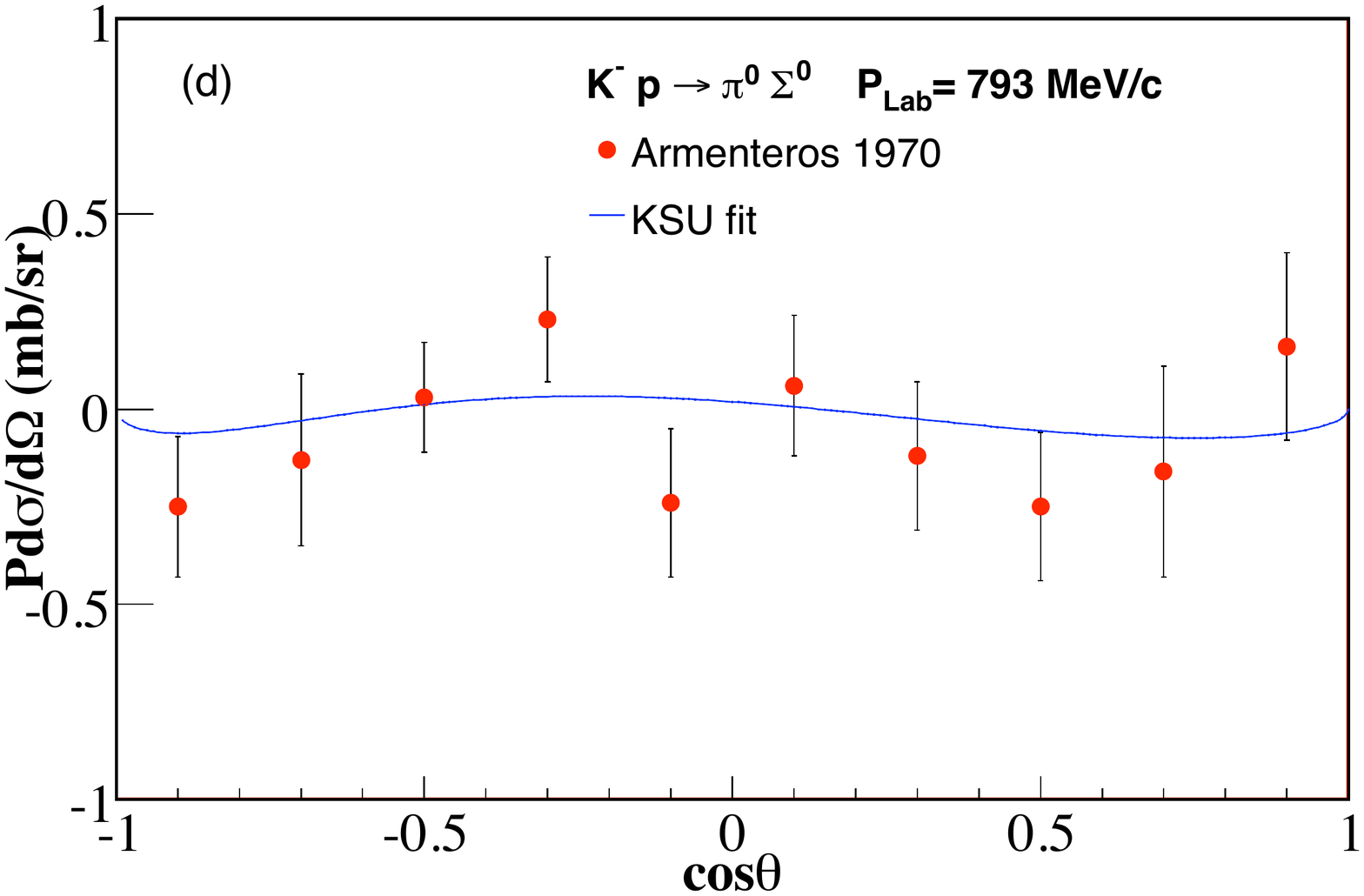}
\caption{(Color online) Representative results of our energy-dependent fit for the $K^- p \rightarrow \pi^0\Sigma^0 $ polarized cross section. Data are from Armenteros 1970 \cite{Armenteros1970}.}
\label{fig:dSigma_11_New}
\end{figure}

\begin{figure}[H]
\vspace{-10mm}
\includegraphics[width=0.51\textwidth]{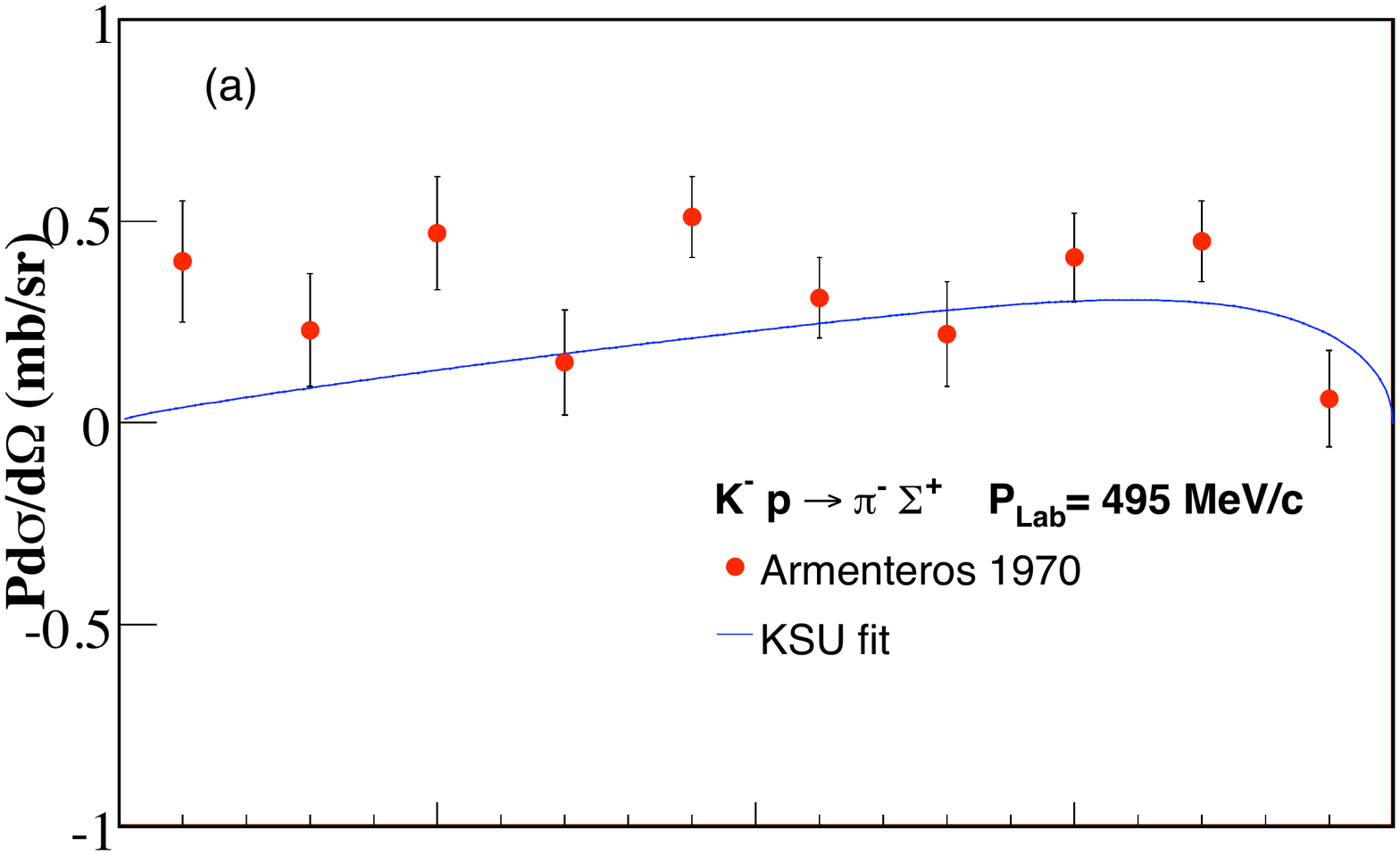}
\vspace{-1mm}
\vspace{-20mm}
\includegraphics[width=0.51\textwidth]{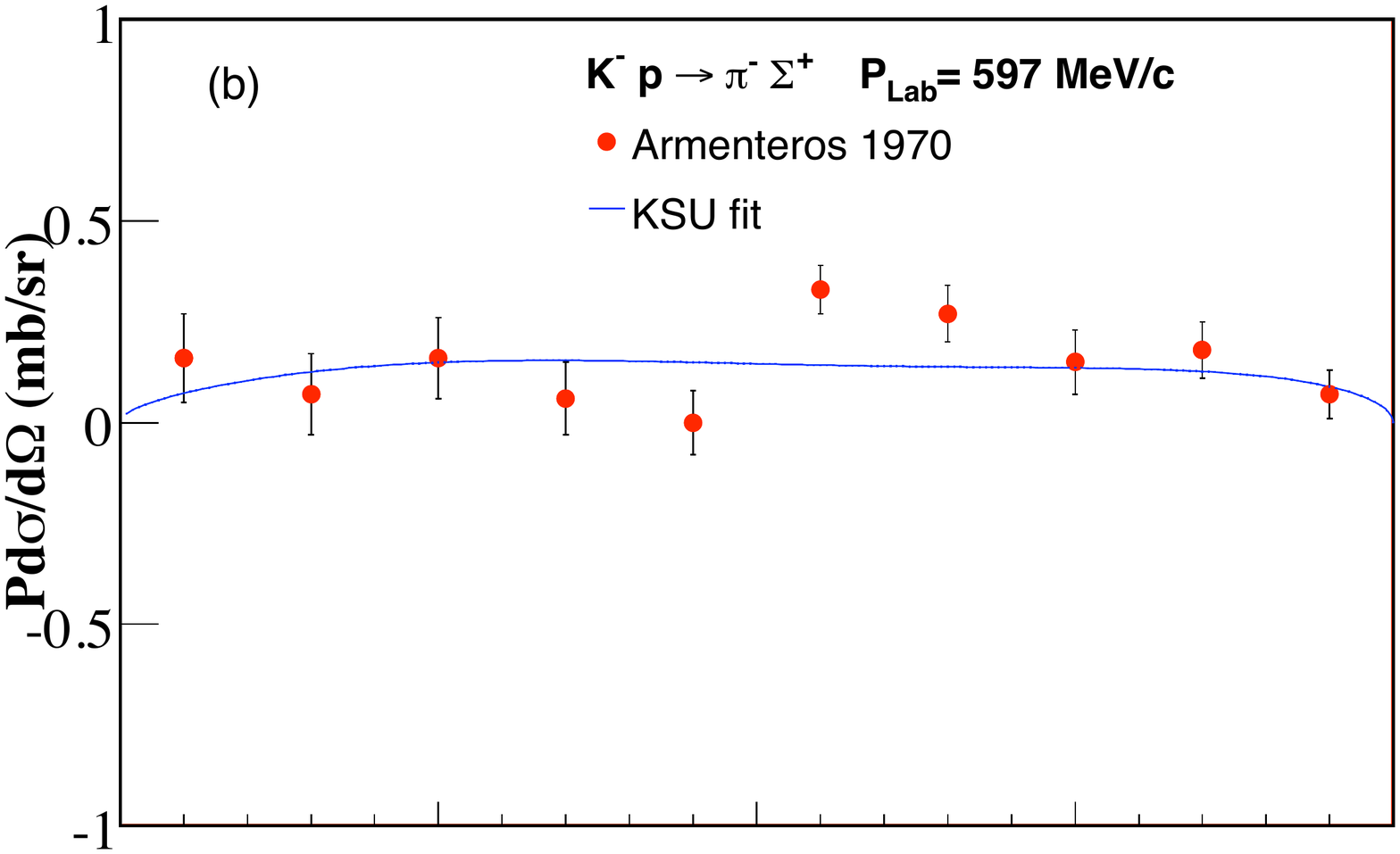}
\vspace{-1mm}
\vspace{-20mm}
\includegraphics[width=0.51\textwidth]{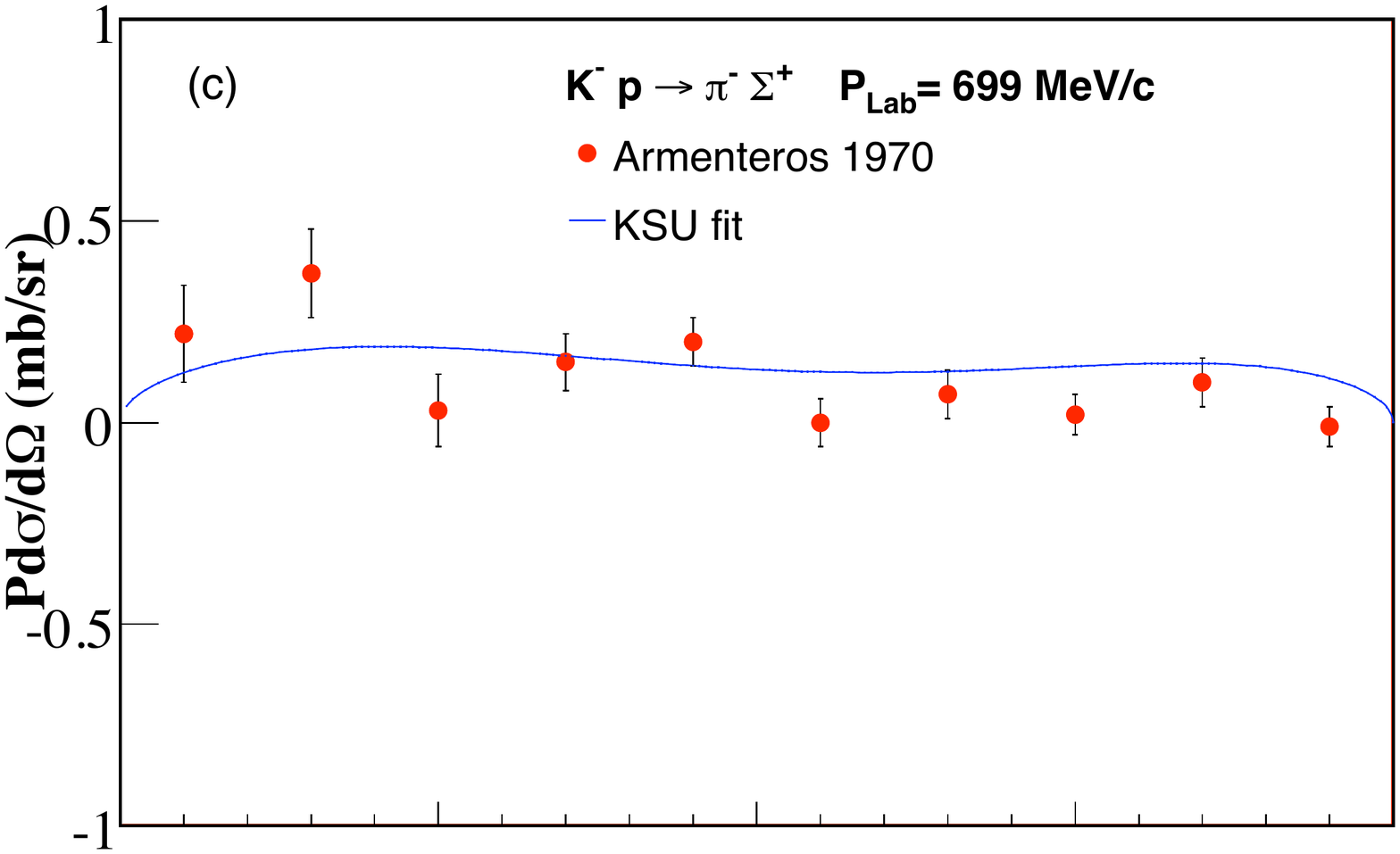}
\vspace{-1mm}
\vspace{-20mm}
\includegraphics[width=0.51\textwidth]{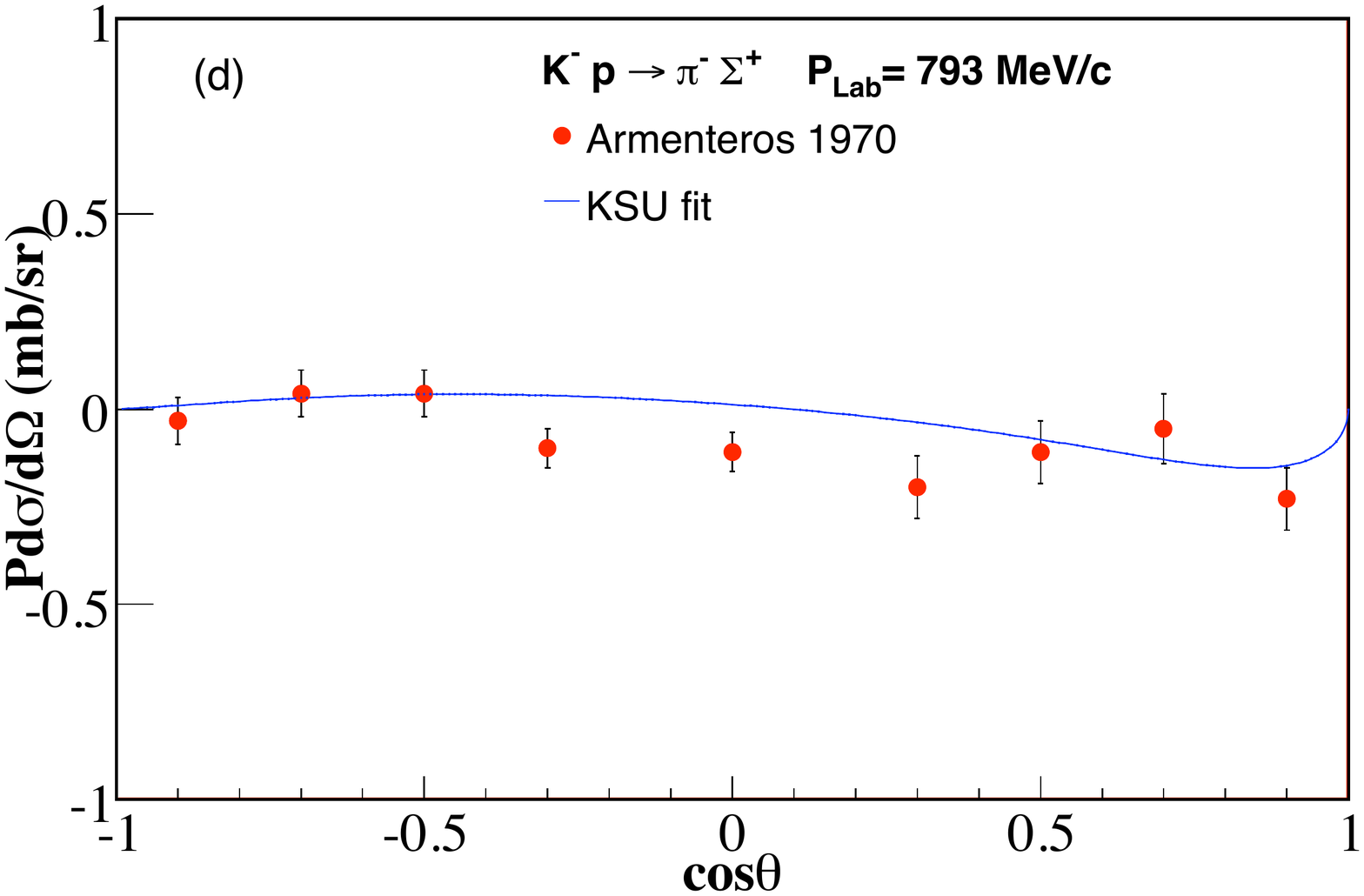}
\caption{(Color online) Representative results of our energy-dependent fit for the $K^- p \rightarrow \pi^- \Sigma^+$ polarized cross section. Data are from Armenteros 1970 \cite{Armenteros1970}}
\label{fig:dSigma_11_New}
\end{figure}

 Figure 16 shows our prediction for the total $K^-p\rightarrow$ cross section.
 Figure 17 shows our prediction for the $K^-p\rightarrow K^-p$ and $K^-p\rightarrow \overline K^0n$ integrated cross sections, Fig.\ 18 shows our prediction for the $K^-p\rightarrow \pi^0\Lambda$ integrated cross section, and Fig. 19 shows our prediction for the $K^-p\rightarrow \pi^+\Sigma^-$, $K^-p\rightarrow \pi^0\Sigma^0$, and $K^-p\rightarrow\pi^-\Sigma^+$ integrated cross sections. Figures 17 and 18 do not show predictions for c.m.\ energies below 1540 MeV because it was not possible to obtain single-energy amplitudes in this region where only integrated cross-section data are available.

\begin{figure}[H]
\begin{center}
\vspace{-1mm}
\scalebox{0.35}{\includegraphics{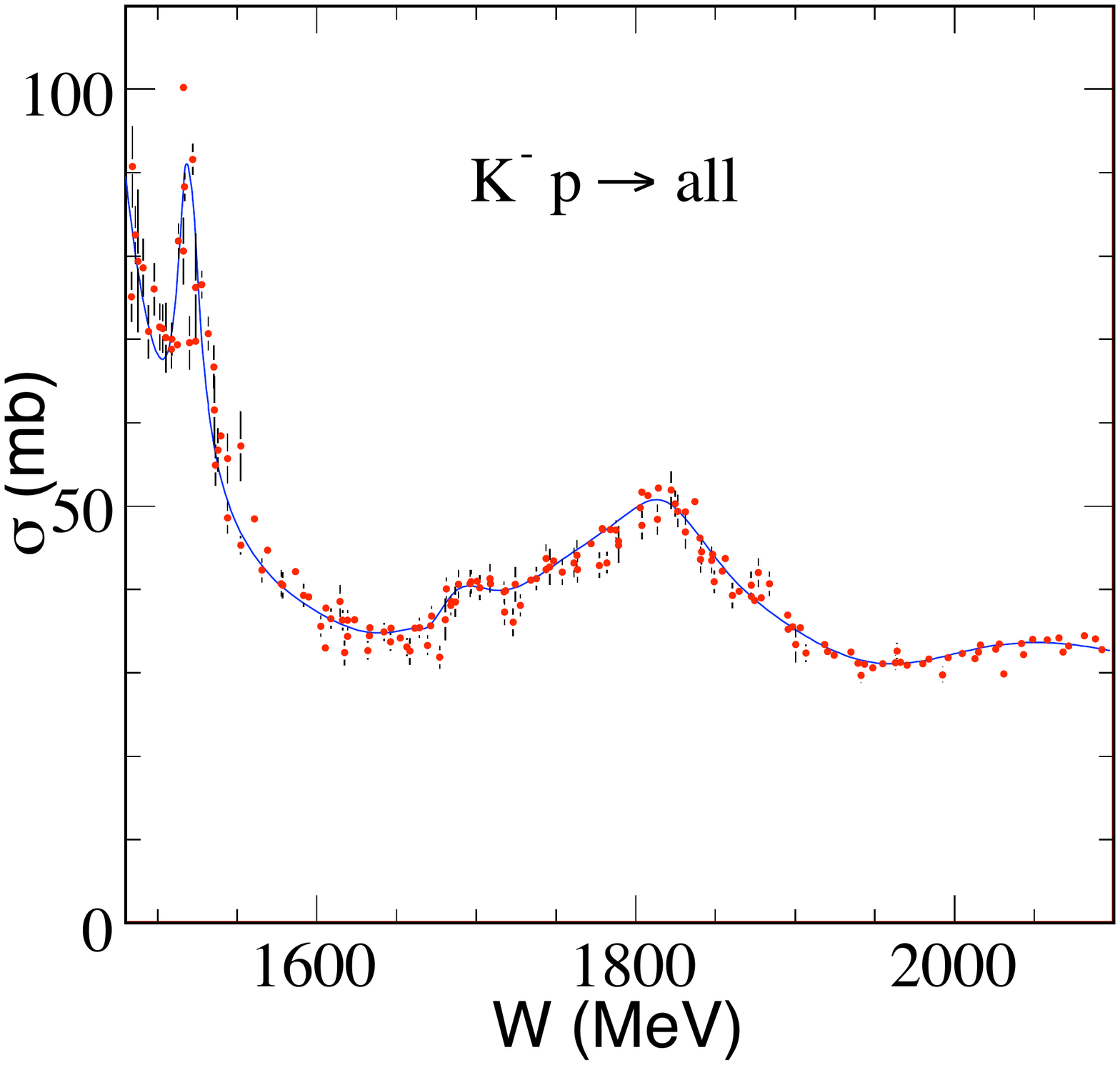}}
\caption{(Color online) Total $K^-p$ cross section compared with the results of our energy-dependent fit. Data are from Baldini 1988 \cite{Baldini1988}}
\label{fig:dSigma_11_New}
\end{center}
\end{figure}
 
\begin{figure}[htpb]
\begin{center}
\vspace{-9.5mm}
\scalebox{0.35}{\includegraphics{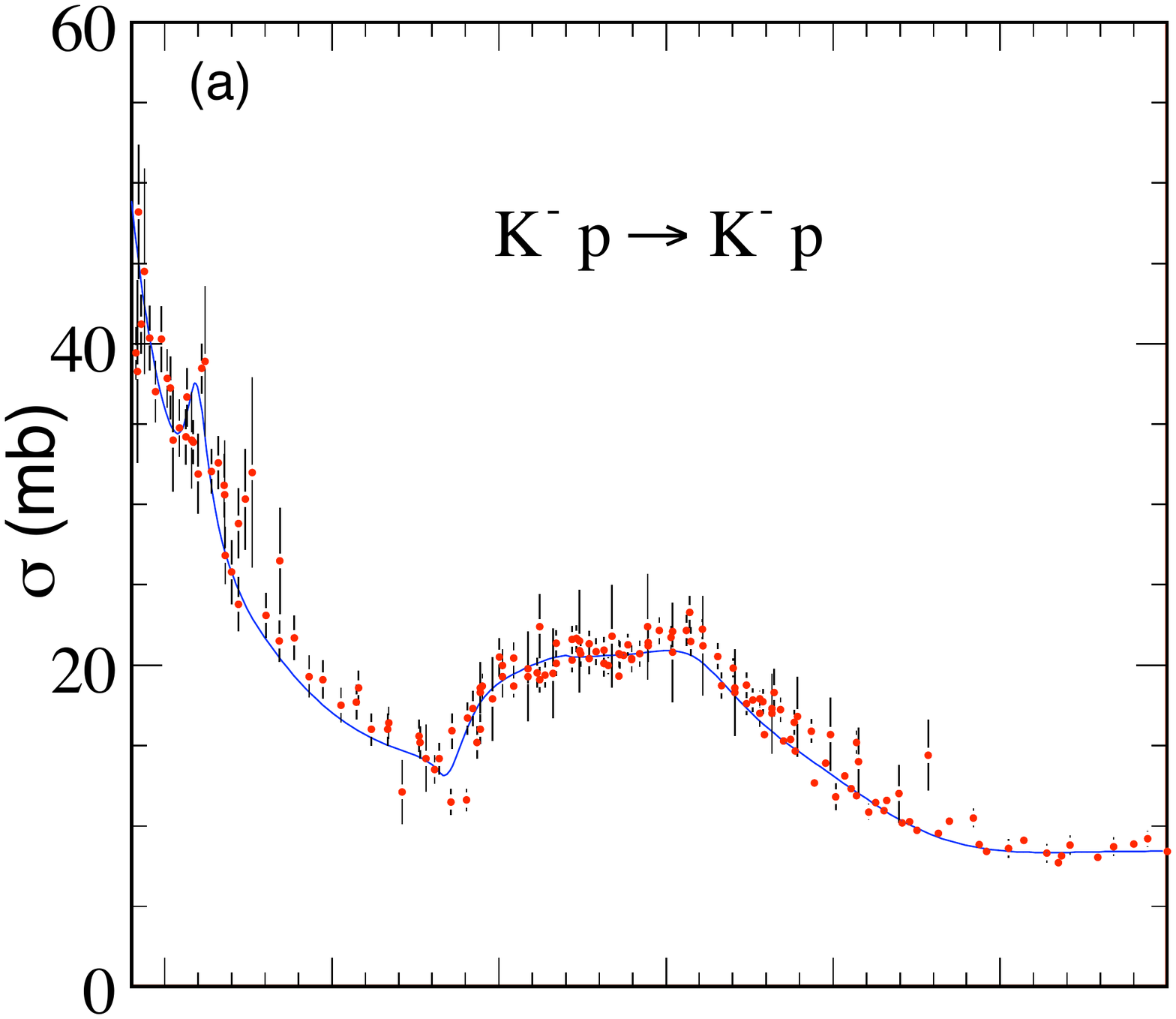}}
\vspace{-10mm}
\vspace{-5mm}
\scalebox{0.35}{\includegraphics{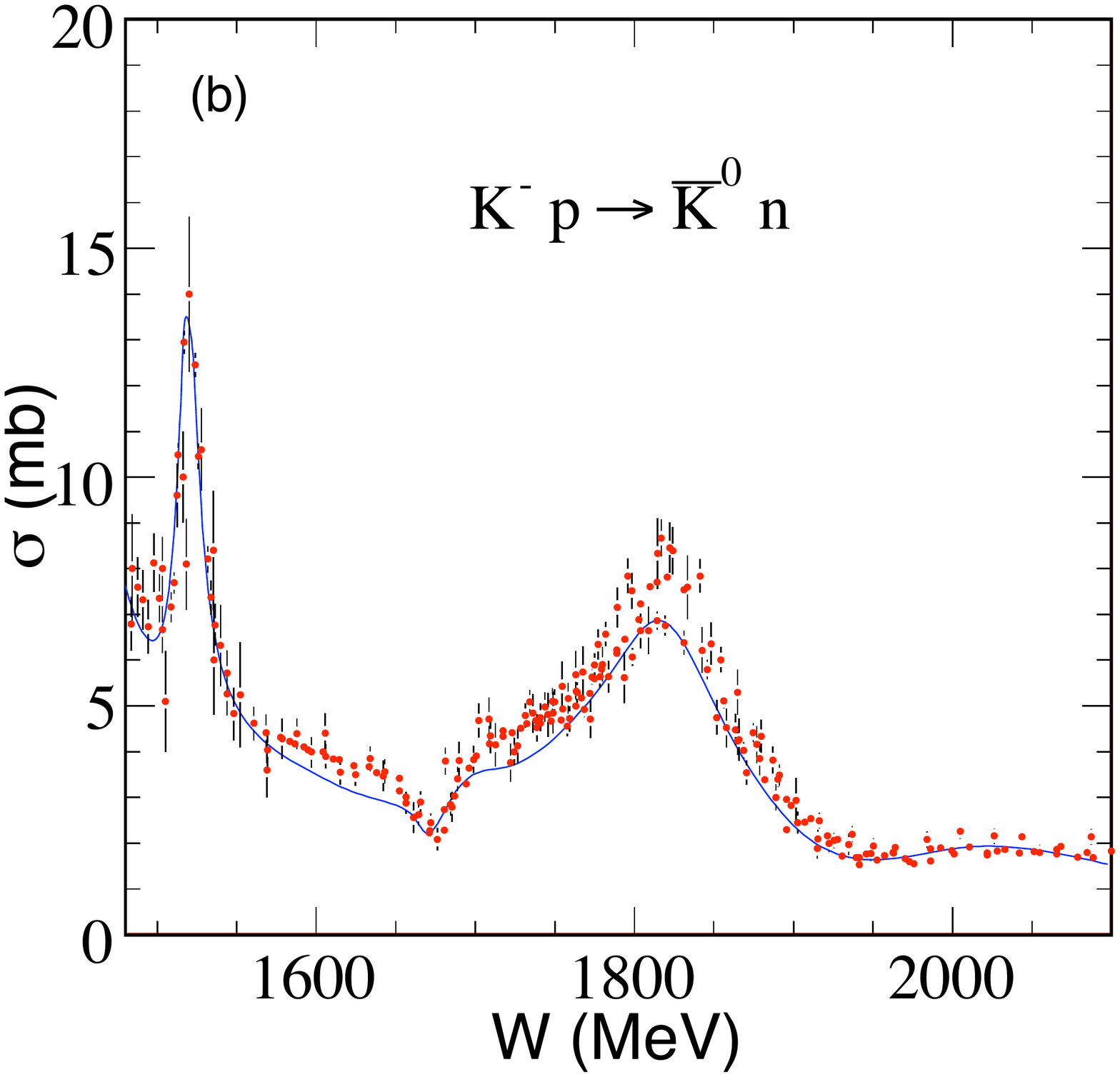}}
\vspace{-10mm}
\caption{(Color online) Integrated cross sections for $K^- p \rightarrow K^-p$ and $K^- p \rightarrow \overline K^0n$ compared with the results of our energy-dependent fit. Data are from Baldini 1988 \cite{Baldini1988}}
\label{fig:dSigma_11_New}
\end{center}
\end{figure}

\begin{figure}[htpb]
\begin{center}
\vspace{-9.08mm}
\scalebox{0.35}{\includegraphics{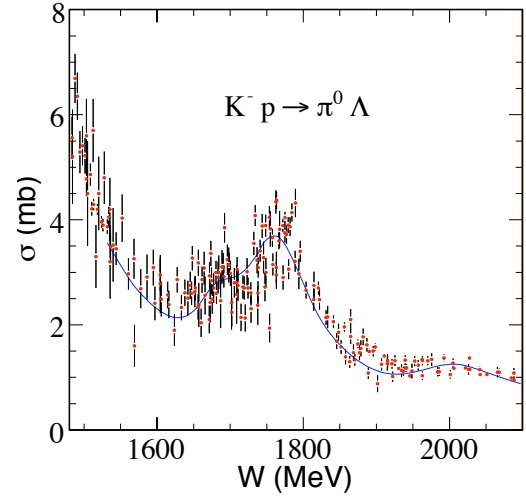}}
\caption{(Color online) Integrated cross section for $K^- p \rightarrow \pi^0\Lambda$ compared with the results of our energy-dependent fit. Data are from Baldini 1988 \cite{Baldini1988}.}
\label{fig:dSigma_11_New}
\end{center}
\end{figure}


\begin{figure}[htpb]
\begin{center}
\vspace{-9.5mm}
\scalebox{0.35}{\includegraphics{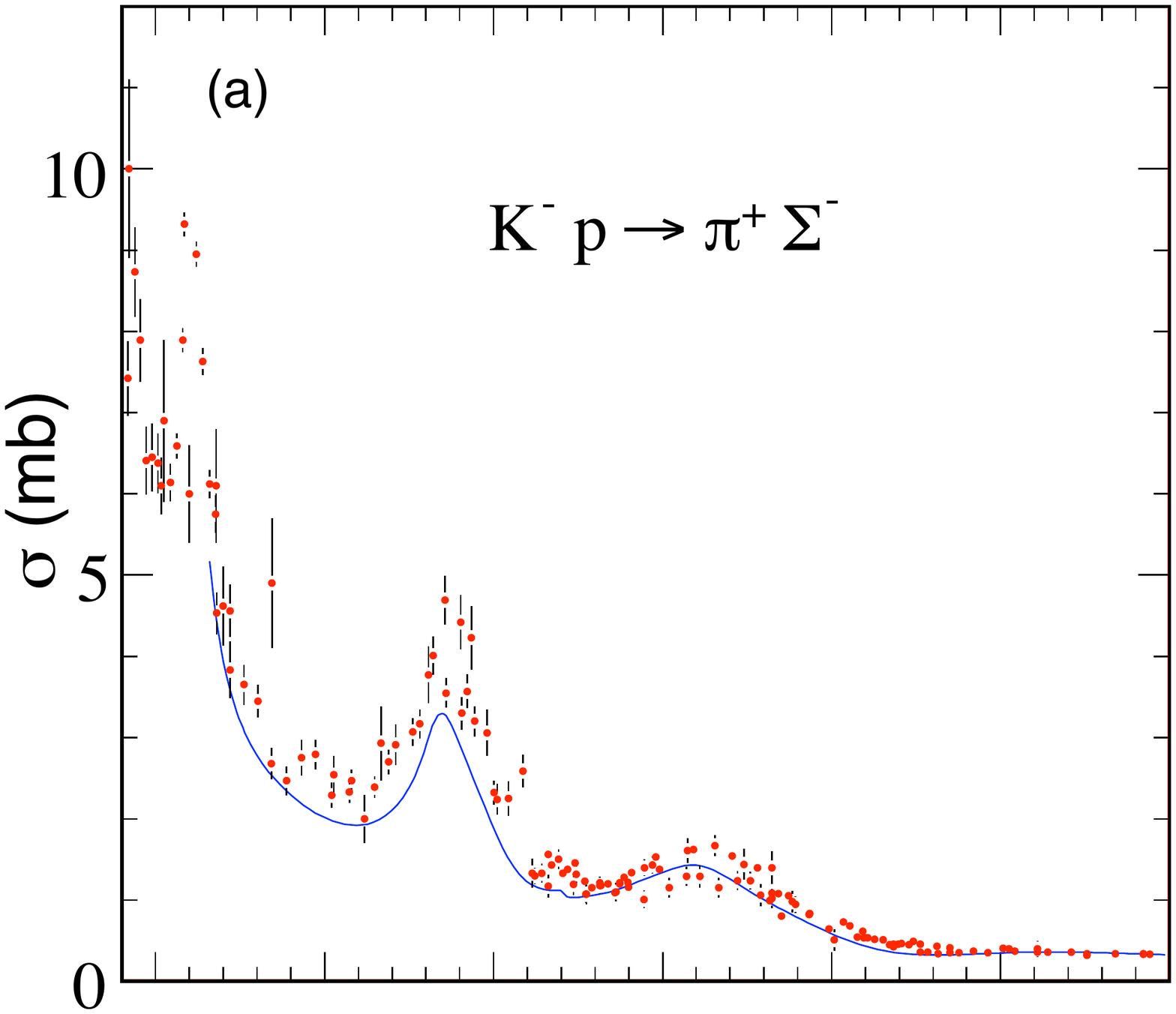}}
\vspace{-10mm}
\vspace{-5mm}
\scalebox{0.35}{\includegraphics{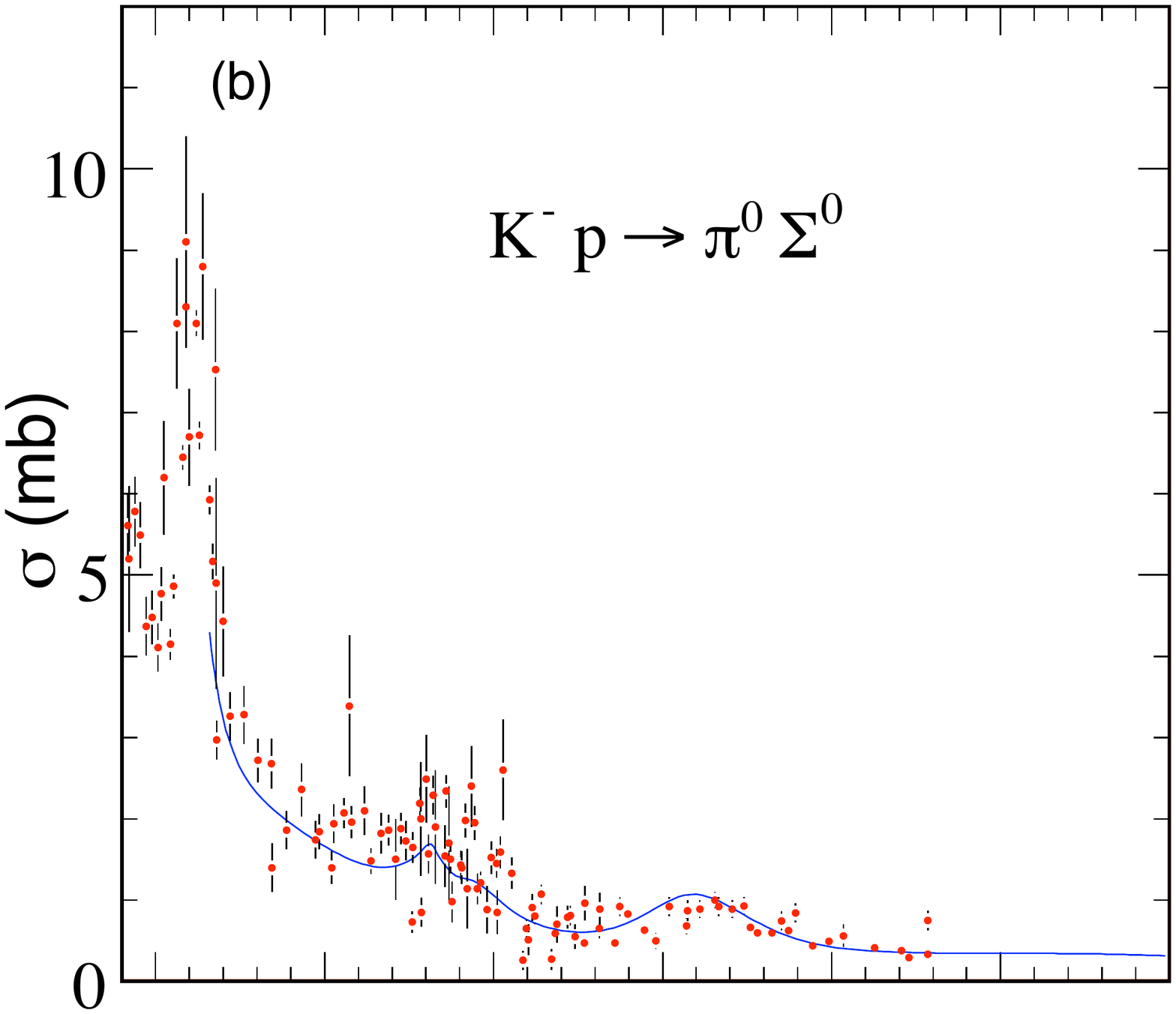}}
\vspace{-10mm}
\vspace{-5mm}
\scalebox{0.35}{\includegraphics{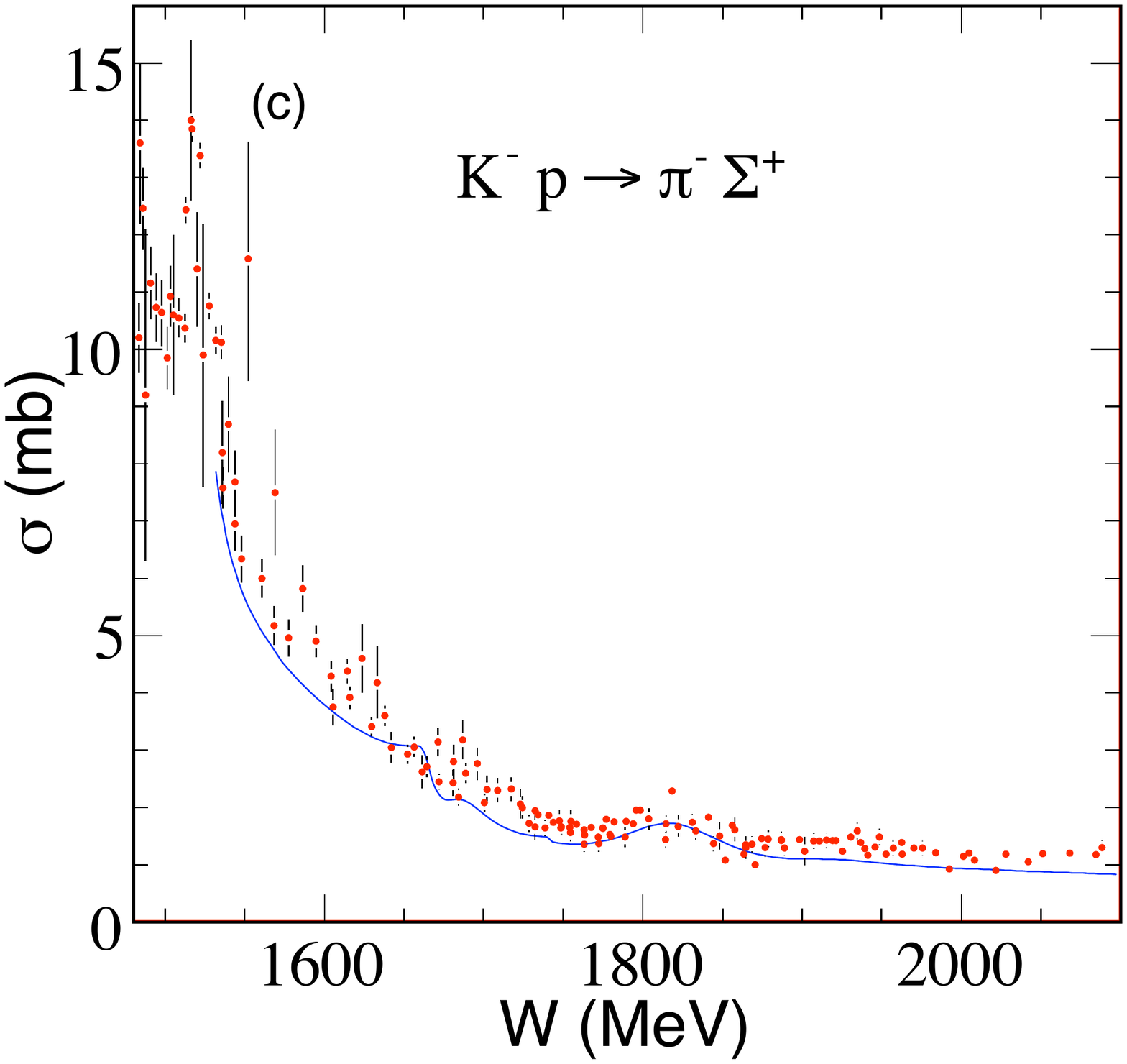}}
\vspace{-10mm}
\caption{(Color online) Integrated cross sections for $K^- p \rightarrow \pi^+\Sigma^-$, $K^- p \rightarrow \pi^0\Sigma^0$, and $K^- p \rightarrow \pi^-\Sigma^+$ compared with the results of our energy-dependent fit. Data are from Baldini 1988 \cite{Baldini1988}.}
\label{fig:dSigma_11_New}
\end{center}
\end{figure}

\clearpage
\section{Summary and Conclusions}
We have investigated $\overline KN\rightarrow \overline KN$, $\overline K N\rightarrow \pi\Lambda$, and $\overline KN\rightarrow \pi\Sigma$ reactions through single-energy analyses constrained by a global unitary energy-dependent fit from threshold to a c.m.\ energy of 2.1 GeV. We found partial waves up to G-waves necessary to describe the available data for the reactions. This work was motivated, in part, by the relatively recent measurements for $K^-p\rightarrow \overline K^0n$, $K^-p\rightarrow \pi^0\Lambda$, $K^-p\rightarrow\pi^0\Sigma^0$, and $K^-p\rightarrow \eta\Lambda$ by the Crystal Ball Collaboration. We were successful in describing these data in addition to older data from constrained single-energy analyses. The partial-wave amplitudes thus extracted were used in our global multichannel fit. A discussion of the resonance parameters from this global fit, which is the most comprehensive multichannel fit to date for $\overline K N$ scattering reactions, is presented in a separate publication \cite{manoj2013}. 

\acknowledgements{This work was supported by the U.S. Department of Energy Grant No. DE-FG02-01ER41194. 

\section{Appendix}
Table IV lists the real and imaginary parts of the $\overline K N\rightarrow \overline K N$ amplitudes tabulated against the central bin energies. Similarly, Table V lists $\overline K N\rightarrow \pi \Lambda$ amplitudes and Table VI lists $\overline K N \rightarrow \pi \Sigma$ amplitudes. The values in these tables represent the final single-energy solutions that were used as input into our subsequent global energy-dependent fits for given partial waves. Plots of all amplitudes in Tables IV - VI are shown in Ref. \cite{manoj2013}. The amplitudes are also available in the form of data files \cite{pub2013}. 

\begin{table*}[bhtp]
\caption{Single-energy amplitudes for $\overline{K} N \rightarrow \overline{K} N$.}
\begin{center}
\begin{ruledtabular}
\begin{tabular}{ccccccc}
\multicolumn{1}{c}{$W_0$} & \multicolumn{2}{c}{$S_{01}$}& \multicolumn{2}{c}{$ P_{01}$} & \multicolumn{2}{c}{$ P_{03}$}  \\ \cline{2-7}
\multicolumn{1}{c}{$\rm (MeV)$} & \multicolumn{1}{c}{Re({$ S_{01}$})} & \multicolumn{1}{c}{ Im({$ S_{01}$})}& \multicolumn{1}{c}{Re({$P_{01}$})} & \multicolumn{1}{c}{ Im({$ P_{01}$})} & \multicolumn{1}{c}{ Re({$ P_{03}$})} & \multicolumn{1}{c}{ Im({$ P_{03}$})}\\
\hline
1480	&	$0.022	\pm	0.037$	&	$0.686	\pm	0.019$  &	$0.057  	\pm	0.039$	&	$0.006	\pm	0.036$	&	$0.065	\pm	0.035$	&	$0.005	\pm	0.023$\\
1500	&	$0.031	\pm	0.028$	&	$0.763	\pm	0.013$  &	$0.098	\pm	0.028$	&	$0.025	\pm	0.027$	&	$0.067	\pm	0.025$	&	$0.010	\pm	0.016$\\
1520	&	$0.038	\pm	0.091$	&	$0.615	\pm	0.052$  &	$0.218	\pm	0.083$	&	$0.034	\pm	0.048$	&	$0.015	\pm	0.050$	&	$0.016	\pm	0.076$\\
1540	&	$0.046	\pm	0.029$	&	$0.785	\pm	0.018$  &	$0.210	\pm	0.029$	&	$0.102	\pm	0.025$	&	$0.082	\pm	0.028$	&	$0.024	\pm	0.021$\\
1560	&	$0.055	\pm	0.025$	&	$0.778	\pm	0.015$  &	$0.205	\pm	0.025$	&	$0.229	\pm	0.024$	&	$0.127	\pm	0.024$	&	$0.033	\pm	0.017$\\
1580	&	$0.051	\pm	0.027$	&	$0.799	\pm	0.017$  &	$0.186	\pm	0.028$	&	$0.245	\pm	0.026$	&	$0.160	\pm	0.026$	&	$0.043	\pm	0.019$\\
\hline \end{tabular}

2000	&	$0.032	\pm	0.028$	&	$0.496	\pm	0.027$  &	$0.131	\pm	0.028$	&	$0.368	\pm	0.028$	&	$0.111	\pm	0.023$	&	$0.431	\pm	0.025$\\
2020	&	$0.050	\pm	0.037$	&	$0.516	\pm	0.036$  &	$0.139	\pm	0.037$	&	$0.334	\pm	0.038$	&	$0.085	\pm	0.036$	&	$0.420	\pm	0.034$\\
2040	&	$0.037	\pm	0.037$	&	$0.494	\pm	0.035$  &	$0.150	\pm	0.037$	&	$0.353	\pm	0.037$	&	$0.074	\pm	0.035$	&	$0.383	\pm	0.034$\\
2060	&	$0.082	\pm	0.032$	&	$0.583	\pm	0.030$  &	$0.157	\pm	0.031$	&	$0.412	\pm	0.032$	&	$0.049	\pm	0.032$	&	$0.440	\pm	0.029$\\
2080	&	$0.078	\pm	0.029$	&	$0.616	\pm	0.027$  &	$0.134	\pm 	0.028$	&	$0.454	\pm	0.029$	&	$0.029	\pm	0.028$	&	$0.449	\pm	0.025$\\
2100	&	$0.057	\pm	0.039$	&	$0.542	\pm	0.033$  &	$0.134	\pm	0.038$	&	$0.482	\pm	0.035$	&	$0.010	\pm	0.038$	&	$0.462	\pm	0.031$\\
2120	&	$0.062	\pm	0.029$	&	$0.650	\pm	0.025$  &	$0.113	\pm	0.027$	&	$0.526	\pm	0.027$	&	$-0.008	\pm	0.027$	&	$0.459	\pm	0.025$\\
2140	&	$0.041	\pm	0.035$	&	$0.651	\pm	0.032$  &	$0.080	\pm	0.034$	&	$0.553	\pm	0.033$	&	$-0.026	\pm	0.033$	&	$0.463	\pm	0.030$\\
2160	&	$0.045	\pm	0.091$	&	$0.440	\pm	0.076$  &	$0.047	\pm	0.091$	&	$0.443	\pm	0.075$	&	$-0.045	\pm	0.089$	&	$0.352	\pm	0.073$\\
2180	&	$0.034	\pm	0.025$	&	$0.670	\pm	0.025$  &	$0.038	\pm	0.025$	&	$0.525	\pm	0.025$	&	$-0.062	\pm	0.025$	&	$0.450	\pm	0.025$\\
 \end{tabular}
\end{ruledtabular}
\end{center}
\label{amplitude:S0_1}
\end{table*}

\begin{table*}[bhtp]
\addtocounter{table}{-1}
\caption{Cont'd.}
\begin{center}
\begin{ruledtabular}
\begin{tabular}{ccccccc}
\multicolumn{1}{c}{$W_{0}$} & \multicolumn{2}{c}{$ D_{03}$} & \multicolumn{2}{c}{$ D_{05}$} & \multicolumn{2}{c}{$ F_{05}$}\\ \cline{2-7}
\multicolumn{1}{c}{$\rm (MeV)$} & \multicolumn{1}{c}{ Re({$ D_{03}$})} & \multicolumn{1}{c}{ Im({$ D_{03}$})} & \multicolumn{1}{c}{ Re({$ D_{05}$})} & \multicolumn{1}{c}{ Im({$ D_{05}$})} & \multicolumn{1}{c}{ Re({$ F_{05}$})} & \multicolumn{1}{c}{ Im({$ F_{05}$})}\\
\hline
1480	&	$0.032	\pm	0.037$	&	$0.003	\pm	0.034$	&	$0.004	\pm	0.034$	&	$0.001	\pm	0.025$   &  $0.000	\pm	0.035$	&	$0.000	\pm	0.018$\\
1500	&	$0.132	\pm	0.021$	&	$0.032	\pm	0.024$	&	$0.007	\pm	0.021$	&	$0.002	\pm	0.016$&	$0.001	\pm	0.021$	&	$0.000	\pm	0.009$	\\
1520	&	$0.032	\pm	0.015$	&	$0.427	\pm	0.030$	&	$0.010	\pm	0.036$	&	$0.003	\pm	0.046$&	$0.002	\pm	0.053$	&	$0.000	\pm	0.051$	\\
1540	&	$-0.164	\pm	0.027$	&	$0.128	\pm	0.021$	&	$0.014	\pm	0.027$	&	$0.003	\pm	0.018$&	$0.003	\pm	0.028$	&	$0.000	\pm	0.022$	\\
1560	&	$-0.137	\pm	0.023$	&	$0.029	\pm	0.022$	&	$0.019	\pm	0.023$	&	$0.003	\pm	0.016$ &	$0.006	\pm	0.022$	&	$0.000	\pm	0.021$	\\
1580	&	$-0.093	\pm	0.026$	&	$0.047	\pm	0.024$	&	$0.024	\pm	0.025$	&	$0.003	\pm	0.018$ &	$0.011	\pm	0.024$	&	$0.000	\pm	0.023$	\\
\hline \end{tabular}
2020	&	$-0.007	\pm	0.037$	&	$0.200	\pm	0.035$	&	$0.147	\pm	0.032$	&	$0.120	\pm	0.035$&	$-0.210	\pm	0.029$	&	$0.138	\pm	0.035$	\\
2040	&	$-0.016	\pm	0.037$	&	$0.207	\pm	0.034$	&	$0.091	\pm	0.033$	&	$0.170	\pm	0.034$&	$-0.207	\pm	0.029$	&	$0.121	\pm	0.034$	\\
2060	&	$-0.029	\pm	0.031$	&	$0.223	\pm	0.030$	&	$0.160	\pm	0.027$	&	$0.178	\pm	0.029$&	$-0.190	\pm	0.026$	&	$0.143	\pm	0.030$	\\
2080	&	$-0.044	\pm	0.028$	&	$0.238	\pm	0.027$	&	$0.142	\pm	0.025$	&	$0.176	\pm	0.026$&	$-0.173	\pm	0.025$	&	$0.141	\pm	0.027$	\\
2100	&	$-0.093	\pm	0.037$	&	$0.221	\pm	0.034$	&	$0.145	\pm	0.033$	&	$0.198	\pm	0.031$&	$-0.153	\pm	0.034$	&	$0.152	\pm	0.034$	\\
2120	&	$-0.126	\pm	0.026$	&	$0.257	\pm	0.026$	&	$0.114	\pm	0.023$	&	$0.201	\pm	0.024$&	$-0.182	\pm	0.023$	&	$0.137	\pm	0.025$	\\
2140	&	$-0.111	\pm	0.032$	&	$0.271	\pm	0.032$	&	$0.086	\pm	0.030$	&	$0.227	\pm	0.030$&	$-0.165	\pm	0.029$	&	$0.142	\pm	0.031$	\\
2160	&	$0.001	\pm	0.090$	&	$0.171	\pm	0.075$	&	$-0.020	\pm	0.090$	&	$0.218	\pm	0.070$&	$-0.055	\pm	0.088$	&	$0.204	\pm	0.075$	\\
2180	&	$-0.126	\pm	0.025$	&	$0.245	\pm	0.025$	&	$0.062	\pm	0.025$	&	$0.224	\pm	0.025$ &	$-0.170	\pm	0.025$	&	$0.113	\pm	0.025$	\\
 \end{tabular}
\end{ruledtabular}
\end{center}
\label{amplitude:D0_1}
\end{table*}

\begin{table*}[bhtp]
\addtocounter{table}{-1}
\caption{Cont'd.}
\begin{center}
\begin{ruledtabular}
\begin{tabular}{ccccc}
\multicolumn{1}{c}{$W_{0}$} & \multicolumn{2}{c}{$ F_{07}$}& \multicolumn{2}{c}{$ G_{07}$} \\
\multicolumn{1}{c}{$\rm (MeV)$} & \multicolumn{1}{c}{ Re({$ G_{07}$})} & \multicolumn{1}{c}{ Im({$ G_{07}$})}& \multicolumn{1}{c}{ Re({ $G_{07}$})} & \multicolumn{1}{c}{ Im({$ G_{07}$})}\\
\hline
1480&	$0.000	\pm	0.032$	&	$0.000	\pm	0.025$&	$0.000	\pm	0.021$	&	$0.000	\pm	0.025$		\\
1500	&	$0.000	\pm	0.014$	&	$0.000	\pm	0.025$  &	$0.000	\pm	0.017$	&	$0.000	\pm	0.005$		\\
1520	&	$0.000	\pm	0.026$	&	$0.000	\pm	0.014$ &	$0.000	\pm	0.043$	&	$0.000	\pm	0.032$		\\
1540	&	$0.001	\pm	0.018$	&	$0.000	\pm	0.017$ &	$0.000	\pm	0.026$	&	$0.000	\pm	0.011$		\\
1560	&	$0.001	\pm	0.017$	&	$0.000	\pm	0.017$ &	$0.000	\pm	0.019$	&	$0.000	\pm	0.011$		\\
1580	&	$0.002	\pm	0.022$	&	$0.000	\pm	0.019$ &	$0.001	\pm	0.016$	&	$0.000	\pm	0.012$		\\
\hline \end{tabular}
2020	&	$0.046	\pm	0.033$	&	$0.028	\pm	0.029$ &	$0.151	\pm	0.028$	&	$0.199	\pm	0.027$		\\
2040&	$0.042	\pm	0.032$	&	$0.030	\pm	0.027$ &	$0.124	\pm	0.027$	&	$0.269	\pm	0.021$		\\
2060	&	$0.038	\pm	0.029$	&	$0.031	\pm	0.025$ &	$0.090	\pm	0.024$	&	$0.234	\pm	0.022$		\\
2080	&	$0.036	\pm	0.026$	&	$0.030	\pm	0.021$ &	$0.069	\pm	0.024$	&	$0.253	\pm	0.021$		\\
2100&	$0.035	\pm	0.035$	&	$0.029	\pm	0.027$ &	$0.020	\pm	0.032$	&	$0.310	\pm	0.022$       	\\
2120&	$0.035	\pm	0.024$	&	$0.028	\pm	0.021$ &	$-0.008	\pm	0.019$	&	$0.268	\pm	0.016$		\\
2140&	$0.035	\pm	0.031$	&	$0.027	\pm	0.027$ &	$-0.032	\pm	0.028$	&	$0.258	\pm	0.022$		\\
2160&	$0.035	\pm	0.087$	&	$0.027	\pm	0.046$ &	$0.049	\pm	0.085$	&	$0.234	\pm	0.057$		\\
2180&	$0.036	\pm	0.025$	&	$0.026	\pm	0.025$ &	$-0.065	\pm	0.025$	&	$0.235	\pm	0.025$		\\
\end{tabular}
\end{ruledtabular}
\end{center}
\label{amplitude:G0_1}
\end{table*}

\begin{table*}[bhtp]
\addtocounter{table}{-1}
\caption{Cont'd.}
\begin{center}
\begin{ruledtabular}
\begin{tabular}{ccccccc}
\multicolumn{1}{c}{$W_{0}$} & \multicolumn{2}{c}{$ S_{11}$}& \multicolumn{2}{c}{$ P_{11}$} & \multicolumn{2}{c|}{$ P_{13}$} \\ \cline{2-7}
\multicolumn{1}{c}{$\rm (MeV)$} & \multicolumn{1}{c}{ Re({$ S_{11}$})} & \multicolumn{1}{c}{ Im({$ S_{11}$})}& \multicolumn{1}{c}{ Re({$ P_{11}$})} & \multicolumn{1}{c}{ Im({$ P_{11}$})} & \multicolumn{1}{c}{ Re({$ P_{13}$})} & \multicolumn{1}{c}{ Im({$ P_{13}$})}\\
\hline
1480	&	$0.232	\pm	0.037$	&	$0.329	\pm	0.019$&	$0.006	\pm	0.039$	&	$0.020	\pm	0.036$	&	$-0.034	\pm	0.035$	&	$0.008	\pm	0.023$	\\
1500	&	$0.191	\pm	0.027$	&	$0.352	\pm	0.012$&	$0.038	\pm	0.027$	&	$0.022	\pm	0.027$	&	$-0.040	\pm	0.024$	&	$0.009	\pm	0.016$	\\
1520&	$0.125	\pm	0.053$	&	$0.360	\pm	0.037$&	$0.168	\pm	0.049$	&	$0.030	\pm	0.061$	&	$-0.041	\pm	0.039$	&	$0.010	\pm	0.039$	\\
1540	&	$0.118	\pm	0.029$	&	$0.385	\pm	0.014$&	$0.029	\pm	0.029$	&	$0.042	\pm	0.025$	&	$-0.036	\pm	0.028$	&	$0.011	\pm	0.019$	\\
1560	&	$0.178	\pm	0.025$	&	$0.357	\pm	0.015$&	$0.085	\pm	0.025$	&	$0.083	\pm	0.023$	&	$-0.028	\pm	0.024$	&	$0.010	\pm	0.017$	\\
1580	&	$0.199	\pm	0.028$	&	$0.339	\pm	0.017$&	$0.084	\pm	0.028$	&	$0.082	\pm	0.026$	&	$-0.020	\pm	0.025$	&	$0.013	\pm	0.019$	\\
\hline \end{tabular}
2020	&	$-0.292	\pm	0.038$	&	$0.389	\pm	0.035$&	$0.021	\pm	0.037$	&	$0.208	\pm	0.038$	&	$0.001	\pm	0.036$	&	$0.284	\pm	0.033$	\\
2040 &	$-0.296	\pm	0.038$	&	$0.298	\pm	0.035$&	$0.014	\pm	0.037$	&	$0.228	\pm	0.038$	&	$-0.011	\pm	0.035$	&	$0.287	\pm	0.033$	\\
2060&	$-0.187	\pm	0.033$	&	$0.314	\pm	0.030$&	$0.010	\pm	0.031$	&	$0.231	\pm	0.032$	&	$-0.021	\pm	0.032$	&	$0.280	\pm	0.028$	\\
2080&	$-0.150	\pm	0.029$	&	$0.331	\pm	0.027$&	$0.006	\pm	0.028$	&	$0.234	\pm	0.029$	&	$-0.030	\pm	0.028$	&	$0.272	\pm	0.025$	\\
2100	&	$-0.173	\pm	0.039$	&	$0.257	\pm	0.034$&	$0.005	\pm	0.038$	&	$0.231	\pm	0.036$	&	$-0.037	\pm	0.038$	&	$0.277	\pm	0.030$	\\
2120	&	$-0.109	\pm	0.029$	&	$0.376	\pm	0.025$&	$0.010	\pm	0.027$	&	$0.246	\pm	0.028$	&	$-0.039	\pm	0.027$	&	$0.277	\pm	0.024$	\\
2140&	$-0.128	\pm	0.035$	&	$0.372	\pm	0.032$&	$0.018	\pm	0.034$	&	$0.230	\pm	0.034$	&	$-0.039	\pm	0.033$	&	$0.252	\pm	0.029$	\\
2160&	$-0.098	\pm	0.091$	&	$0.439	 \pm	0.076$&	$0.023	\pm	0.091$	&	$0.271	\pm	0.075$	&	$-0.038	\pm	0.089$	&	$0.346	\pm	0.073$	\\
2180&	$-0.115	\pm	0.025$	&	$0.428	\pm	0.025$&	$0.025	\pm	0.025$	&	$0.208	\pm	0.025$	&	$-0.040	\pm	0.025$	&	$0.263	\pm	0.025$	\\
 \end{tabular}
\end{ruledtabular}
\end{center}
\label{amplitude:P1_1}
\end{table*}

\begin{table*}[bhtp]
\addtocounter{table}{-1}
\caption{Cont'd.}
\begin{center}
\begin{ruledtabular}
\begin{tabular}{ccccccc}
\multicolumn{1}{c}{$W_{0}$} & \multicolumn{2}{c}{$ D_{13}$} & \multicolumn{2}{c}{$ D_{15}$} & \multicolumn{2}{c}{$ F_{15}$}\\ \cline{2-7}
\multicolumn{1}{c}{$\rm (MeV)$} & \multicolumn{1}{c}{ Re({$ D_{13}$})} & \multicolumn{1}{c}{ Im({$ D_{13}$})} & \multicolumn{1}{c}{ Re({$ D_{15}$})} & \multicolumn{1}{c}{ Im({$ D_{15}$})} & \multicolumn{1}{c}{ Re({$ F_{15}$})} & \multicolumn{1}{c}{ Im({$ F_{15}$})}\\
\hline
1480	&	$0.000	\pm	0.025$	&	$0.000	\pm	0.025$	&	$0.004	\pm	0.033$	&	$0.000	\pm	0.018$&	$0.000	\pm	0.025$	&	$0.000	\pm	0.025$	\\
1500	&	$0.000	\pm	0.025$	&	$0.000	\pm	0.025$	&	$0.008	\pm	0.018$	&	$0.000	\pm	0.009$&	$0.000	\pm	0.025$	&	$0.000	\pm	0.025$	\\
1520	&	$0.000	\pm	0.025$	&	$0.000	\pm	0.025$	&	$0.015	\pm	0.031$	&	$0.001	\pm	0.014$&	$0.000	\pm	0.025$	&	$0.000	\pm	0.025$	\\
1540	&	$0.000	\pm	0.025$	&	$0.000	\pm	0.025$	&	$0.024	\pm	0.020$	&	$0.001	\pm	0.011$&	$0.000	\pm	0.025$	&	$0.000	\pm	0.025$	\\
1560	&	$0.002	\pm	0.023$	&	$0.014	\pm	0.022$	&	$0.035	\pm	0.018$	&	$0.002	\pm	0.016$&	$0.000	\pm	0.025$	&	$0.000	\pm	0.025$	\\
1580	&	$0.007	\pm	0.026$	&	$0.023	\pm	0.024$	&	$0.048	\pm	0.023$	&	$0.005	\pm	0.018$ &	$0.000	\pm	0.025$	&	$0.000	\pm	0.025$	\\
\hline \end{tabular}
2020	&	$0.038	\pm	0.037$	&	$0.141	\pm	0.035$	&	$0.012	\pm	0.032$	&	$0.116	\pm	0.035$&	$0.029	\pm	0.029$	&	$0.057	\pm	0.035$	\\
2040	&	$0.037	\pm	0.037$	&	$0.154	\pm	0.034$	&	$-0.007	\pm	0.033$	&	$0.132	\pm	0.034$&	$0.024	\pm	0.028$	&	$0.048	\pm	0.034$	\\
2060	&	$0.037	\pm	0.031$	&	$0.134	\pm	0.030$	&	$-0.045	\pm	0.027$	&	$0.123	\pm	0.029$&	$0.015	\pm	0.026$	&	$0.100	\pm	0.030$	\\
2080	&	$0.037	\pm	0.028$	&	$0.136	\pm	0.027$	&	$-0.041	\pm	0.025$	&	$0.109	\pm	0.026$&	$0.003	\pm	0.025$	&	$0.105	\pm	0.027$	\\
2100	&	$0.038	\pm	0.037$	&	$0.101	\pm	0.035$	&	$-0.037	\pm	0.033$	&	$0.120	\pm	0.031$&	$-0.009	\pm	0.033$	&	$0.104	\pm	0.033$	\\
2120	&	$0.040	\pm	0.026$	&	$0.144	\pm	0.026$	&	$-0.034	\pm	0.023$	&	$0.121	\pm	0.024$&	$-0.020	\pm	0.023$	&	$0.110	\pm	0.025$	\\
2140	&	$0.044	\pm	0.032$	&	$0.140	\pm	0.032$	&	$-0.033	\pm	0.030$	&	$0.126	\pm	0.030$&	$-0.030	\pm	0.029$	&	$0.109	\pm	0.031$	\\
2160	&	$0.048	\pm	0.090$	&	$0.093	\pm	0.076$	&	$-0.037	\pm	0.090$	&	$0.141	\pm	0.070$&	$-0.037	\pm	0.087$	&	$0.121	\pm	0.074$	\\
2180	&	$0.050	\pm	0.025$	&	$0.137	\pm	0.025$	&	$-0.046	\pm	0.025$	&	$0.141	\pm	0.025$&	$-0.043	\pm	0.025$	&	$0.083	\pm	0.025$	\\
 \end{tabular}
\end{ruledtabular}
\end{center}
\label{amplitude:D1_1}
\end{table*}

\begin{table*}[bhtp]
\addtocounter{table}{-1}
\caption{Cont'd.}
\begin{center}
\begin{ruledtabular}
\begin{tabular}{ccccc}
\multicolumn{1}{c}{$W_{0}$} & \multicolumn{2}{c}{$ F_{17}$} & \multicolumn{2}{c}{$ G_{17}$}\\
\multicolumn{1}{c}{$\rm (MeV)$} & \multicolumn{1}{c}{ Re({$ F_{17}$})} & \multicolumn{1}{c}{ Im({$ F_{17}$})} & \multicolumn{1}{c}{ Re({$ G_{17}$})} & \multicolumn{1}{c}{ Im({$ G_{17}$})}\\
\hline
1480	&	$0.000	\pm	0.031$	&	$0.000	\pm	0.014$&	0.000			&	0.000				\\
1500	&	$0.000	\pm	0.013$	&	$0.000	\pm	0.008$&	0.000			&	0.000				\\
1520	&	$0.001	\pm	0.026$	&	$0.000	\pm	0.015$&	0.000			&	0.000				\\
1540&	$0.002	\pm	0.015$	&	$0.000	\pm	0.009$&	0.000			&	0.000				\\
1560&	$0.003	\pm	0.016$	&	$0.000	\pm	0.008$&	0.001			&	0.000		          	\\
1580&	$0.005	\pm	0.020$	&	$0.000	\pm	0.009$&	0.001			&	0.000				\\
\hline \end{tabular}
2020&	$0.094	\pm	0.033$	&	$0.164	\pm	0.031$&	0.041			&	0.036				\\
2040&	$0.055	\pm	0.032$	&	$0.177	\pm	0.030$&	0.046			&	0.043				\\
2060&	$0.031	\pm	0.029$	&	$0.177	\pm	0.026$&	0.056			&	0.052				\\
2080&	$0.009	\pm	0.026$	&	$0.202	\pm	0.023$&	0.068			&	0.053		          	\\
2100	&	$-0.007	\pm	0.035$	&	$0.168	\pm	0.029$&	0.029			&	0.105		          	\\
2120	&	$-0.016	\pm	0.024$	&	$0.155	\pm	0.022$&	0.058			&	0.093				\\
2140	&	$-0.022	\pm	0.030$	&	$0.139	\pm	0.028$&	0.045			&	0.117		          	\\
2160	&	$-0.025	\pm	0.087$	&	$0.187	\pm	0.050$&	0.033			&	0.185				\\
2180&	$-0.028	\pm	0.025$	&	$0.117	\pm	0.025$&	0.017			&	0.137				\\
\end{tabular}
\end{ruledtabular}
\end{center}
\label{amplitude:G1_1}
\end{table*}

\clearpage

\begin{table*}[bhtp]
\caption{Single-energy amplitudes for $\overline{K} N \rightarrow \pi \Lambda$.} 
\begin{center}
\begin{ruledtabular}
\begin{tabular}{ccccccc}
\multicolumn{1}{c}{$W_{0}$} & \multicolumn{2}{c}{$ S_{11}$} & \multicolumn{2}{c}{$ P_{11}$} & \multicolumn{2}{c}{$ P_{13}$} \\ \cline{2-7}
\multicolumn{1}{c}{$\rm (MeV)$} & \multicolumn{1}{c}{ Re({$ S_{11}$})} & \multicolumn{1}{c}{ Im({ $S_{11}$})} & \multicolumn{1}{c}{ Re({$ P_{11}$})} & \multicolumn{1}{c}{ Im({$ P_{11}$})} & \multicolumn{1}{c}{ Re({$ P_{13}$})} & \multicolumn{1}{c}{ Im({$ P_{13}$})}\\
\hline
1480	&	$-0.218	\pm	0.035$	&	$-0.108	\pm	0.035$&	$-0.055	\pm	0.035$	&	$-0.020	\pm	0.035$	&	$-0.071	\pm	0.035$	&	$-0.016	\pm	0.035$	\\
1500	&	$-0.209	\pm	0.035$	&	$-0.119	\pm	0.035$&	$-0.071	\pm	0.035$	&	$-0.031	\pm	0.035$	&	$-0.076	\pm	0.035$	&	$-0.019	\pm	0.035$	\\
1520	&	$-0.206	\pm	0.035$	&	$-0.121	\pm	0.035$&	$-0.094	\pm	0.035$	&	$-0.049	\pm	0.035$	&	$-0.076	\pm	0.035$	&	$-0.019	\pm	0.035$	\\
1540	&	$-0.237	\pm	0.023$	&	$-0.149	\pm	0.031$&	$-0.093	\pm	0.031$	&	$-0.098	\pm	0.031$	&	$-0.058	\pm	0.019$	&	$-0.019	\pm	0.025$	\\
1560&	$-0.239	\pm	0.025$	&	$-0.165	\pm	0.033$&	$-0.079	\pm	0.032$	&	$-0.060	\pm	0.029$	&	$-0.042	\pm	0.020$	&	$-0.016	\pm	0.029$	\\
1580	&	$-0.252	\pm	0.027$	&	$-0.216	\pm	0.031$&	$-0.042	\pm	0.031$	&	$-0.039	\pm	0.032$	&	$-0.049	\pm	0.021$	&	$-0.012	\pm	0.028$	\\
\hline \end{tabular}
2000	&	$-0.107	\pm	0.042$	&	$0.174	\pm	0.029$&	$-0.150	\pm	0.033$	&	$-0.027	\pm	0.037$	&	$0.030	\pm	0.033$	&	$0.037	\pm	0.024$	\\
2020	&	$-0.063	\pm	0.035$	&	$0.080	\pm	0.035$&	$-0.129	\pm	0.035$	&	$-0.052	\pm	0.035$	&	$0.079	\pm	0.035$	&	$0.038	\pm	0.035$	\\
2040	&	$-0.089	\pm	0.035$	&	$0.145	\pm	0.035$&	$-0.185	\pm	0.035$	&	$-0.048	\pm	0.035$	&	$0.041	\pm	0.035$	&	$0.038	\pm	0.035$	\\
2060	&	$-0.164	\pm	0.057$	&	$0.129	\pm	0.045$&	$-0.228	\pm	0.052$	&	$-0.044	\pm	0.053$	&	$0.050	\pm	0.047$	&	$0.039	\pm	0.042$	\\
2080	&	$-0.120	\pm	0.067$	&	$0.117	\pm	0.046$&	$-0.209	\pm	0.053$	&	$-0.039	\pm	0.060$	&	$0.047	\pm	0.052$	&	$0.039	\pm	0.048$	\\
2100	&	$-0.075	\pm	0.035$	&	$0.010	\pm	0.035$&	$-0.058	\pm	0.035$	&	$-0.036	\pm	0.035$	&	$0.046	\pm	0.035$	&	$0.038	\pm	0.035$	\\
2120	&	$-0.122	\pm	0.072$	&	$0.141	\pm	0.047$&	$-0.205	\pm	0.058$	&	$-0.038	\pm	0.065$	&	$0.046	\pm	0.045$	&	$0.037	\pm	0.049$	\\
2140	&	$-0.113	\pm	0.035$	&	$0.135	\pm	0.035$&	$-0.164	\pm	0.035$	&	$-0.040	\pm	0.035$	&	$0.045	\pm	0.035$	&	$0.034	\pm	0.035$	\\
2160	&	$0.040	\pm	0.038$	&	$0.090	\pm	0.035$&	$-0.170	\pm	0.041$	&	$-0.041	\pm	0.039$	&	$0.045	\pm	0.022$	&	$0.032	\pm	0.036$	\\
2180	&	$-0.116	\pm	0.035$	&	$0.135	\pm	0.035$&	$-0.164	\pm	0.035$	&	$-0.039	\pm	0.035$	&	$0.045	\pm	0.035$	&	$0.031	\pm	0.035$	\\
 \end{tabular}
 \end{ruledtabular}
\end{center}
\label{amplitude:P1_2}
\end{table*}

\begin{table*}[bhtp]
\caption{Summary of single-energy amplitudes for $\overline{K} N \rightarrow \pi \Lambda$.} 
\begin{center}
\begin{ruledtabular}
\begin{tabular}{ccccccc}
\multicolumn{1}{c}{$W_{0}$} & \multicolumn{2}{c}{$ D_{13}$} & \multicolumn{2}{c}{$ D_{15}$} & \multicolumn{2}{c}{$ F_{15}$}\\ \cline{2-7}
\multicolumn{1}{c}{$\rm (MeV)$} & \multicolumn{1}{c}{ Re({$ D_{13}$})} & \multicolumn{1}{c}{ Im({$ D_{13}$})} & \multicolumn{1}{c}{ Re({$ D_{15}$})} & \multicolumn{1}{c}{ Im({$ D_{15}$})} & \multicolumn{1}{c}{ Re({$ F_{15}$})} & \multicolumn{1}{c}{ Im({$ F_{15}$})}\\
\hline
1480	&	$0.000	\pm	0.035$	&	$0.000	\pm	0.035$	&	$-0.006	\pm	0.035$	&	$-0.001	\pm	0.035$&	$0.000	\pm	0.035$	&	$0.000	\pm	0.035$	\\
1500	&	$0.000	\pm	0.035$	&	$0.000	\pm	0.035$	&	$-0.010	\pm	0.035$	&	$-0.002	\pm	0.035$&	$0.000	\pm	0.035$	&	$0.000	\pm	0.035$	\\
1520	&	$0.000	\pm	0.035$	&	$0.000	\pm	0.035$	&	$-0.015	\pm	0.035$	&	$-0.003	\pm	0.035$&	$0.000	\pm	0.035$	&	$0.000	\pm	0.035$	\\
1540	&	$0.000	\pm	0.035$	&	$0.000	\pm	0.035$	&	$-0.021	\pm	0.013$	&	$-0.004	\pm	0.018$&	$0.000	\pm	0.035$	&	$0.000	\pm	0.035$	\\
1560	&	$-0.006	\pm	0.018$	&	$0.005	\pm	0.018$	&	$-0.029	\pm	0.012$	&	$-0.005	\pm	0.012$&	$0.000	\pm	0.035$	&	$0.000	\pm	0.035$	\\
1580	&	$-0.008	\pm	0.016$	&	$0.010	\pm	0.018$	&	$-0.038	\pm	0.014$	&	$-0.008	\pm	0.011$&	$0.000	\pm	0.035$	&	$0.000	\pm	0.035$	\\
\hline \end{tabular}
2000	&	$-0.073	\pm	0.029$	&	$-0.014	\pm	0.027$	&	$0.094	\pm	0.027$	&	$-0.028	\pm	0.024$ &	$0.006	\pm	0.023$	&	$0.015	\pm	0.019$	\\
2020	&	$-0.052	\pm	0.035$	&	$-0.015	\pm	0.035$	&	$0.121	\pm	0.035$	&	$0.013	\pm	0.035$ &	$-0.002	\pm	0.035$	&	$0.020	\pm	0.035$	\\
2040	&	$-0.036	\pm	0.035$	&	$-0.016	\pm	0.035$	&	$0.087	\pm	0.035$	&	$-0.005	\pm	0.035$ &	$-0.011	\pm	0.035$	&	$0.022	\pm	0.035$	\\
2060	&	$-0.065	\pm	0.051$	&	$-0.018	\pm	0.041$	&	$0.061	\pm	0.044$	&	$-0.050	\pm	0.038$&	$-0.021	\pm	0.044$	&	$0.021	\pm	0.031$	\\
2080	&	$-0.095	\pm	0.045$	&	$-0.019	\pm	0.043$	&	$0.083	\pm	0.048$	&	$-0.049	\pm	0.042$&	$-0.030	\pm	0.041$	&	$0.017	\pm	0.032$	\\
2100	&	$-0.120	\pm	0.035$	&	$-0.020	\pm	0.035$	&	$0.189	\pm	0.035$	&	$-0.048	\pm	0.035$&	$-0.034	\pm	0.035$	&	$0.014	\pm	0.035$	\\
2120	&	$-0.079	\pm	0.048$	&	$-0.019	\pm	0.051$	&	$0.054	\pm	0.044$	&	$-0.046	\pm	0.043$&	$-0.035	\pm	0.045$	&	$0.013	\pm	0.037$	\\
2140	&	$-0.081	\pm	0.035$	&	$-0.018	\pm	0.035$	&	$0.070	\pm	0.035$	&	$-0.044	\pm	0.035$&	$-0.036	\pm	0.035$	&	$0.012	\pm	0.035$	\\
2160	&	$-0.191	\pm	0.022$	&	$-0.017	\pm	0.038$	&	$-0.058	\pm	0.023$	&	$-0.043	\pm	0.030$&	$-0.037	\pm	0.022$	&	$0.009	\pm	0.026$	\\
2180	&	$-0.080	\pm	0.035$	&	$-0.017	\pm	0.035$	&	$0.071	\pm	0.035$	&	$-0.041	\pm	0.035$&	$-0.039	\pm	0.035$	&	$0.004	\pm	0.035$	\\
 \end{tabular}
 \end{ruledtabular}
\end{center}
\label{amplitude:D1_2}
\end{table*}

\begin{table*}[bhtp]
\addtocounter{table}{-1}
\caption{Cont'd.} 
\begin{center}
\begin{ruledtabular}
\begin{tabular}{ccccc}
\multicolumn{1}{c}{$W_{0}$} & \multicolumn{2}{c}{$ F_{17}$}& \multicolumn{2}{c}{$ G_{17}$}\\
\multicolumn{1}{c}{$\rm (MeV)$} & \multicolumn{1}{c}{ Re({$ F_{17}$})} & \multicolumn{1}{c}{ Im({$ F_{17}$})} & \multicolumn{1}{c}{ Re({$ G_{17}$})} & \multicolumn{1}{c}{ Im({$ G_{17}$})}\\
\hline
1480&	$0.000	\pm	0.035$	&	$0.000	\pm	0.035$	&	$0.000$			&	0.000 		\\
1500&	$0.000	\pm	0.035$	&	$0.000	\pm	0.035$	&	$0.000$ 			&	0.000 		\\
1520	&	$0.001	\pm	0.035$	&	$0.000	\pm	0.035$	&	$0.000$ 			&	0.000 		\\
1540&	$0.001	\pm	0.011$	&	$0.000	\pm	0.014$	&	$0.000$ 			&	0.000 		\\
1560	&	$0.002	\pm	0.005$	&	$0.000	\pm	0.007$	&	$0.000$ 			&	0.000 		\\
1580&	$0.003	\pm	0.005$	&	$0.000	\pm	0.007$	&	$0.000$			&	0.000 		\\
\hline \end{tabular}
2020	&	$-0.021	\pm	0.035$	&	$0.182	\pm	0.035$	&	$-0.011$ 			&	-0.007 			\\
2040	&	$0.018	\pm	0.035$	&	$0.170	\pm	0.035$	&	$-0.012$ 			&	-0.008 			\\
2060&	$-0.008	\pm	0.044$	&	$0.147	\pm	0.029$	&	$-0.013$ 			&	-0.010 			\\
2080	&	$-0.026	\pm	0.051$	&	$0.148	\pm	0.032$	&	$-0.014$ 			&	-0.012 			\\
2100	&	$-0.036	\pm	0.035$	&	$0.159	\pm	0.035$	&	$-0.015$ 			&	-0.015 			\\
2120	&	$-0.040	\pm	0.045$	&	$0.094	\pm	0.033$	&	$-0.015$ 			&	-0.017 			\\
2140&	$-0.042	\pm	0.035$	&	$0.102	\pm	0.035$	&	$-0.014$ 			&	-0.020 			\\
2160	&	$-0.042	\pm	0.020$	&	$0.059	\pm	0.021$	&	$-0.013$			&	-0.022 			\\
2180	&	$-0.042	\pm	0.035$	&	$0.084	\pm	0.035$	&	$-0.012$ 			&	-0.025 			\\
 \end{tabular}
 \end{ruledtabular}
\end{center}
\label{amplitude:G1_2}
\end{table*}


\begin{table*}[bhtp]
\caption{Summary of single-energy amplitudes for $\overline{K} N \rightarrow \pi \Sigma$.} 
\begin{center}
\begin{ruledtabular}
\begin{tabular}{ccccccc}
\multicolumn{1}{c}{$W_{0}$} & \multicolumn{2}{c}{$ S_{01}$}& \multicolumn{2}{c}{$ P_{01}$} & \multicolumn{2}{c}{$ P_{03}$} \\ \cline{2-7}
\multicolumn{1}{c}{$\rm (MeV)$} & \multicolumn{1}{c}{ Re({$ S_{01}$})} & \multicolumn{1}{c}{ Im({$ S_{01}$})}& \multicolumn{1}{c}{ Re({$ P_{01}$})} & \multicolumn{1}{c}{ Im({$ P_{01}$})} & \multicolumn{1}{c}{ Re({$ P_{03}$})} & \multicolumn{1}{c}{ Im({$ P_{03}$})}\\
\hline
1480	&	$-0.308    \pm	0.035$ 	&	$0.316 	 \pm	0.03$&	$-0.036 	 \pm	0.035$ 	&	$-0.019 	 \pm	0.035$ 	&	$0.058 	 \pm	0.035$ 	&	$-0.001 	 \pm	0.035 $	\\
1500	&	$-0.285 	 \pm	0.035$ 	&	$0.321 	 \pm	0.035$&	$-0.064 	 \pm	0.035 $	&	$-0.027 	 \pm	0.035 $	&	$0.062 	 \pm	0.035 $	&	$0.001 	 \pm	0.035 $	\\
1520	&	$-0.268 	 \pm	0.035$ 	&	$0.318 	 \pm	0.035$&	$-0.104 	 \pm	0.035 $	&	$-0.050 	 \pm	0.035 $	&	$0.063 	 \pm	0.035 $	&	$0.004 	 \pm	0.035 $	\\
1540	&	$-0.206 	 \pm	0.045$ 	&	$0.224 	 \pm	0.040$&	$-0.147  	 \pm	0.039 $	&	$-0.230 	 \pm	0.044 $	&	$0.147 	 \pm	0.034$ 	&	$0.006 	 \pm	0.041 $	\\
1560&	$-0.280 	 \pm	0.031$ 	&	$0.282 	 \pm	0.032$&	$-0.131 	 \pm	0.031 $	&	$-0.133 	 \pm	0.037 $	&	$0.119 	 \pm	0.027$ 	&	$0.009 	 \pm	0.030 $	\\
1580&	$-0.242 	 \pm	0.030$ 	&	$0.294 	 \pm	0.028$&	$-0.095 	 \pm	0.028 $	&	$-0.195 	 \pm	0.033 $	&	$0.121 	 \pm	0.025$ 	&	$0.012 	 \pm	0.027 $	\\
\hline \end{tabular}
2000&	$0.099 	 \pm	0.062$ 	&	$-0.022 	 \pm	0.054$&	$0.001 	 \pm	0.051 $	&	$-0.008 	 \pm	0.059 $	&	$0.176 	 \pm	0.056 $	&	$0.051 	 \pm	0.052 $	\\
2020&	$0.076 	 \pm	0.033$ 	&	$-0.017 	 \pm	0.029$ &	$0.029 	 \pm	0.029 $	&	$0.004 	 \pm	0.032 $	&	$0.160 	 \pm	0.029 $	&	$0.086 	 \pm	0.029 $	\\
2040	&	$0.064 	 \pm	0.035$ 	&	$-0.013 	 \pm	0.035$ &	$0.006 	 \pm	0.035 $	&	$0.011 	 \pm	0.035 $	&	$0.131 	 \pm	0.035 $	&	$0.098 	 \pm	0.035 $	\\
2060	&	$0.092 	 \pm	0.045$ 	&	$-0.006 	 \pm	0.047$&	$-0.010 	 \pm	0.043 $	&	$0.016 	 \pm	0.048 $	&	$0.159 	 \pm	0.040 $	&	$0.158 	 \pm	0.041 $	\\
2080&	$0.116 	 \pm	0.058$ 	&	$-0.002 	 \pm	0.058$&	$-0.028 	 \pm	0.055 $	&	$0.019 	 \pm	0.059 $	&	$0.120 	 \pm	0.049 $	&	$0.089 	 \pm	0.048 $	\\
2100	&	$0.031 	 \pm	0.035 $	&	$0.002 	 \pm	0.035$&	$-0.044 	 \pm	0.035 $	&	$0.021 	 \pm	0.035 $	&	$0.204 	 \pm	0.035 $	&	$0.176 	 \pm	0.035 $	\\
2120	&	$0.053 	 \pm	0.043$ 	&	$0.007 	 \pm	0.041$&	$-0.024 	 \pm	0.044 $	&	$0.025 	 \pm	0.044 $	&	$0.072 	 \pm	0.038 $	&	$0.120 	 \pm	0.034 $	\\
2140	&	$0.050 	 \pm	0.035 $	&	$0.012 	 \pm	0.035$&	$-0.065 	 \pm	0.035 $	&	$0.028 	 \pm	0.035 $	&	$0.104 	 \pm	0.035 $	&	$0.193 	 \pm	0.035 $	\\
\end{tabular}
\end{ruledtabular}
\end{center}
\label{amplitude:P0_3}
\end{table*}


\begin{table*}[bhtp]
\caption{Cont'd.}
\begin{center}
\begin{ruledtabular}
\begin{tabular}{ccccccc}
\multicolumn{1}{c}{$W_{0}$} & \multicolumn{2}{c}{$ D_{03}$} & \multicolumn{2}{c}{$ D_{05}$} & \multicolumn{2}{c}{$ F_{05}$} \\ \cline{2-7}
\multicolumn{1}{c}{$\rm (MeV)$} & \multicolumn{1}{c}{ Re({$ D_{03}$})} & \multicolumn{1}{c}{ Im({$ D_{03}$})} & \multicolumn{1}{c}{ Re({$ D_{05}$})} & \multicolumn{1}{c}{ Im({$ D_{05}$})} & \multicolumn{1}{c}{ Re({$ F_{05}$})} & \multicolumn{1}{c}{ Im({$ F_{05}$})}\\
\hline
1480	&	$0.038 	\pm	0.035$ 	&	$-0.005 	\pm	0.035$ 	&	$0.003 	\pm	0.035$ 	&	$0.002 	\pm	0.035$	&	$0.000 	\pm	0.035$ 	&	$0.000 	\pm	0.035$  	\\
1500	&	$0.113 	\pm	0.035$ 	&	$0.023 	\pm	0.035$ 	&	$0.002 	\pm	0.035 $	&	$0.004 	\pm	0.035 $	&	$0.000 	\pm	0.035 $	&	$0.000 	\pm	0.035 $\\
1520	&	$-0.025 	\pm	0.035$ 	&	$0.445 	\pm	0.035$ 	&	$0.000 	\pm	0.035$ 	&	$0.006 	\pm	0.035$	 &	$-0.001 	\pm	0.035 $	&	$0.000 	\pm	0.035$ 	\\
1540	&	$-0.213 	\pm	0.029$ 	&	$0.098 	\pm	0.038$ 	&	$-0.003 	\pm	0.028$ 	&	$0.006 	\pm	0.028 $	&	$-0.002 	\pm	0.031 $	&	$0.000 	\pm	0.035$ \\
1560	&	$-0.109 	\pm	0.027$ 	&	$0.028 	\pm	0.034$ 	&	$-0.006 	\pm	0.025$ 	&	$0.005 	\pm	0.026 $	&	$-0.003 	\pm	0.023 $	&	$0.000 	\pm	0.029$\\
1580	&	$-0.117 	\pm	0.024$ 	&	$0.011 	\pm	0.031$ 	&	$-0.009 	\pm	0.022$ 	&	$0.004 	\pm	0.023 $	&	$-0.004 	\pm	0.021 $	&	$0.000 	\pm	0.024$\\
\hline \end{tabular}
2000	&	$0.113 	\pm	0.049$ 	&	$0.084 	\pm	0.058 $	&	$0.090 	\pm	0.052 $	&	$-0.020 	\pm	0.052 $	&	$0.143 	\pm	0.045 $	&	$-0.005 	\pm	0.047$\\
2020	&	$0.103 	\pm	0.031$ 	&	$0.065 	\pm	0.032 $	&	$0.081 	\pm	0.028 $	&	$-0.016 	\pm	0.026 $	&	$0.155 	\pm	0.027 $	&	$0.004 	\pm	0.027$\\
2040	&	$0.105 	\pm	0.035$ 	&	$0.041 	\pm	0.035 $	&	$0.043 	\pm	0.035 $	&	$-0.013 	\pm	0.035 $	&	$0.126 	\pm	0.035 $	&	$0.011 	\pm	0.035$\\
2060	&	$0.105 	\pm	0.043$ 	&	$0.054 	\pm	0.046 $	&	$0.043 	\pm	0.037 $	&	$-0.011 	\pm	0.039 $	&	$0.178 	\pm	0.036 $	&	$0.017 	\pm	0.040$\\
2080	&	$0.085 	\pm	0.052$ 	&	$0.084 	\pm	0.054 $	&	$0.049 	\pm	0.048 $	&	$-0.009 	\pm	0.046 $	&	$0.126 	\pm	0.048 $	&	$0.021 	\pm	0.046$\\
2100	&	$-0.040 	\pm	0.035$ 	&	$0.134 	\pm	0.035 $	&	$0.046 	\pm	0.035 $	&	$-0.007 	\pm	0.035 $	&	$0.171 	\pm	0.035 $	&	$0.023 	\pm	0.035$\\
2120	&	$0.072 	\pm	0.042$ 	&	$0.046 	\pm	0.038 $	&	$0.044 	\pm	0.037 $	&	$-0.005 	\pm	0.033 $	&	$0.102 	\pm	0.036 $	&	$0.026 	\pm	0.030$\\
2140	&	$0.043 	\pm	0.035$ 	&	$0.132 	\pm	0.035 $	&	$0.042 	\pm	0.035 $	&	$-0.004 	\pm	0.035 $	&	$0.080 	\pm	0.035 $	&	$0.027 	\pm	0.035$\\
2160	&	$0.036 	\pm	0.042$ 	&	$0.056 	\pm	0.041 $	&	$0.040 	\pm	0.040 $	&	$-0.003 	\pm	0.037 $	&	$0.120 	\pm	0.036 $	&	$0.027 	\pm	0.037$\\
2180	&	$0.028 	\pm	0.035$ 	&	$0.134 	\pm	0.035 $	&	$0.038 	\pm	0.035 $	&	$-0.002 	\pm	0.035 $	&	$0.069 	\pm	0.035 $	&	$0.027 	\pm	0.035$\\
 \end{tabular}
 \end{ruledtabular}
\end{center}
\label{amplitude:D0_3}
\end{table*}

\begin{table*}[bhtp]
\addtocounter{table}{-1}
\caption{Cont'd.}
\begin{center}
\begin{ruledtabular}
\begin{tabular}{ccccccc}
\multicolumn{1}{c}{$W_{0}$} & \multicolumn{2}{c}{$ F_{07}$}& \multicolumn{2}{c}{$ G_{07}$} \\
\multicolumn{1}{c}{$\rm (MeV)$} & \multicolumn{1}{c}{ Re({$ F_{07}$})} && \multicolumn{1}{c}{ Re({$ F_{07}$})} & \multicolumn{1}{c}{ Im({$ G_{07}$})} \\
\hline
1480&	$0.000 	\pm	0.035 $	&	$0.000 	\pm	0.035$	&	$0.000 	\pm	0.035$ 	&	$0.000 	\pm	0.035$ 		\\
1500	&	$0.000 	\pm	0.035 $	&	$0.000 	\pm	0.035$	&	$0.000 	\pm	0.035 $	&	$0.000 	\pm	0.035 $		\\
1520	&	$0.000 	\pm	0.035 $	&	$0.000 	\pm	0.035$	&	$0.000 	\pm	0.035 $	&	$0.000 	\pm	0.035 $		\\
1540	&	$0.000 	\pm	0.022 $	&	$0.000 	\pm	0.027 $	&	$0.000 	\pm	0.035 $	&	$0.000 	\pm	0.035 $	         \\
1560	&	$0.000 	\pm	0.011 $	&	$0.000 	\pm	0.015 $	&	$0.000 	\pm	0.035 $	&	$0.000 	\pm	0.035 $	         \\
1580	&	$0.000 	\pm	0.011 $	&	$0.000 	\pm	0.014 $	&	$0.000 	\pm	0.035 $	&	$0.000 	\pm	0.035 $	         \\
\hline \end{tabular}
2020	&	$0.088 	\pm	0.027 $	&	$0.045 	\pm	0.019 $	&	$0.000 	\pm	0.035 $	&	$0.000 	\pm	0.035 $	\\
2040	&	$0.111 	\pm	0.035 $	&	$0.053 	\pm	0.035 $	&	$0.000 	\pm	0.035 $	&	$0.000 	\pm	0.035 $	\\
2060	&	$0.069 	\pm	0.038 $	&	$0.073 	\pm	0.031 $	&	$0.000 	\pm	0.035 $	&	$0.000 	\pm	0.035 $	\\
2080	&	$0.115 	\pm	0.049 $	&	$0.130 	\pm	0.041 $	&	$0.000 	\pm	0.035 $	&	$0.000 	\pm	0.035 $     \\
2100	&	$0.145 	\pm	0.035 $	&	$0.118 	\pm	0.035 $	&	$0.000 	\pm	0.035 $	&	$0.000 	\pm	0.035 $	\\
2120	&	$0.100 	\pm	0.038 $	&	$0.126 	\pm	0.025 $	&	$0.000 	\pm	0.035 $	&	$0.000 	\pm	0.035 $	\\
2140	&	$0.042 	\pm	0.035 $	&	$0.101 	\pm	0.035 $	&	$0.000 	\pm	0.035 $	&	$0.000 	\pm	0.035 $	\\
2160	&	$0.036 	\pm	0.035 $	&	$0.152 	\pm	0.032 $	&	$0.000 	\pm	0.035 $	&	$0.000 	\pm	0.035 $	\\
2180	&	$0.033 	\pm	0.035 $	&	$0.094 	\pm	0.035 $	&	$0.000 	\pm	0.035 $	&	$0.000 	\pm	0.035 $	\\
 \end{tabular}
 \end{ruledtabular}
\end{center}
\label{amplitude:G0_3}
\end{table*}

\begin{table*}[bhtp]
\caption{Summary of single-energy amplitudes for $\overline{K} N \rightarrow \pi \Sigma$.} 
\begin{center}
\begin{ruledtabular}
\begin{tabular}{ccccccc}
\multicolumn{1}{c}{$W_{0}$} & \multicolumn{2}{c}{$ S_{11}$} & \multicolumn{2}{c}{$ P_{11}$} & \multicolumn{2}{c}{$ P_{13}$} \\ \cline{2-7}
\multicolumn{1}{c}{$\rm (MeV)$} & \multicolumn{1}{c}{ Re({$ S_{11}$})} & \multicolumn{1}{c}{ Im({$ S_{11}$})}& \multicolumn{1}{c}{ Re({$ P_{11}$})} & \multicolumn{1}{c}{ Im({$ P_{11}$})} & \multicolumn{1}{c}{ Re({$ P_{13}$})} & \multicolumn{1}{c}{ Im({$ P_{13}$})}\\
\hline
1480	&	$0.188 	\pm	0.035$ 	&	$0.282 	\pm	0.035$	&	$0.001 	\pm	0.035$ 	&	$-0.012 	\pm	0.035$ 	&	$0.060 	\pm	0.035 $	&	$-0.011 	\pm	0.035$ 	\\
1500	&	$0.174 	\pm	0.035 $	&	$0.290 	\pm	0.035$	&	$-0.010 	\pm	0.035 $	&	$-0.015 	\pm	0.035 $	&	$0.068 	\pm	0.035 $	&	$-0.014 	\pm	0.035 $	\\
1520&	$0.170 	\pm	0.035 $	&	$0.288 	\pm	0.035$	&	$-0.027 	\pm	0.035 $	&	$-0.022 	\pm	0.035 $	&	$0.070 	\pm	0.035 $	&	$-0.017 	\pm	0.035 $	\\
1540	&	$0.151 	\pm	0.039 $	&	$0.316 	\pm	0.026$	&	$-0.041 	\pm	0.039 $	&	$-0.032 	\pm	0.041 $	&	$0.035 	\pm	0.033 $	&	$-0.019 	\pm	0.034 $	\\
1560	&	$0.168 	\pm	0.027 $	&	$0.283 	\pm	0.025$	&	$-0.049 	\pm	0.029 $	&	$-0.041 	\pm	0.037 $	&	$0.066 	\pm	0.023 $	&	$-0.021 	\pm	0.027 $	\\
1580 &	$0.190 	\pm	0.028 $	&	$0.332 	\pm	0.022$	&	$0.006 	\pm	0.027 $	&	$-0.077 	\pm	0.033 $	&	$0.034 	\pm	0.023 $	&	$-0.023 	\pm	0.026 $	\\
\hline \end{tabular}
2000	&	$-0.044 	\pm	0.059 $	&	$-0.010 	\pm	0.050 $	&	$0.038 	\pm	0.044 $	&	$-0.001 	\pm	0.056 $	&	$0.145 	\pm	0.050 $	&	$-0.037 	\pm	0.045 $	\\
2020	&	$-0.035 	\pm	0.031 $	&	$-0.041 	\pm	0.026$	&	$0.038 	\pm	0.026 $	&	$0.001 	\pm	0.029 $	&	$0.081 	\pm	0.026 $	&	$-0.036 	\pm	0.026 $	\\
2040	&	$-0.023 	\pm	0.035 $	&	$-0.102 	\pm	0.035$	&	$0.037 	\pm	0.035 $	&	$0.003 	\pm	0.035 $	&	$0.113 	\pm	0.035 $	&	$-0.035 	\pm	0.035 $	\\
2060	&	$-0.011 	\pm	0.043 $	&	$-0.058 	\pm	0.046$	&	$0.033 	\pm	0.039 $	&	$0.005 	\pm	0.047 $	&	$0.052 	\pm	0.035 $	&	$-0.032 	\pm	0.035 $	\\
2080	&	$-0.001 	\pm	0.056 $	&	$-0.110 	\pm	0.054$	&	$0.030 	\pm	0.050 $	&	$0.008 	\pm	0.055 $	&	$0.055 	\pm	0.049 $	&	$-0.031 	\pm	0.043 $	\\
2100	&	$0.008 	\pm	0.035 $	&	$-0.051 	\pm	0.035$	&	$0.026 	\pm	0.035 $	&	$0.009 	\pm	0.035 $	&	$0.100 	\pm	0.035 $	&	$-0.030 	\pm	0.035 $	\\
2120	&	$0.015 	\pm	0.042 $	&	$-0.108 	\pm	0.035$	&	$0.019 	\pm	0.041 $	&	$0.010 	\pm	0.042 $	&	$0.067 	\pm	0.035 $	&	$-0.028 	\pm	0.029 $	\\
2140	&	$0.023 	\pm	0.035 $	&	$-0.096 	\pm	0.035$	&	$0.011 	\pm	0.035 $	&	$0.011 	\pm	0.035 $	&	$0.096 	\pm	0.035 $	&	$-0.026 	\pm	0.035 $	\\
\end{tabular}
\end{ruledtabular}
\end{center}
\label{amplitude:P1_3}
\end{table*}

\begin{table*}[bhtp]
\addtocounter{table}{-1}
\caption{Cont'd.}
\begin{center}
\begin{ruledtabular}
\begin{tabular}{ccccccc}
\multicolumn{1}{c}{$W_{0}$} & \multicolumn{2}{c}{$ D_{13}$} & \multicolumn{2}{c}{$ D_{15}$} & \multicolumn{2}{c}{$ F_{15}$}\\ \cline{2-7}
\multicolumn{1}{c}{$\rm (MeV)$} & \multicolumn{1}{c}{ Re({$ D_{13}$})} & \multicolumn{1}{c}{ Im({$ D_{13}$})} & \multicolumn{1}{c}{ Re({$ D_{15}$})} & \multicolumn{1}{c}{ Im({$ D_{15}$})}& \multicolumn{1}{c}{ Re({$ F_{15}$})} & \multicolumn{1}{c}{ Im({$ F_{15}$})}\\
\hline
1480	&	$0.000 	\pm	0.035$ 	&	$0.000 	\pm	0.035 $	&	$0.003 	\pm	0.035$ 	&	$-0.002 	\pm	0.035 $		&	$0.000 	\pm	0.035 $	&	$0.000 	\pm	0.035$\\
1500	&	$0.000 	\pm	0.035 $	&	$0.000 	\pm	0.035 $	&	$0.005 	\pm	0.035 $	&	$-0.003 	\pm	0.035 $		&	$0.000 	\pm	0.035 $	&	$0.000 	\pm	0.035$\\
1520	&	$0.000 	\pm	0.035 $	&	$0.000 	\pm	0.035 $	&	$0.007 	\pm	0.035 $	&	$-0.004 	\pm	0.035 $		&	$0.000 	\pm	0.035 $	&	$0.000 	\pm	0.035$\\
1540	&	$0.000 	\pm	0.035 $	&	$0.000 	\pm	0.035 $	&	$0.009 	\pm	0.028 $	&	$-0.006 	\pm	0.023 $		&	$0.000 	\pm	0.035 $	&	$0.000 	\pm	0.035$\\
1560	&	$0.006 	\pm	0.025 $	&	$0.001 	\pm	0.033 $	&	$0.012 	\pm	0.021 $	&	$-0.007 	\pm	0.024 $		&	$0.000 	\pm	0.035 $	&	$0.000 	\pm	0.035$\\
1580	&	$0.023 	\pm	0.024 $	&	$0.002 	\pm	0.031 $	&	$0.016 	\pm	0.021 $	&	$-0.009 	\pm	0.023 $		&	$0.000 	\pm	0.035 $	&	$0.000 	\pm	0.035$\\
\end{tabular}
2000	&	$-0.002 	\pm	0.044 $	&	$0.006 	\pm	0.052 $	&	$0.000 	\pm	0.042 $	&	$0.007 	\pm	0.045 $		&	$0.089 	\pm	0.041 $	&	$-0.085 	\pm	0.041$\\
2020	&	$0.001 	\pm	0.028 $	&	$0.007 	\pm	0.031 $	&	$0.005 	\pm	0.026 $	&	$0.003 	\pm	0.024 $		&	$0.122 	\pm	0.024 $	&	$-0.089 	\pm	0.025$\\
2040	&	$0.003 	\pm	0.035 $	&	$0.007 	\pm	0.035 $	&	$0.011 	\pm	0.035 $	&	$0.000 	\pm	0.035 $		&	$0.126 	\pm	0.035 $	&	$-0.038 	\pm	0.035$\\
2060	&	$0.003 	\pm	0.040 $	&	$0.007 	\pm	0.044 $	&	$0.018 	\pm	0.031 $	&	$-0.004 	\pm	0.034 $		&	$0.162 	\pm	0.032 $	&	$-0.022 	\pm	0.036$\\
2080	&	$0.003        \pm	0.049 $	&	$0.008 	\pm	0.051 $	&	$0.026 	\pm	0.040 $	&	$-0.006 	\pm	0.038 $		&	$0.105 	\pm	0.044 $	&	$-0.013 	\pm	0.040$\\
2100	&	$0.004 	\pm	0.035 $	&	$0.008 	\pm	0.035 $	&	$0.033 	\pm	0.035 $	&	$-0.006 	\pm	0.035 $		&	$0.089 	\pm	0.035 $	&	$-0.005 	\pm	0.035$ \\
2120	&	$0.007 	\pm	0.039 $	&	$0.007 	\pm	0.036 $	&	$0.040 	\pm	0.030 $	&	$-0.004 	\pm	0.025 $		&	$0.120 	\pm	0.034 $	&	$0.007 	\pm	0.029$ \\
2140	&	$0.009 	\pm	0.035 $	&	$0.006 	\pm	0.035 $	&	$0.049 	\pm	0.035 $	&	$-0.001 	\pm	0.035 $		&	$0.104 	\pm	0.035 $	&	$0.017 	\pm	0.035$ \\
\end{tabular}
\end{ruledtabular}
\end{center}
\label{amplitude:D1_3}
\end{table*}

\begin{table*}[bhtp]
\caption{Cont'd.}
\begin{center}
\begin{ruledtabular}
\begin{tabular}{ccccc}
\multicolumn{1}{c}{$W_{0}$} & \multicolumn{2}{c}{$ F_{17}$} & \multicolumn{2}{c}{$ G_{17}$} \\ \cline{2-5} 
\multicolumn{1}{c}{$\rm (MeV)$} & \multicolumn{1}{c}{ Re({$ F_{17}$})} & \multicolumn{1}{c}{ Im({$ F_{17}$})} & \multicolumn{1}{c}{ Re({$ G_{17}$})} & \multicolumn{1}{c}{ Im({$ G_{17}$})}\\ 
\hline
1480	&	$0.000 	\pm	0.035 $	&	$0.000 	\pm	0.035$ 	&	0.000 			&	0.000 			\\
1500&	$0.000 	\pm	0.035 $	&	$0.000 	\pm	0.035 $	&	0.000 			&	0.000 		\\
1520&	$0.000 	\pm	0.035 $	&	$0.000 	\pm	0.035 $	&	0.000 			&	0.000 			\\
1540	&	$0.000 	\pm	0.017 $	&	$0.000 	\pm	0.020$	&	0.000 			&	0.000 		\\
1560&	$0.000 	\pm	0.014 $	&	$0.000 	\pm	0.014$	&	$-0.001$ 			&	0.000 				\\
1580	&	$0.000 	\pm	0.015 $	&	$0.000 	\pm	0.014$	&	$-0.001$ 			&	0.000 			\\
\hline \end{tabular}
2020&	$-0.012 	\pm	0.023 $	&	$-0.082 	\pm	0.020$	&	0.050 			&	$-0.004$ 		\\
2040&	$0.010 	\pm	0.035 $	&	$-0.103 	\pm	0.035$	&	0.098 			&	$-0.001$ 		\\
2060	&	$0.026 	\pm	0.029 $	&	$-0.077 	\pm	0.030 $	&	0.070 			&	0.004 		\\
2080	&	$0.035 	\pm	0.037 $	&	$-0.073 	\pm	0.038$	&	0.079 			&	0.009 		\\
2100	&	$0.040 	\pm	0.035 $	&	$-0.026 	\pm	0.035$	&	0.057 			&	0.017         	\\
2120&	$0.041 	\pm	0.031 $	&	$-0.057 	\pm	0.026$	&	0.080 			&	0.026 		\\
2140	&	$0.041 	\pm	0.035 $	&	$-0.056 	\pm	0.035$	&	0.088 			&	0.037 		\\
2160	&	$0.039 	\pm	0.033 $	&	$-0.050 	\pm	0.026 $	&	0.078 			&	0.050         	\\
2180	&	$0.038 	\pm	0.035 $	&	$-0.045 	\pm	0.035$	&	0.090 			&	0.062 		\\
 \end{tabular}
\end{ruledtabular}
\end{center}
\label{amplitude:G1_3}
\end{table*} 

  \end{document}